\documentclass[useAMS,usenatbib,a4paper]{mn2e}

\pdfoutput=1

\newcommand{\lyat}{Ly$\alpha$}
\newcommand{\lya}{Ly$\alpha~$} 
\newcommand{\msun}{\ifmmode M_{\odot} \else M$_{\odot}$\fi}
\newcommand{\msunyr}{\ifmmode M_{\odot}.{\rm yr}^{-1} \else
M$_{\odot}$ .yr$^{-1}$\fi}
\newcommand{\kms}{\ifmmode {\rm km s}^{-1} \else km s$^{-1}$\fi}
\newcommand{\ergs}{\ifmmode {\rm erg s}^{-1} \else erg s$^{-1}$\fi}
\newcommand{\ergscm}{\ifmmode {\rm erg s}^{-1}{\rm cm}^{-2} \else erg s$^{-1}$ cm$^{-2}$\fi}
\newcommand{\sqarcm}{\ifmmode {\rm arcmin}^{2} \else arcmin$^{2} \:$\fi}
\newcommand{\hi}{\ifmmode {\textrm{H\textsc{i}}} \else H\textsc{i} \fi}
\newcommand{\hii}{\ifmmode {\textrm{H\textsc{ii}}} \else H\textsc{ii} \fi}

\newcommand{\affil}[1]{$^{\rm #1}$}

\usepackage{times}
\usepackage[pdftex]{graphicx}
\usepackage{epsfig}
\usepackage{amssymb}
\usepackage{natbib}
\usepackage[latin1]{inputenc}
\graphicspath{plots/}
\usepackage{amsmath}
\usepackage{fixltx2e}
\usepackage{array}
\usepackage{wrapfig}
\usepackage{relsize}
\usepackage{color}
\usepackage{lipsum}
\usepackage{subfig}
\usepackage[section] {placeins}
\usepackage[cyr]{aeguill}     
\usepackage[margin=2cm]{geometry} 
\usepackage{caption}
\usepackage{hhline}
\usepackage{multirow}
\usepackage{arydshln}
\usepackage{colortbl}
\usepackage[table]{xcolor}
\usepackage{afterpage}
\usepackage{ftnxtra}
\usepackage{dblfloatfix}
\usepackage[colorlinks=true,linkcolor=blue,citecolor=blue,urlcolor=blue]{hyperref}

\title[Lyman-$\alpha$ Emitters in the context of hierarchical galaxy formation]{Lyman-$\alpha$ Emitters in the context of hierarchical galaxy formation: predictions for VLT/MUSE surveys}
\author[Garel et al.]{{T. Garel\affil{1,2}, B. Guiderdoni\affil{2} and J. Blaizot\affil{2}}\\
\vspace{-0.3cm}
{\affil{1}\, Centre for Astrophysics and Supercomputing, Swinburne University of Technology, Hawthorn, Victoria 3122, Australia}\\
\vspace{-0.25cm}
{\affil{2}\, Centre de Recherche Astrophysique de Lyon, Universit\'e de Lyon, Universit\'e Lyon 1, CNRS, Observatoire de Lyon;}\\
\vspace{-0.3cm}
{\affil{}\,\,\, 9 avenue Charles Andr\'e, 69561 Saint-Genis Laval Cedex, France}}

\begin{document}

\pdfminorversion=5
\pdfobjcompresslevel=2

\pagerange{\pageref{firstpage}--\pageref{lastpage}} \pubyear{2015}

\maketitle
\label{firstpage}

\begin{abstract}
The VLT Multi Unit Spectroscopic Explorer (MUSE) integral-field spectrograph can detect \lya emitters (LAE) in the redshift range 2.8 $\lesssim$ $z$ $\lesssim$ 6.7 in a homogeneous way. Ongoing MUSE surveys will notably probe faint \lya sources that are usually missed by current narrow-band surveys. We provide quantitative predictions for a typical wedding-cake observing strategy with MUSE based on mock catalogs generated with a semi-analytic model of galaxy formation coupled to numerical \lya radiation transfer models in gas outflows. We expect $\approx$ 1500 bright LAEs ($F_{\rm Ly\alpha}\gtrsim10^{-17}$ erg s$^{-1}$ cm$^{-2}$) in a typical Shallow Field (SF) survey carried over $\approx$ 100 \sqarcm, and $\approx$ 2,000 sources as faint as $10^{-18}$ erg s$^{-1}$ cm$^{-2}$ in a Medium-Deep Field (MDF) survey over 10 \sqarcm. In a typical Deep Field (DF) survey of 1 \sqarcm, we predict that $\approx$ 500 extremely faint LAEs ($F_{\rm Ly\alpha} \gtrsim 4 \times 10^{-19}$ erg s$^{-1}$ cm$^{-2}$) will be found. Our results suggest that faint \lya sources contribute significantly to the cosmic \lya luminosity and SFR budget. While the host halos of bright LAEs at $z$ $\approx$ 3 and 6 have descendants with median masses of $2 \times 10^{12}$ and $5 \times 10^{13}$ \msun{} respectively, the faintest sources detectable by MUSE at these redshifts are predicted to reside in halos which evolve into typical sub-L$^{*}$ and L$^{*}$ galaxy halos at $z$ $=$ 0. We expect typical DF and MDF surveys to uncover the building blocks of Milky Way-like objects, even probing the bulk of the stellar mass content of LAEs located in their progenitor halos at $z$ $\approx$ 3.
\end{abstract}


\section{Introduction}
\label{sec:intro}
{\let\thefootnote\relax\footnotetext{Email: \href{mailto:thibault.garel@univ-lyon1.fr}{thibault.garel@univ-lyon1.fr}}}

Since the late nineties, the \lya emission line has become increasingly efficient at detecting high-redshift star-forming galaxies. Lyman-alpha emitters (LAE) are now commonly found up to a redshift of seven, allowing us to study the formation and evolution of galaxies in the early Universe. 
Most LAEs have been extensively probed in narrow-band (NB) imaging surveys \citep[e.g.][]{hu98, rhoads00,shima06,ouch08}, and blind spectroscopic searches have led to hundreds of detections, especially in the last years \citep{rauch08, cassata2011a, blanc2011a}. These observations have mainly put statistical constraints on the LAE population at $F_{\rm Ly\alpha} \gtrsim 10^{-17}$ erg s$^{-1}$ cm$^{-2}$ (e.g. \lya luminosity functions and clustering) and they tend to show that LAEs are slightly less massive, bluer and more metal-poor than the other well-studied high-redshift galaxy population, the Lyman-Break galaxies \citep{shapley2001a,shapley03,pentericci2007a,bouwens09}. However, the existing \lya data remains somewhat more inhomogeneous than that of dropout galaxies, due to the different selection methods used in various surveys, potential significant contamination and rather small statistics. 

The acquisition of large, homogeneous, spectroscopic samples of \lya emitting galaxies is one of the main objectives of the VLT Multi Unit Spectroscopic Explorer \citep[MUSE;][]{bacon06} which started to operate in 2014. The MUSE integral-field spectrograph, which has a field-of-view of 1 $\sqarcm$, will probe the \lya emission line from $z$ $\approx $ 2.8 to $z$ $\approx$ 6.7. MUSE has been optimised for performing deep field observations, and it will thus enable to detect very faint LAEs at high redshift.

A few tens of objects have been observed previously at $F_{\rm Ly\alpha} \gtrsim 10^{-18}$ erg s$^{-1}$ cm$^{-2}$ \citep{rauch08,cassata2011a,dressler2015a}. MUSE is expected to dramatically increase the statistics at these fluxes, and furthermore explore an uncharted territory with LAEs as faint as $\approx 4 \times10^{-19}$ \ergscm{} \citep[][]{bacon2010a}. 

These unprecedentedly low \lya detection limits will offer a glimpse of the population of dwarf star-forming galaxies in the early Universe, unveiling objects with star formation rates (SFRs) much lower than current LAE and LBG surveys. This will therefore provide fundamental knowledge on the properties of galaxies at high redshift that will put tight constraints on models of galaxy formation. As high redshift sources are the building blocks of local galaxies in the hierarchical merging scenario, these faint LAEs are natural candidates to be the progenitors of local late-type galaxies. MUSE will help constrain the abundance of the population of faint galaxies and their contribution to the global SFR density from $z$ $\approx$ 3 to $z$ $\approx$ 7, allowing us to investigate the mass assembly of our Galaxy.

Besides, in order to help refine the observing strategy for MUSE surveys and interpret forthcoming data, it is essential to develop theoretical tools able to predict the expected number counts as a function of \lya flux and redshift, and to quantify the effect of cosmic variance. Simple models have been developed in order to interpret the existing observational constraints at high-redshift. Using cosmological simulations, \citet{nag} explore a stochastic scenario, in which galaxies undergo a \lyat-bright phase of finite duration, and adjust the \lya luminosity functions at $z$ $\approx$ 3-6 assuming all \lya photons can escape the galaxy. It is however well-known that interstellar/circumgalactic gas kinematics and distribution strongly affect the \lya line profile and escape fraction \citep[$f_{\rm esc}$;][]{neufeld1990a,tenorio-tagle1999a,shapley03,mas-hesse2003a,steidel2010a}, so the complex radiative transfer (RT) of resonant \lya photons must be accounted for. While \citet{le-delliou2005a} adopted a simple, constant $f_{\rm esc}$ model to match the \lya LF \citep[see also][]{dayal08,nag}, more refined models were investigated to describe $f_{\rm esc}$ for various interstellar medium configurations (slab geometry, clumpy dust distribution, static/outflow phases, etc) using phenomenological recipes \citep[e.g.][]{haiman99,koba07,dayal10b,koba10,dayal2011a,shimizu2011a}.

Yet, the accurate treatment of the \lya RT in galaxies requires numerical Monte Carlo calculations, that can be performed as a post-processing step of hydrodynamical simulation runs \citep{laursen09,zheng2010a,yajima2012a,verhamme2012}. These are highly (CPU-)time-consuming and a trade-off must be found between the size of the galaxy sample and the need for sufficient spatial resolution at the galaxy scale, preventing the use of \lya RT algorithms on to statistical galaxy samples in high-resolution simulations. To bypass this issue, semi-analytic models or hydrodynamical simulations can be coupled with results of \lya RT experiments in idealised geometries, like a slab-like configuration \citep{forero-romero2011}, or the so-called shell model \citep{garel2012a,orsi2012a}. This method provides a very suitable alternative due to much smaller computing time requirements, although their description of galaxies is more idealised than in high-resolution hydrodynamical simulations.

Here, we use the model of \citet{garel2015a} which couples the GALICS hybrid of model of galaxy formation \citep{hatton} with a grid of numerical \lya RT calculations through gas outflows \citep{schaerer2011a}. Using this model, we create mock lightcones to make quantitative predictions for typical MUSE surveys of LAEs, and intend to assess the role of these objects in the hierarchical scenario of galaxy formation. Our paper is laid out as follows. In Section \ref{sec:model}, we describe our model and the mock catalogs of LAEs. Section \ref{sec:literature} gives a brief overview of the existing data sets of LAEs. In Section \ref{sec:predictions}, we present our predictions in terms of number counts, \lya luminosity and SFR budget that can be probed by typical MUSE surveys. In Section \ref{sec:desc-prog}, we investigate the descendant/progenitor link between high redshift LAEs and present day objects, and discuss our results on Section \ref{sec:discussion}. Finally, we give a summary in Section \ref{sec:summary}.

Throughout this paper, we assume the following set of cosmological parameters: $h=H_0/(100 \: {\rm km \: s}^{-1} {\rm \: Mpc}^{-1}) = 0.70$, $\Omega_\Lambda = 0.72$, $\Omega_{\rm m} = 0.28$, $\Omega_{\rm b} = 0.046$, and $\sigma_8 = 0.82$, consistent with the $WMAP$-5 results \citep{komatsu09}. All magnitudes are expressed in the AB system.
\vskip-4ex

\section{Semi-analytic modelling and mock catalogues}
\label{sec:model}

In this paper, we use mock catalogues of \lyat-emitting galaxies computed with the model set out in \citet{garel2015a} \citep[see also][]{garel2012a} which combines a hybrid approach for the formation of galaxies in the cosmological context with a simple model of \lya emission and transfer.

We describe the formation and evolution of galaxies with GALICS \citep[\textit{GALaxies In Cosmological Simulations};][]{hatton}. The GALICS hybrid model includes (i) the hierarchical growth of dark matter (DM) structures described by a $N$-body cosmological simulation, and (ii) semi-analytic prescriptions to model the evolution of the baryonic component within virialised dark matter halos. The GALICS version that we use is based on the original model of \citet{hatton} and subsequent updates presented in \citet{blaizot04}, \citet{lanzoni2005}, and \citet{cattaneo2006a} \citep[see also][]{garel2012a,garel2015a}. The output of GALICS is combined in post-processing with a \textit{shell} model \citep{verh08} which describes the radiative transfer of \lya photons through thin expanding shells of hydrogen gas homogeneously mixed with dust, used as a proxy for outflows triggered by supernovae. Below, we outline the main features of our model.
 
\subsection{Cosmological N-body simulation}
Our N-body simulation has been run with GADGET \citep[][]{gadget2} using 1024$^3$ dark matter (DM) particles in a cubic periodic (comoving) volume of 100$h^{-1}$ Mpc on a side. We assume a standard $\Lambda$CDM \textit{concordance} cosmology in agreement with the WMAP-5 data release \citep[][]{komatsu09}, which parameter values are given in Section \ref{sec:intro}. Halo identification is performed with a Friends-of-Friends algorithm \citep[FOF;][]{davis85} and we follow \citet{tweed09} to compute the merging histories of the DM halos. The FOF links together groups of particles with overdensity of $\sim$ 200 times the mean density (which translates into a linking-length $b$ of 0.2). Bound groups of $\geq$ 20 particles are then identified as halos \citep[see][for more details]{hatton}, hence the minimum halo mass we can resolve in our simulation is $M_{\rm halo}^{\rm min} = 2$ $\times$ 10$^9$ \msun.

\subsection{Baryonic prescriptions}

\begin{figure*}
\vskip-2ex
\hskip-10ex
\includegraphics[width=18.2cm,height=5.9cm]{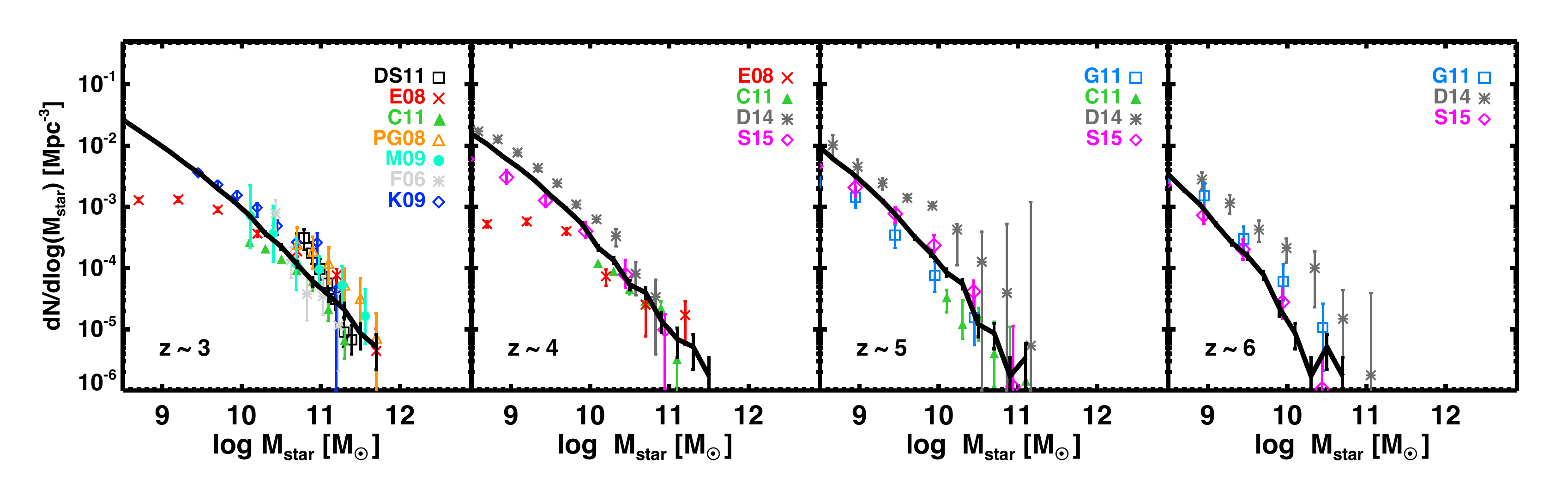}
\vskip-3ex
\caption{Stellar mass functions at $z$ $\approx$ 3, 4, 5, and 6. The solid black line with Poisson error bars corresponds to our model and the symbols are observational estimates (converted to our initial mass function) from \citet[][DS11]{dominguez-sanchez2011a}, \citet[][E08]{elsner2008a}, \citet[][C11]{caputi2011a}, \citet[][PG08]{perez-gonzalez2008a}, \citet[][M09]{marchesini2009a}, \citet[][F06]{fontana2006a}, \citet[][K09]{kajisawa2009a}, \citet[][G11]{gonzalez2011a}, \citet[][D14]{duncan2014}, and \citet[][S15]{song2015a}.}
\label{fig:smfs}
\end{figure*}

In GALICS, galaxies are evolved through the DM halo merger trees using physically motivated and phenomenological semi-analytic prescriptions. We refer to \citet{hatton} for a more complete description of the physical recipes and free parameters implemented in GALICS.  Below, we only highlight the main ingredients as well as the departures from the original version.\\

In the original version of \citet{hatton}, a mass of hot gas $M_{\rm hot}$ was assigned to each DM halo when first identified, consistently with the primordial baryonic fraction (i.e. $M_{\rm hot} = \Omega_b / \Omega_m M_{\rm halo}$). As the DM halo subsequently accreted mass, the hot gas reservoir was increased accordingly. At each timestep, the hot gas was able to cool and form stars at the centre of the DM halo. This scheme was replaced in \citet{cattaneo2006a} by a bimodal mode of accretion in high-redshift galaxies \citep[e.g.][]{birnboim2003,ocvirk08,dekel_nature}. In this scenario, gas from the intergalactic medium is shock-heated to the virial temperature in massive haloes, while cold gas can accrete along filaments at a rate set by the free-fall time below a critical halo mass set to $10^{12}$ \msun{} at $z$ $=$ 3.

Unlike \citet{hatton} who inferred the SFR directly from the mass of cold gas of the galaxy, $M_{\rm cold}$, we have now implemented the Kennicutt-Schmidt law which computes the SFR surface density from the cold gas mass surface density: $\Sigma_{\rm SFR} = \epsilon \Sigma_{\rm cold}^{1.4}$. Here, $\epsilon=1$ gives the z=0 normalisation of \citet{kennicutt98} in code units. As discussed in \citet{garel2015a}, we require $\epsilon=5$ to reproduce observational constraints (i.e. luminosity functions) at the redshifts we are focusing on in this study, namely $z$ $\gtrsim$ 3. Newly formed stars are distributed according to the \citet{kenn83} initial mass function (IMF) and their evolution is followed over substeps of 1 Myr.

We describe metal enrichment of the interstellar medium and supernovae feedback in a similar fashion as \citet{hatton}. Following \citet{silk03}, the gas ejection rate is proportional to $\alpha_{\rm SN}$SFR$/ v_{\rm esc}^2$ where $v_{\rm esc}$ is the escape velocity and $\alpha_{\rm SN}$ is the feedback efficiency, set to 0.2 as in \citet{cattaneo2006a}. The ejected (cold gas and metals) material can start being re-accreted at a constant rate through a galactic fountain after a time $\tau_{\rm delay}$ (set to half a halo dynamical time).

When two DM halos merge, the galaxies they host are placed in the descendant halo. As we do not follow substructures, we decide that a satellite can either merge with the central galaxy over a free-fall time\footnote{Whereas the dynamical friction time was used in the original version of \citet{hatton}, we now merge satellites with central galaxies over a free-fall time to be consistent with the cold filamentary accretion mode.}, or it may collide with another satellite (satellite-satellite encounters), following the procedure described in \citet[][Section 5]{hatton}.

The spectral energy distributions (SEDs) are computed from the star formation histories of galaxies using the STARDUST libraries \citep{devriendt} for a Kennicutt IMF. The effect of dust attenuation is given by Equation 3 of \citet{garel2012a} assuming a spherical geometry, consistent with the shell approximation described in the next paragraphs.

\subsection{Model calibration and comparison to data}
\label{subsec:data_comparison}

\begin{figure*}
\vskip-4ex
\hspace{-1.9cm}
\includegraphics[width=18.77cm,height=6.58cm]{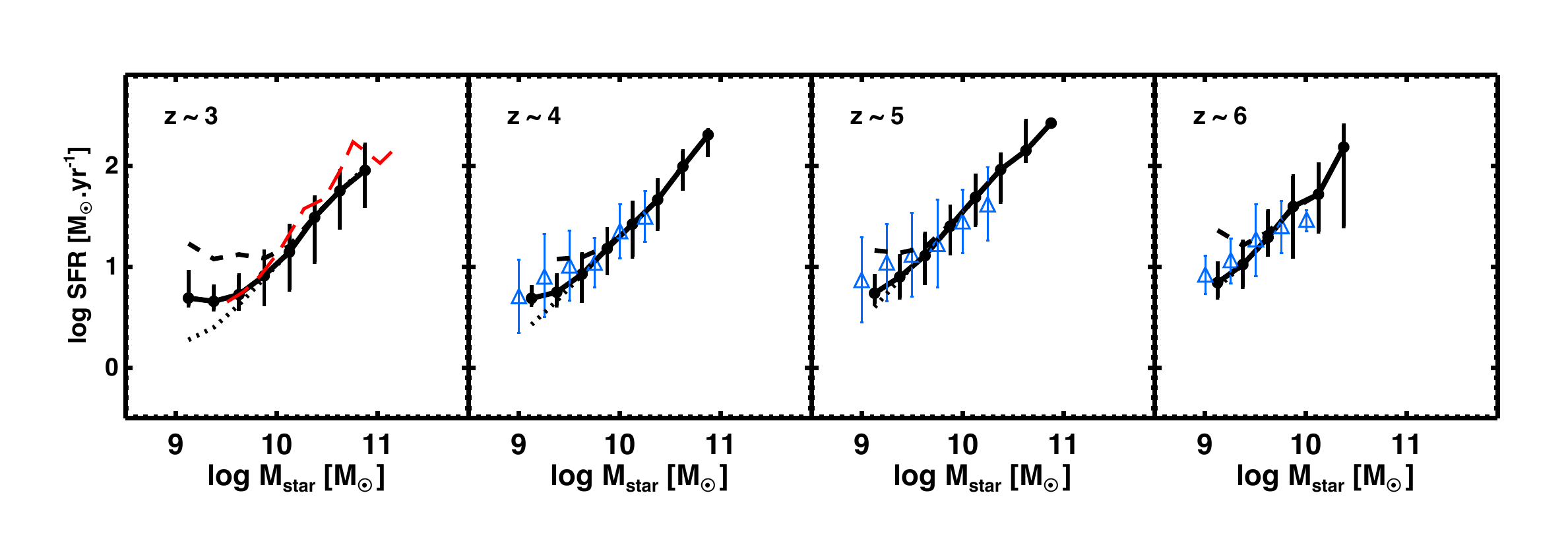}
\vskip-5ex
\caption{Relation between stellar mass ($M_{\rm star}$) and star formation rate (SFR) at $z$ $\approx$ 3, 4, 5, and 6. In black, we show the median SFR per bin of stellar mass along with the 10th-90th percentiles for galaxies with $10^9 < (M_{\rm star}/\msun) < 10^{11}$. The dotted, solid, and dashed curves correspond to UV magnitude cuts of $M_{\rm 1500} < -19$, $M_{\rm 1500} < -20$, and $M_{\rm 1500} < -21$ respectively. The red dashed line and the blue triangles correspond to the data from \citet{kajisawa2010a} and \citet{salmon2015a} respectively.}
\label{fig:sfr_ms}
\end{figure*}

In \citet{garel2015a}, our model was adjusted against observational constraints by choosing a set of reasonable model parameter values able to reproduce the luminosity functions of LBGs and LAEs at 3 $\lesssim$ $z$ $\lesssim$ 7. The UV luminosity function is a major constraint at high redshift  and it is now reasonably well measured at $z$ $\approx$ 3-7 \citep[e.g.][]{sawicki,reddy08,bouwens2015a}.  It traces the star formation rate of galaxies over a timescale of $\approx$ 100 Myr (modulo the effect of dust) and our model can reproduce it at various redshifts \citep[see Section 3.1 in][]{garel2015a}. Here, we show in addition the stellar mass functions (SMF) from our model and compare them to observational estimates. As can be seen on Figure \ref{fig:smfs}, the predicted $M_{\rm star}$ distributions are in good agreement with the observations, considering the large scatter between the different estimates. The best match is obtained when comparing with the recent CANDELS data from \citet{song2015a} at $z$ $\approx$ 4, 5, and 6. In Figure \ref{fig:sfr_ms}, we explore the positive correlation between stellar mass and SFR at high redshifts. Here, we use three different cuts in absolute UV magnitude, $M_{\rm 1500}=-19, -20,$ and $-21$, to try to mimic the observational selection of galaxies. We find a reasonable agreement between the model and the observational estimates, and this result appears to be weakly sensitive to the value of our UV magnitude cut.

It is important to stress that the derivation of physical quantities such as stellar masses and SFRs is subject to large uncertainties not always reflected by the error bars of data points in Figures \ref{fig:smfs} and \ref{fig:sfr_ms}, such as SED modelling assumptions, dust correction, or photometric redshift errors \citep[e.g.][]{marchesini2009a,wilkins2012a,schaerer2013a,stark2013a}. Nonetheless, our model appears well calibrated against existing observations describing the build up of galaxies at high redshift.

\subsection{Emission and radiative transfer of the \lya line}
\label{subsec:lya_model}
Under the case B approximation \citep{osterbrock2006}, the \lya emission line is powered by the reprocessing of two-thirds of the ionising photons through a radiative cascade in the \hii regions. The intrinsic \lya luminosity is thus given by: $L_{\rm Ly\alpha}^{\rm intr}= \frac{2}{3}Q(H)\frac{h_{\rm p} c}{\lambda_\alpha}$, where $Q(H)$ is the production rate of hydrogen-ionizing photons computed from the SEDs, $\lambda_\alpha = 1215.67$ \AA{} is the Ly$\alpha$ wavelength at line centre, $c$ is the speed of light, $h_{\rm p}$ the Planck constant. The intrinsic \lya line is described by a Gaussian centered on $\lambda_\alpha$ with a width given by the rotational velocity of the emission sources in the galaxy \citep[see Section 3.1 in][]{garel2012a}.

To account for the \lya radiation transfer (RT) and dust extinction, we compute the escape of \lya photons through galactic outflows. To do so, we combine the output of GALICS with the grid of \lya RT models in spherical expanding shells presented in \citet{schaerer2011a}. In these simulations, run with a 3D Monte Carlo code \citep[MCLya;][]{verh06}, the thin spherical expanding shells of gas and dust are characterised by four parameters: the expansion velocity, the gas column density, the internal velocity dispersion, and the dust opacity. These parameters are estimated for each galaxy using simple scaling arguments connected to the output of GALICS as described in Section 3.2.2 of \citet{garel2012a} and Section 2 of \citet{garel2015a}. We then compute the \lya escape fraction by interpolating the shell parameters predicted by GALICS on to the MCLya grid to obtain the \textit{observed} \lya luminosity, $L_{\rm Ly\alpha}$, and \lya flux, $F_{\rm Ly\alpha} = L_{\rm Ly\alpha} / (4\pi d_{\rm L}^2(z))$ where $d_{\rm L}(z)$, is the luminosity distance at redshift $z$.

The above escape fraction only accounts for internal attenuation of \lya photons (i.e. dust absorption in the shell). Nevertheless, interactions with \hi gas along the line of sight may affect the blue side of the \lya line, and then reduce the transmitted \lya flux, especially at the highest redshifts. We have tested the effect of IGM on the \lya lines using the prescriptions of \citet{madau} and \citet{inoue2014a} which compute the mean \lya transmission from observational statistics of intergalactic absorbers. In our model, the \lya lines are Doppler-shifted away from line centre due to radiative transfer in the shell, such that most photons emerging from our galaxies have $\lambda > 1215.67$ \AA{} in the rest-frame of the source. The intervening neutral gas is transparent to these photons, and we find that the IGM has no noticeable impact on our \lya fluxes even at $z$ $\approx$ 7 (see Section 3.2 of \citet{garel2015a} and Section 4.4 of \citet{garel2012a} for more details). This modelling of the effect of IGM remains somehow crude, and a more realistic scenario would require a detailed description of the gas distribution, kinematics, or ionisation state, which is beyond the capabilities of our semi-analytic approach. We note that the \hi opacity may also affect the red side of the \lya line due to peculiar gas motions in the surroundings of galaxies (e.g. infalls), or strong damping wings in a highly neutral Universe (i.e. before reionisation is complete), which can thus reduce the overall transmitted \lya fluxes \citep[e.g.][]{dijk07a,iliev2008a,laursen2011a,dayal2011a,jensen2013a}. We also note that faint LAEs might be more strongly attenuated than bright LAEs in inhomogeneously ionised IGM models at $z$ $>$ 6 since bright sources are thought to sit in larger \hii bubbles at the EoR, which may flatten the \lya LF towards faint luminosities \citep{furlanetto2006a,mcquinn2007a}.

\subsection{Mass resolution of the simulation}
\label{subsec:resol}
\begin{figure}
\begin{minipage}[]{0.99\textwidth}
\includegraphics[width=8.8cm,height=6.9cm]{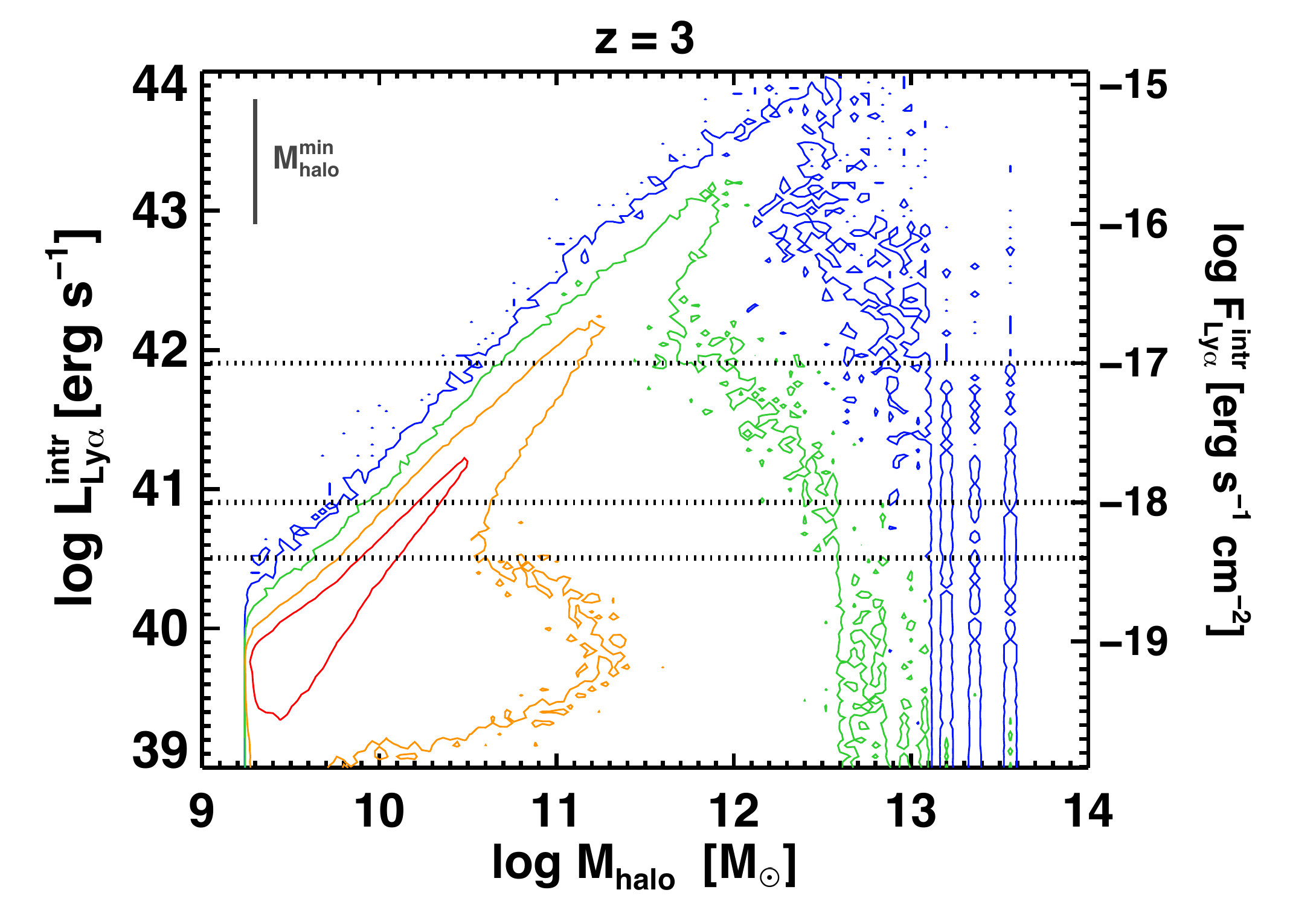}
\end{minipage}
\begin{minipage}[]{0.99\textwidth}
\includegraphics[width=8.79cm,height=7.05cm]{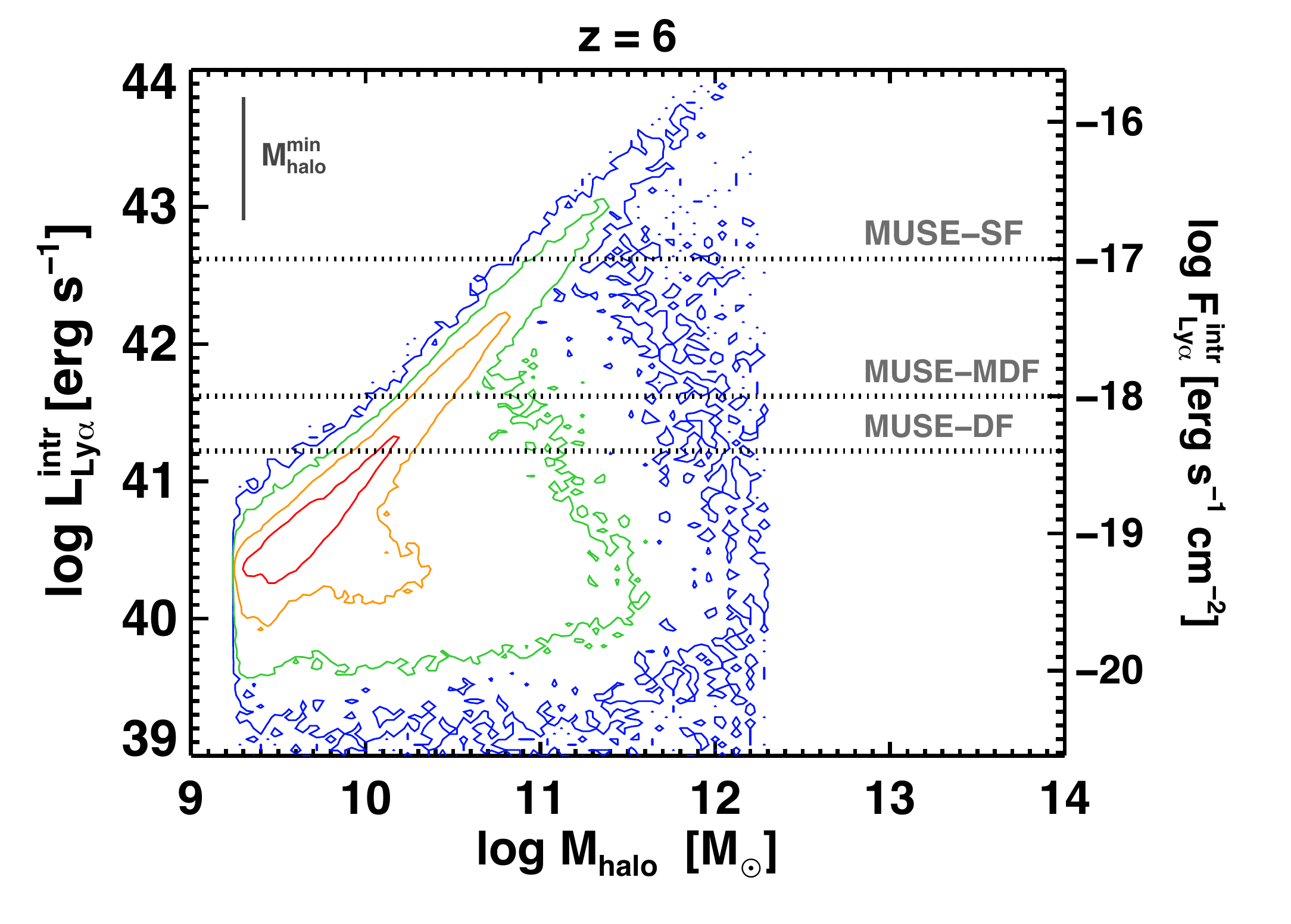}
\end{minipage}
\vskip-1ex 
\caption{Intrinsic \lya luminosity versus halo mass at $z$ $=$ 3 (top panel) and $z$ $=$ 6 (bottom panel). Contours show the number distribution of galaxies in the model. The vertical line in the top-left corner illustrates the halo mass resolution limit of the simulation ($M_{\rm halo}^{\rm min}=2 \times 10^9$ \msun). The dotted lines correspond to the \lya flux limits of typical MUSE Shallow Field (SF), Medium-Deep Field (MDF) and Deep Field (DF) surveys. The brightest \lyat-emitting galaxies residing in the least massive dark matter halos in our model have an approximate intrinsic \lya luminosity of $2 \times 10^{40}$ \ergs{} and $7 \times 10^{40}$ \ergs{} at $z$ $=$ 3 and $z$ $=$ 6 respectively (namely, a \lya flux of $\approx 2 \times 10^{-19}$ \ergscm{} at both redshifts). We consider that the sample of \lyat-emitting galaxies is complete above these values.}
\label{fig:lya_mh_resol}
\vskip-2ex 
\end{figure}

MUSE is expected to carry out very deep \lya observations, down to $F_{\rm Ly\alpha}^{\rm limit} \approx 4 \times 10^{-19}$ \ergscm. In order to make reliable statistical predictions, we want to ensure that we have sufficient mass resolution to produce complete samples of LAEs with $F_{\rm Ly\alpha}$ $\geq$ $F_{\rm Ly\alpha}^{\rm limit}$. In Figure \ref{fig:lya_mh_resol}, we show the predicted intrinsic \lya luminosity/flux of galaxies at $z$ $=$ 3 (top panel) and $z$ $=$ 6 (bottom panel) as a function of the mass of their host halo. The vertical line illustrates the halo mass resolution limit of our simulation, $M_{\rm halo}^{\rm min}$. Galaxies can thus only form in halos more massive than $M_{\rm halo}^{\rm min}$. For a given halo mass, galaxies can span a wide range of properties, i.e. stellar mass or \lya emission, depending on their own accretion and star formation history. Hence, it is not straightforward to assess the galaxy mass or \lya luminosity resolution limit. For the purpose of this paper, we consider the brightest intrinsic \lya luminosity displayed by galaxies residing in the least massive halos as a proxy for the \lya luminosity resolution limit. From Figure \ref{fig:lya_mh_resol}, we find this value to be $\approx 2 \times10^{40}$ \ergs{} at $z$ $=$ 3 and $\approx7 \times 10^{40}$ \ergs{} at $z$ $=$ 6, corresponding approximatively to the same \lya flux of $\approx 2 \times10^{-19}$ \ergscm{} at both redshifts. Thus, we expect our samples of mock LAEs to be statistically complete for this current study.

In addition, we note that gas accretion can be suppressed within low-mass DM halos as a result of photoheating of the IGM by a UV background during reionisation \citep[e.g.][]{ef}. Using high-resolution hydrodynamic simulations, \citet{okamoto08} have shown that this effect becomes significant for halos below a characteristic mass, $M_{\rm C}(z)$. $M_{\rm C}(z)\approx10^9$ and $M_{\rm C}(z)\approx2 \times 10^8$ \msun{} at $z$ $=$ 3 and $z$ $=$ 6 respectively. These values are below the minimum halo mass we can resolve in our simulation, so we assume that photoheating of the IGM would have a negligible impact on the baryonic content of our halos, and we do not take it into account in our model.

\subsection{Mock catalogues}
\label{subsec:mocks}

In order to produce mock observations easily comparable to real surveys, we convert the output of our semi-analytic model into lightcones with the MOMAF tool \citep[\textit{Mock Map Facility};][]{blaizot05}. MOMAF performs the (random) tiling of the simulation box snapshots and computes the apparent properties of galaxies in a cone-like geometry. Thus, in addition to the physical properties of galaxies predicted by GALICS (star formation rates, stellar masses, host halo masses, metallicity, gas content, etc), MOMAF provides an extra set of \textit{observables}: apparent redshifts/positions/velocities/sizes, and \lya fluxes.

In this paper, we assume an observing strategy with MUSE which consists of three typical surveys: a Deep Field (DF), a Medium-Deep Field (MDF), and a Shallow Field (SF) survey that reach \lya fluxes of $4 \times 10^{-19}$, $10^{-18}$, and $10^{-17}$ \ergscm{} respectively, corresponding approximatively to 1, 10, and 80 hours per exposure \citep{bacon2010a}. We consider the DF, MDF, and SF surveys to cover a sky area of 1, 10, 100 \sqarcm respectively.

To assess the variance on the number counts, we generate a large number of each set of lightcones filled with mock galaxies in the redshift range where \lya can be probed by MUSE (2.8 $\lesssim$ $z$ $\lesssim$ 6.7). We note that the effect of cosmic variance is inevitably underestimated here because we miss the fluctuations on the very large scales due to the finite comoving volume of our simulation box ($\approx 3 \times 10^6$ Mpc$^{3}$).

\section{Review of the literature}
\label{sec:literature}

\begin{figure*}
\vskip-24ex
\hskip-14ex
\begin{minipage}[]{0.45\textwidth}
\centering
\includegraphics[width=8.9cm,height=11.75cm]{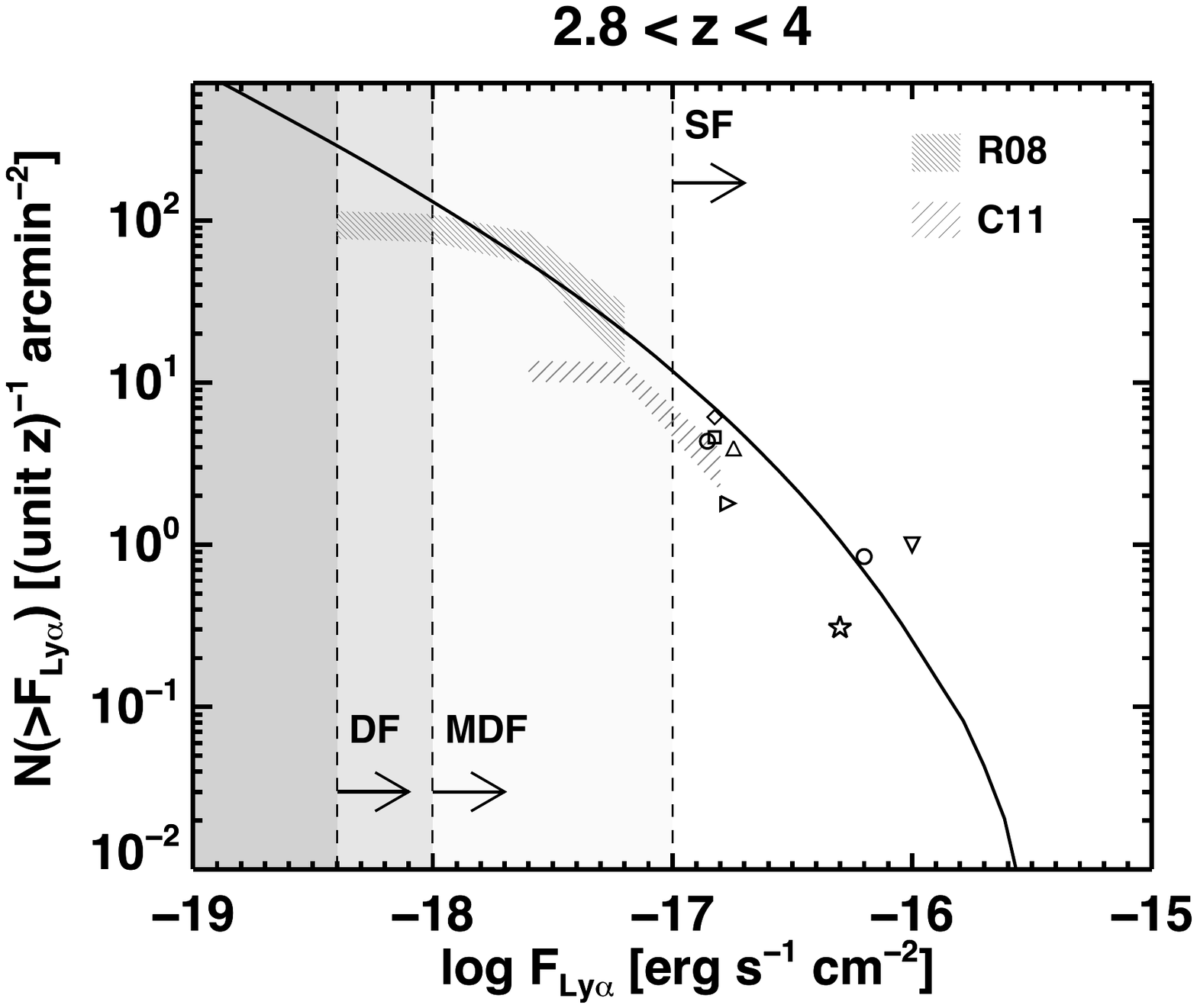}
\end{minipage}
\hskip10ex
\begin{minipage}[]{0.45\textwidth}
\centering
\includegraphics[width=8.9cm,height=11.75cm]{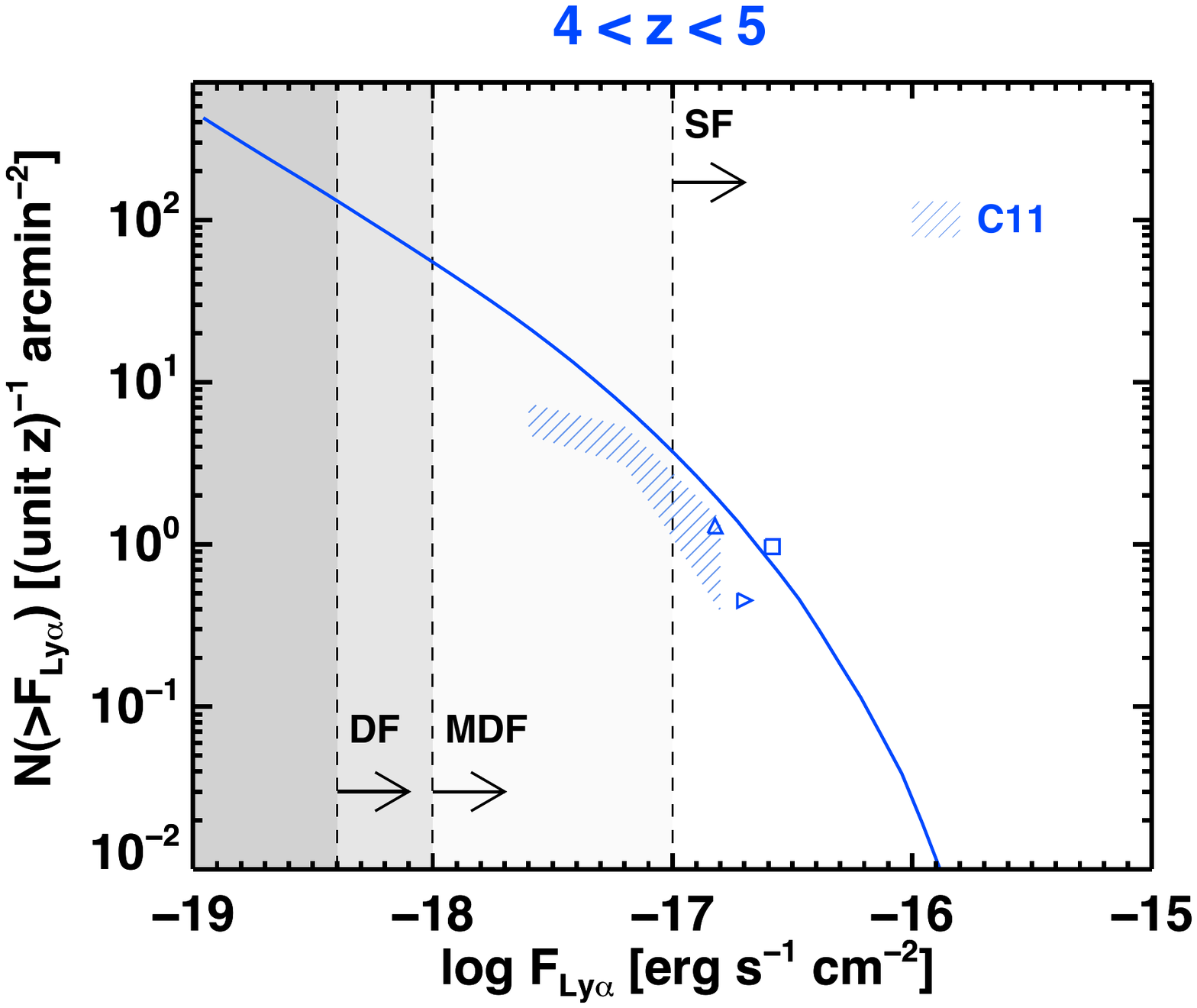}\\
\end{minipage}
\vskip-40ex
\hskip-14ex
\begin{minipage}[]{0.45\textwidth}
\centering
\includegraphics[width=8.9cm,height=11.75cm]{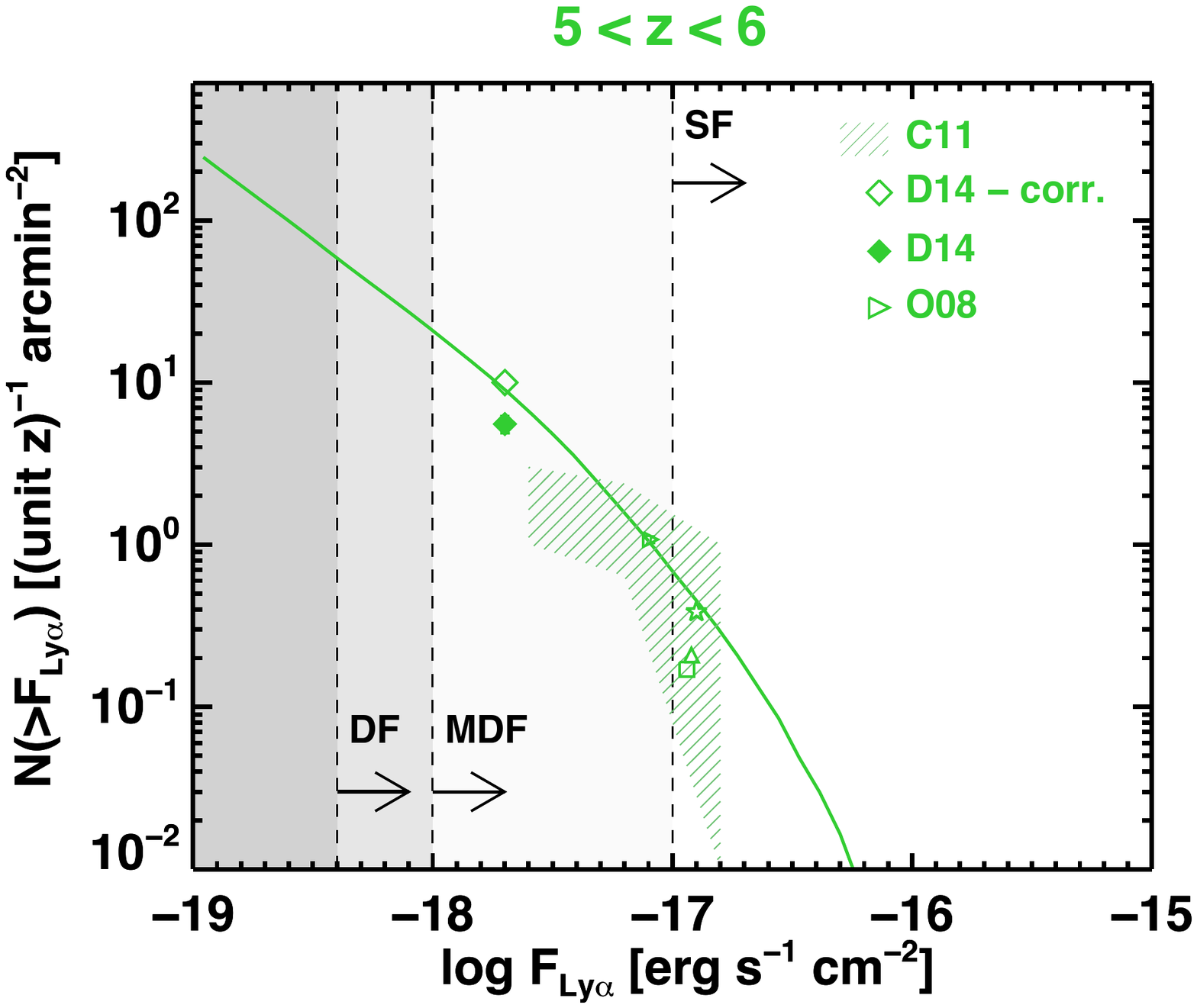}
\end{minipage}
\hskip10ex
\begin{minipage}[]{0.45\textwidth}
\centering
\includegraphics[width=8.9cm,height=11.75cm]{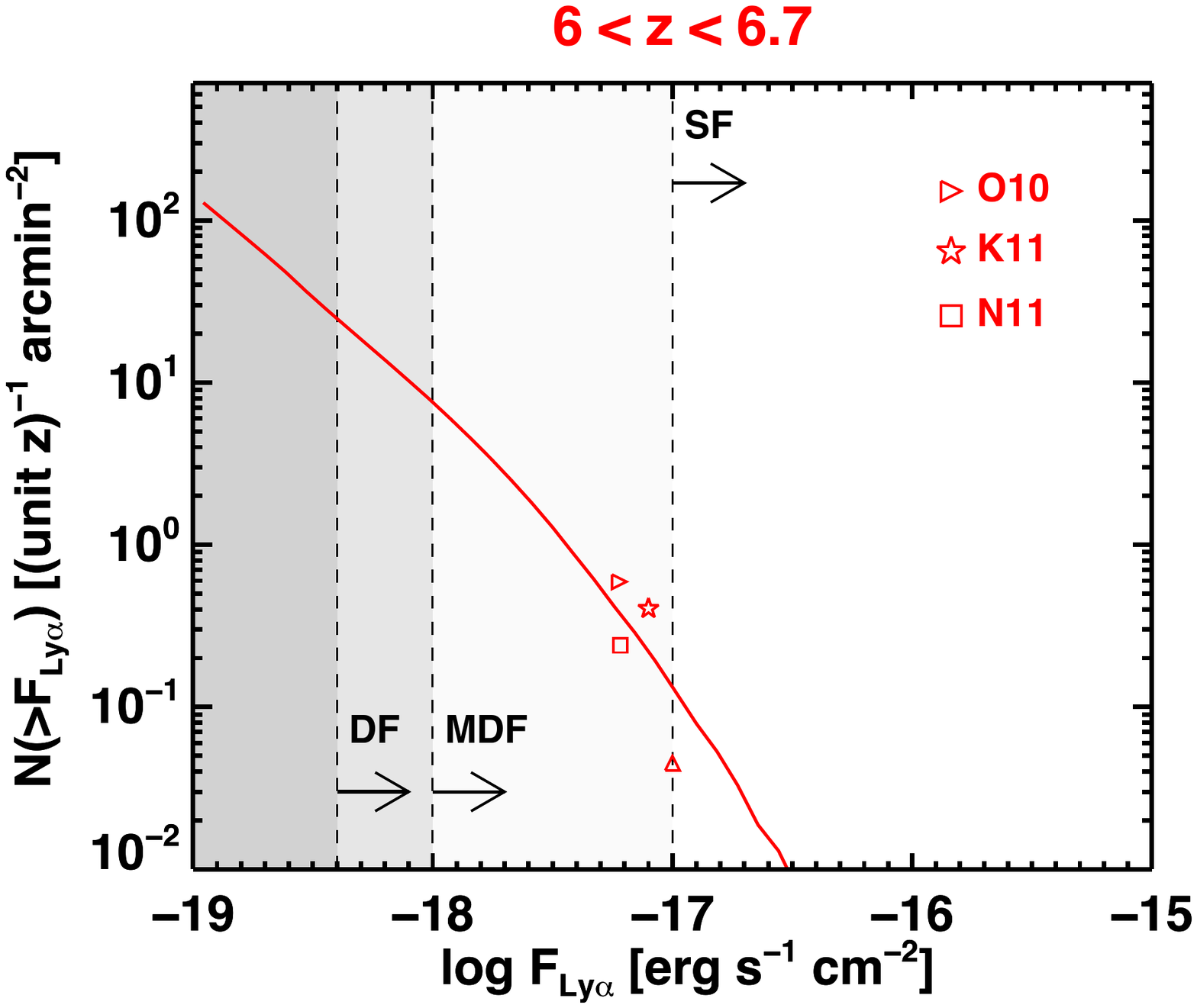}
\end{minipage}
\vskip-20ex
\caption{
Mean number count predictions (curves) at four different redshifts: $2.8 < z < 4$, $4 < z < 5$, $5 < z < 6$, and $6 < z < 6.7$. The arrows show the limiting fluxes for typical MUSE Deep, Medium-Deep, and Shallow field surveys (DF, MDF and SF respectively). The dashed areas encompass the error bars (Poisson statistics) of the data of \citet{rauch08} (R11) and the VVDS Ultra-Deep serendipitous detections of \citet{cassata2011a} (C11). The filled (empty) green diamond correspond to the number counts of \citet{dressler2015a} without (with) incompleteness correction (D14). The faint data ($F_{\rm Ly\alpha} \lesssim 10^{-17}$ \ergscm) of \citet{ouch08}, \citet{ouchi2010a}, \citet{nakamura2011a}, and \citet{kashikawa2011a} are labelled O08, O10, N11 and K11 respectively. For the sake of clarity, the references for shallower surveys ($F_{\rm Ly\alpha} \gtrsim 10^{-17}$ \ergscm) are not shown in the legend: \citet{vanb05} (black circle), \citet{yamada2012} (black upward triangle), \citet{hayashino2004a} (black diamond), \citet{ouch08} (black rightward triangle), \citet{kud00} (black downward triangle), \citet{gronwall07} (black square), \citet{blanc2011a} (black star), \citet{hu98} (blue upward triangle), \citet{rhoads00} (blue square), \citet{malhotra04} (blue rightward triangle), \citet{murayama07} (green square), \citet{kashikawa2011a} (green and red star) and \citet{hu2010a} (green and red upward triangle). Unless stated otherwise, the data points plotted here correspond to the number of detections at the flux limit of a given survey, which may not be the limit of completeness.}
\label{fig:counts1deg}
\end{figure*}

In Figure \ref{fig:counts1deg}, we show LAE number counts reported by previous surveys at various redshifts in terms of LAE number density per unit redshift in four redshift intervals, i.e. $2.8 < z < 4$, $4 < z < 5$, $5 < z < 6$, and $6 < z < 6.7$. The flux limits of typical MUSE DF, MDF, and SF surveys are illustrated by arrows. 

With a Deep-field survey, MUSE could collect a sample of extremely faint galaxies, with a \lya flux limit of $4 \times 10^{-19}$ \ergscm{} in about 80 hours over one \sqarcm. Similar \lya fluxes have already been reached by \citet{santos04} in a spectroscopic blind survey using the strong lensing technique, but they only discovered a handful of objects at $z$ $=$4-6. A few years ago, \citet{rauch08} found 27 LAEs as part of a 92-hour long-slit spectroscopy search with FORS2 at $z$ $\approx$ 3, which translates into a number density of objects at $F_{\rm Ly\alpha} \gtrsim 10^{-18}$ \ergscm{} as high as $\approx 100$ LAE per \sqarcm per unit redshift. Although the faintest source reported by \citet{rauch08} has a flux of $\approx 7 \times 10^{-19}$ \ergscm, their distribution starts to flatten at $\approx 10^{-18}$ \ergscm, probably due to incompleteness issues. 


\begin{table*}
\vskip-1ex
\hskip10ex
\centering
\vskip1ex
\begin{tabular}{ccccc}
\hline
\hline
 & & \:  Deep field  \:  & \:  Medium-Deep field \:  & \:  Shallow field \\
\hline
\multicolumn{1}{c}{\multirow{5}{*}{Mean counts \: }} & $2.8 < z < 4$ \:  \: \:  & 317$ \: {\scriptstyle\pm} \: $56 & 1417$ \: {\scriptstyle\pm} \: $208 & 1241$ \: {\scriptstyle\pm} \: $183 \\
& $4 < z < 5$ \:  \: \:  & 119$ \: {\scriptstyle\pm} \: $31 &  505$ \: {\scriptstyle\pm} \: $110 & 327$ \: {\scriptstyle\pm} \: $69 \\
& $5 < z < 6$ \:  \: \:   & 53$ \: {\scriptstyle\pm} \: $19  & 185$ \: {\scriptstyle\pm} \: $55 &  58$ \: {\scriptstyle\pm} \: $19 \\
& $6 < z < 6.7$ \:  \: \:   & 15$ \: {\scriptstyle\pm} \: $9 & 47$ \: {\scriptstyle\pm} \: $23 &  7$ \: {\scriptstyle\pm} \: $5 \\
\cdashline{2-5}
& $2.8 < z < 6.7$ \:  \: \:   & 504$ \: {\scriptstyle\pm} \: $67 & 2155$ \: {\scriptstyle\pm} \: $241 & 1633$ \: {\scriptstyle\pm} \: $194 \\
\hline
\multicolumn{1}{c}{\multirow{5}{*}{Median counts \: }} & $2.8 < z < 4$ \:  \:  \:  &  314 (248$/$389) & 1411 (1158$/$1687) & 1234 (1011$/$1477) \\
& $4 < z < 5$ \:  \:  \:  & 117 (82$/$159) & 497 (371$/$648) &  325 (242$/$416) \\
& $5 < z < 6$ \:  \:  \:  & 50 (30$/$78) & 178 (120$/$260) &  56 (35$/$83) \\
& $6 < z < 6.7$ \:  \:  \:  & 14 (6$/$27) & 42 (22$/$77) &  6 (2$/$14) \\
\cdashline{2-5}
& $2.8 < z < 6.7$ \:  \:  \:  & 501 (418$/$590) & 2146 (1852$/$2466) & 1632 (1391$/$1883) \\
\hline
\end{tabular}
\vskip1ex 
\caption{Mean number counts with standard deviation \& median number counts with 10th/90th percentiles predicted for typical MUSE surveys: a Deep Field ($F_{\rm Ly\alpha} \geq 4 \times 10^{-19}$ \ergscm{} - 1 \sqarcm), a Medium-Deep Field ($F_{\rm Ly\alpha} \geq 10^{-18}$ \ergscm{} - 10 \sqarcm), and a Shallow Field ($F_{\rm Ly\alpha} \geq 10^{-17}$ \ergscm{} - 100 \sqarcm) survey.}
\vskip-2ex
\label{tab:surveys_counts} 
\end{table*}

The \lya detection limit of a MUSE Medium-Deep field survey ($F_{\rm Ly\alpha} \approx 10^{-18}$ \ergscm) will be comparable to the VVDS Ultra-Deep survey \citep[$2 \lesssim z \lesssim 6.6$;][]{cassata2011a} and slightly deeper than the spectroscopic sample of \citet{dressler2015a} at $z$ $\approx$ 5.7. We show in Figure \ref{fig:counts1deg} the number density of \lya sources at $2.8 \lesssim z \lesssim 4$ and $4 \lesssim z \lesssim 5$ from the VVDS Ultra-Deep survey (serendipitous), including the slit losses x1.8 flux correction quoted by \citet{cassata2011a}. We note that the number counts at $z$ $=$ 2.8-4 seem slightly less than those reported by \citet{rauch08}, although the two measurements roughly remain in the (Poisson) error bars of one another. Also, while the detection limit of the VVDS Ultra-Deep survey is $\approx 1.5 \times 10^{-18}$ \ergscm, the $\approx 100\%$ completeness level is reached at about $F_{\rm Ly\alpha} = 4-7 \times 10^{-18}$ \ergscm{} \citep[see Figure 9 in][]{cassata2011a} so their actual surface density of LAEs should be larger than what is shown in Figure \ref{fig:counts1deg} at fainter fluxes. Furthermore, the volumes probed by these two surveys are rather small, so part of the difference may be due to cosmic variance effects. 

At $z$ $\approx$ 6, the abundance of faint LAEs has recently been investigated by \citet{dressler2015a} using high-resolution IMACS observations, as a follow-up of a previous survey \citep{dressler2011a}. Targeting 110 out of their 210 LAE candidates, \citet{dressler2015a} spectroscopically confirmed about one third of the sources as genuine high redshift LAEs. Extrapolating this confirmation rate to the whole sample of candidates, the surface density of LAEs with $F_{\rm Ly\alpha} \gtrsim 2 \times 10^{-18}$ \ergscm{} is $\approx$ 6 arcmin$^{-2}$ per unit redshift, and $\approx$ 10 arcmin$^{-2}$ per unit redshift once corrected for incompleteness (A. Dressler, private communication), which suggests a very steep faint end slope of the \lya luminosity function at $z$ $=$ 5.7.

The detection limit of a Shallow field survey would be of the same order of magnitude as most existing \lya data sets ($F_{\rm Ly\alpha}^{\rm limit} \approx 10^{-17}$ \ergscm), which sample the bright end of the \lya LF, i.e. $L_{\rm Ly\alpha} \gtrsim 1-5 \times 10^{42}$ \ergs{} at $z$ $=$ 3-6.
Wide-field narrow-band surveys usually span a large area on the sky allowing to obtain large samples of candidates within large volumes \citep[up to a few $\sim 10^6$ Mpc$^3$;][]{ouch08,yamada2012} and to minimise the effect of cosmic variance. They nevertheless can only select LAEs in a rather restricted redshift window ($\Delta{\rm z} \lesssim 0.1$), and usually necessitate extensive amounts of telescope time for spectroscopic follow-up observations, required to remove low-redshift interlopers. Alternatively, blind spectroscopic surveys can easily detect line emitters over a wider redshift range, but they usually cannot probe large volumes due to the small area sampled by the slit \citep[$\sim 7 \times 10^4$ Mpc$^3$;][]{sawicki2008a}, or small IFUs field-of-view \citep[$\sim 10^4$ Mpc$^3$;][]{vanb05}. Yet the Hobby-Eberly Telescope Dark Energy Experiment \citep[HETDEX;][]{hill2008a}, a blind spectroscopic survey making use of the wide field-of-view VIRUS integral field spectrograph, is expected to detect up to one million bright LAEs ($F_{\rm Ly\alpha} \gtrsim 3.5 \times 10^{-17}$ \ergscm) over a 60 deg$^2$ sky area between $z$ $\approx$ 1.9 and 3.8, which corresponds to a volume of almost 9 Gpc$^3$. The HETDEX survey will take years to complete, but first observations of LAEs have already been released as part of the pilot survey \citep[e.g.][]{blanc2011a}. Despite the much smaller area covered by a typical MUSE SF survey ($\approx$ 100 $\sqarcm$), it will be very complementary to HETDEX, as it will be slightly deeper, able to probe LAEs at much higher redshift and at higher spectral resolution.

Our number count predictions, represented by the curves in Figure \ref{fig:counts1deg}, are computed over the full sample of objects at each timestep in our simulation, using mock lightcones of  $1 \times 1$ deg$^2$ which roughly corresponds to the angular size of our 100$h^{-1}$ Mpc box at $z$ $\sim$ 3-6. They are in very good agreement with the faint LAE number counts ($F_{\rm Ly\alpha} \gtrsim 10^{-18}$ \ergscm) reported by \citet{rauch08} at $z$ $\approx$ 3 and \citet{dressler2015a} at $z$ $\approx$ 5.7. At $F_{\rm Ly\alpha} \gtrsim 3 \times 10^{-18}$ \ergscm, they are slightly higher than the projected densities of serendipitous LAEs measured by \citet{cassata2011a} in the VVDS Ultra-Deep survey at $z$ $\lesssim$ 5, albeit the agreement is reasonable at $z$ $=$ 5-6. Our model roughly matches number counts from shallower observations shown as symbols in Figure \ref{fig:counts1deg}. These correspond to the number of detections at the flux limit of each given survey, which may not be the limit of completeness. A more reliable comparison of our model with observed bright LAE abundances can be found in Figure 2 of \citet{garel2015a} where we plot the predicted luminosity functions against observed ones from $z$ $\approx$ 3 to 7. They reasonably agree over this redshift range but scatter remains in the \lya LF data, and we note that our model better matches the higher (lower) end of the envelope of data points at $z$ $\approx$ 3 (z $\approx$ 6).

\section{Model predictions}
\label{sec:predictions}

\begin{figure*}
\vskip-18ex
\hskip-10ex
\begin{minipage}[]{0.31\textwidth}
\includegraphics[width=7cm,height=10.cm]{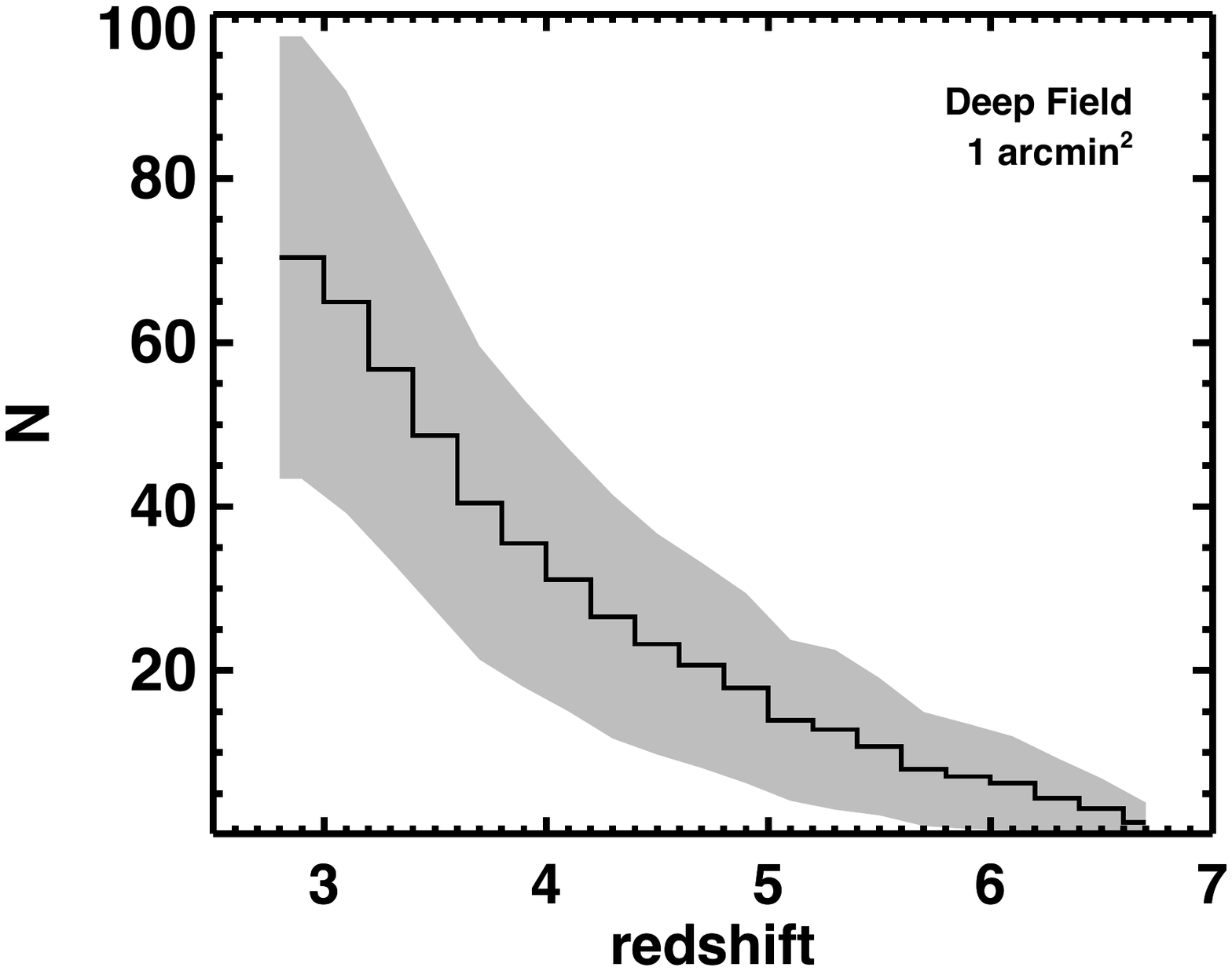}
\end{minipage}
\hskip6ex
\begin{minipage}[]{0.31\textwidth}
\includegraphics[width=7cm,height=10.cm]{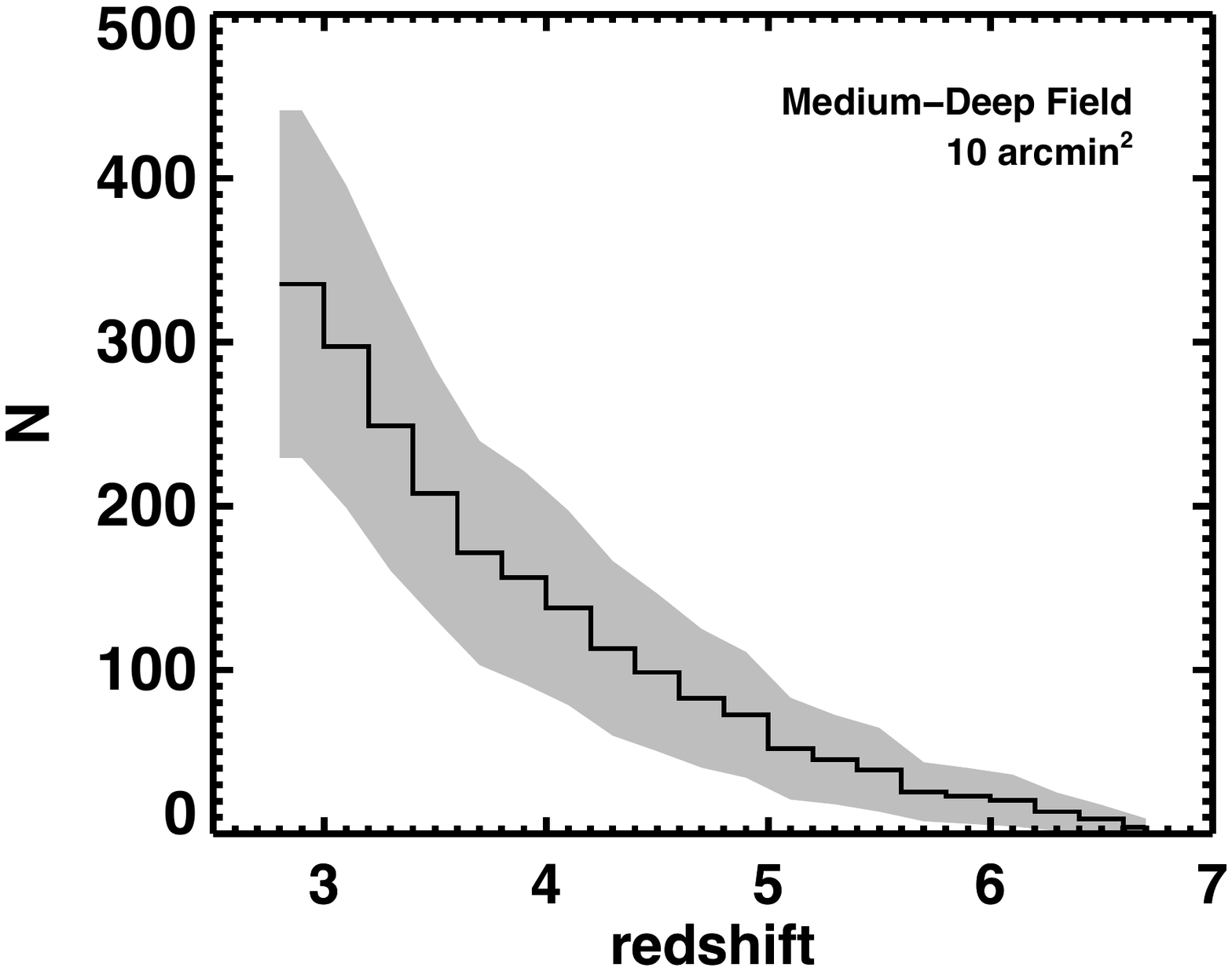}
\end{minipage}
\hskip6ex
\begin{minipage}[]{0.31\textwidth}
\includegraphics[width=7cm,height=10.cm]{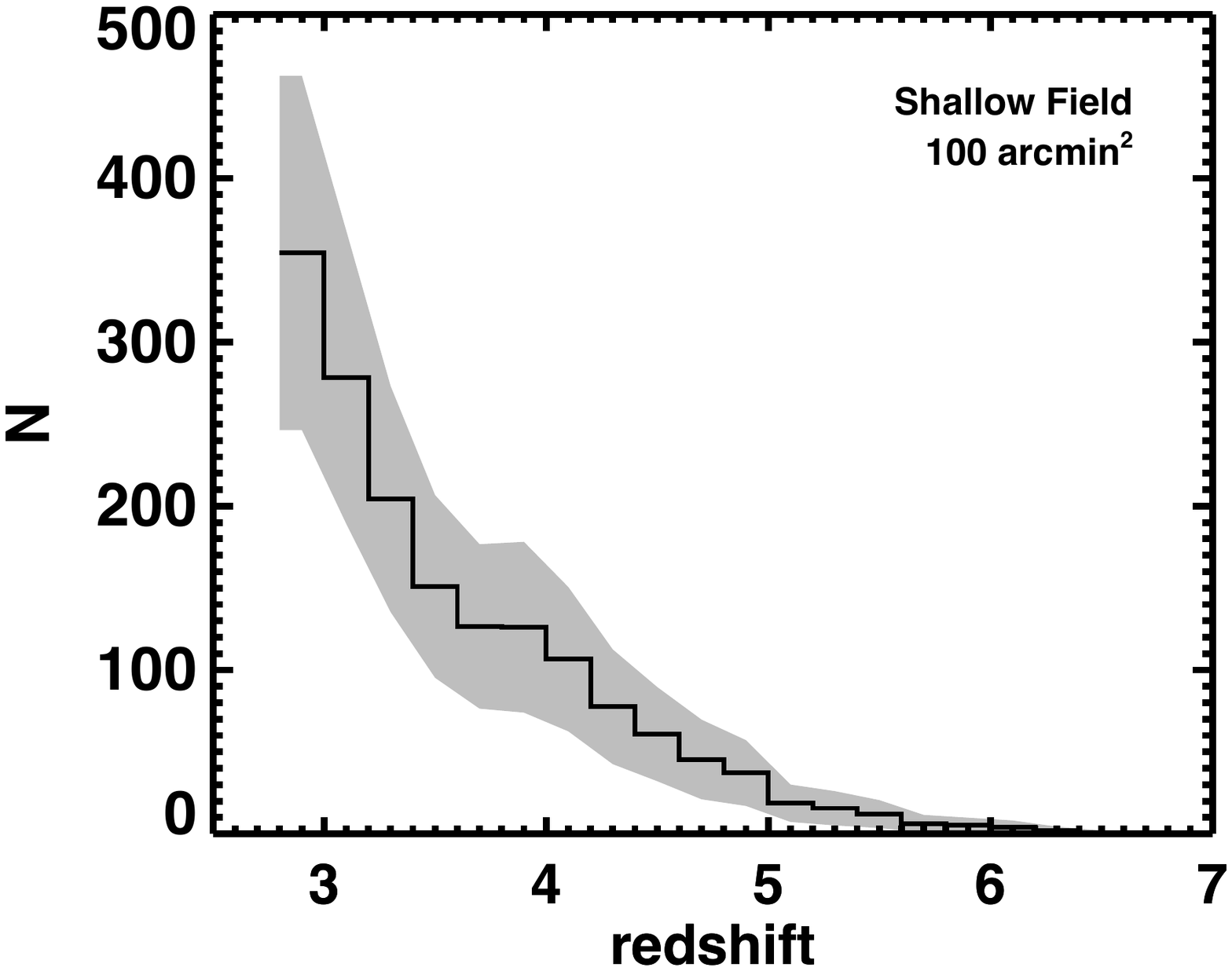} 
\end{minipage}
\vskip-18ex
\caption{Predicted redshift distributions of \lyat-emitting galaxies for typical MUSE Deep Field (left: $F_{\rm Ly\alpha} \geq 4 \times 10^{-19}$ \ergscm), Medium-Deep Field (centre: $F_{\rm Ly\alpha} \geq 10^{-18}$ \ergscm), and Shallow Field (right: $F_{\rm Ly\alpha} \geq 10^{-17}$ \ergscm) surveys between $z$ $=$ 2.8 and $z$ $=$ 6.7. The histograms show the mean number of objects in redshift bins of 0.2dex, except the last one which is 0.1dex wide, i.e. 6.6 $<$ $z$ $<$ 6.7. The grey shaded area illustrates the expected standard deviation computed from large numbers of lightcones.}
\label{fig:z_dist}
\vskip-2ex
\end{figure*}

\begin{figure*}
\vskip-18ex
\hskip-16ex
\begin{minipage}[]{0.3\textwidth}
\includegraphics[width=7.5cm,height=10.5cm]{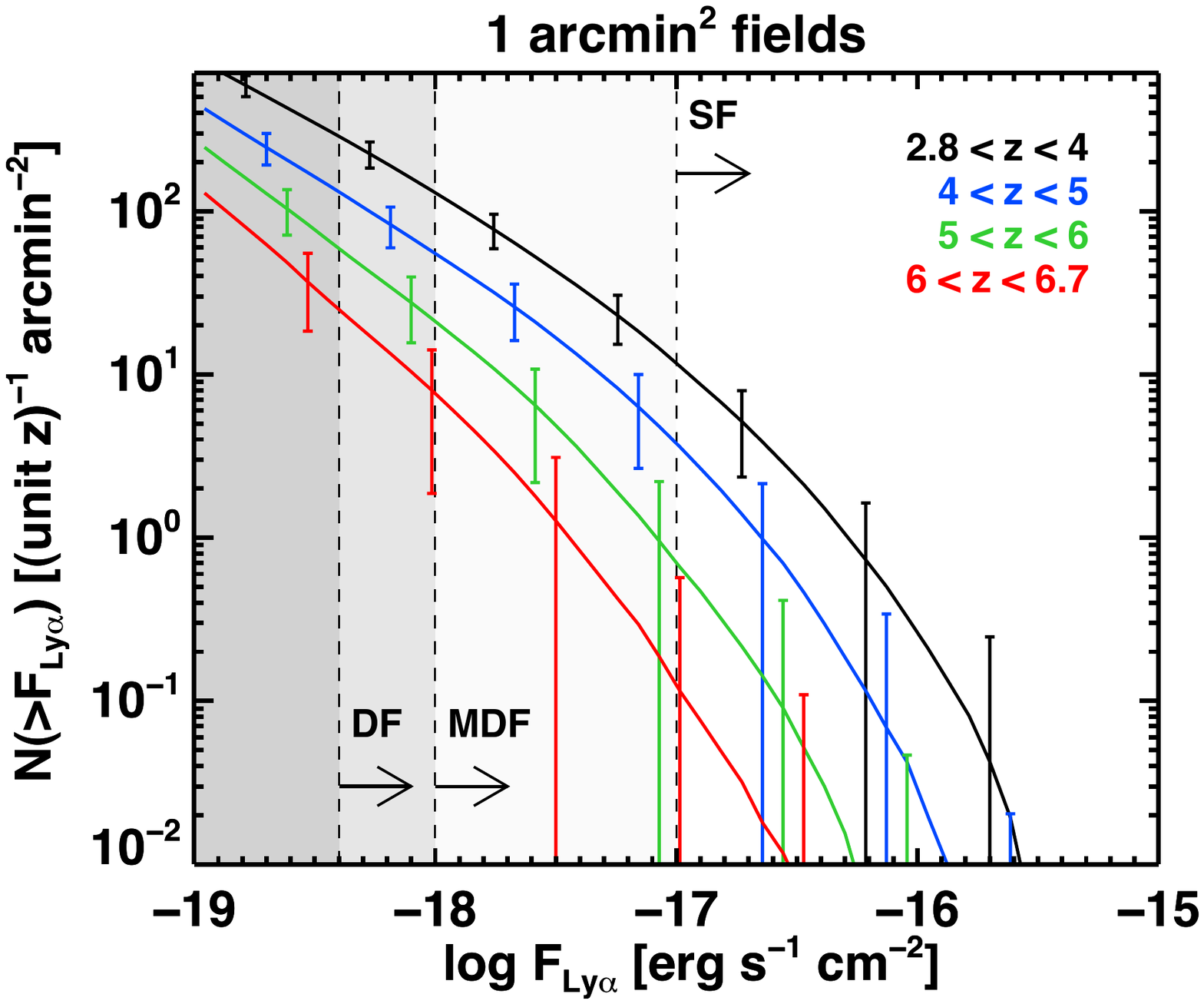}
\end{minipage}
\hskip-1ex
\begin{minipage}[]{0.3\textwidth}
\includegraphics[width=7.5cm,height=10.5cm]{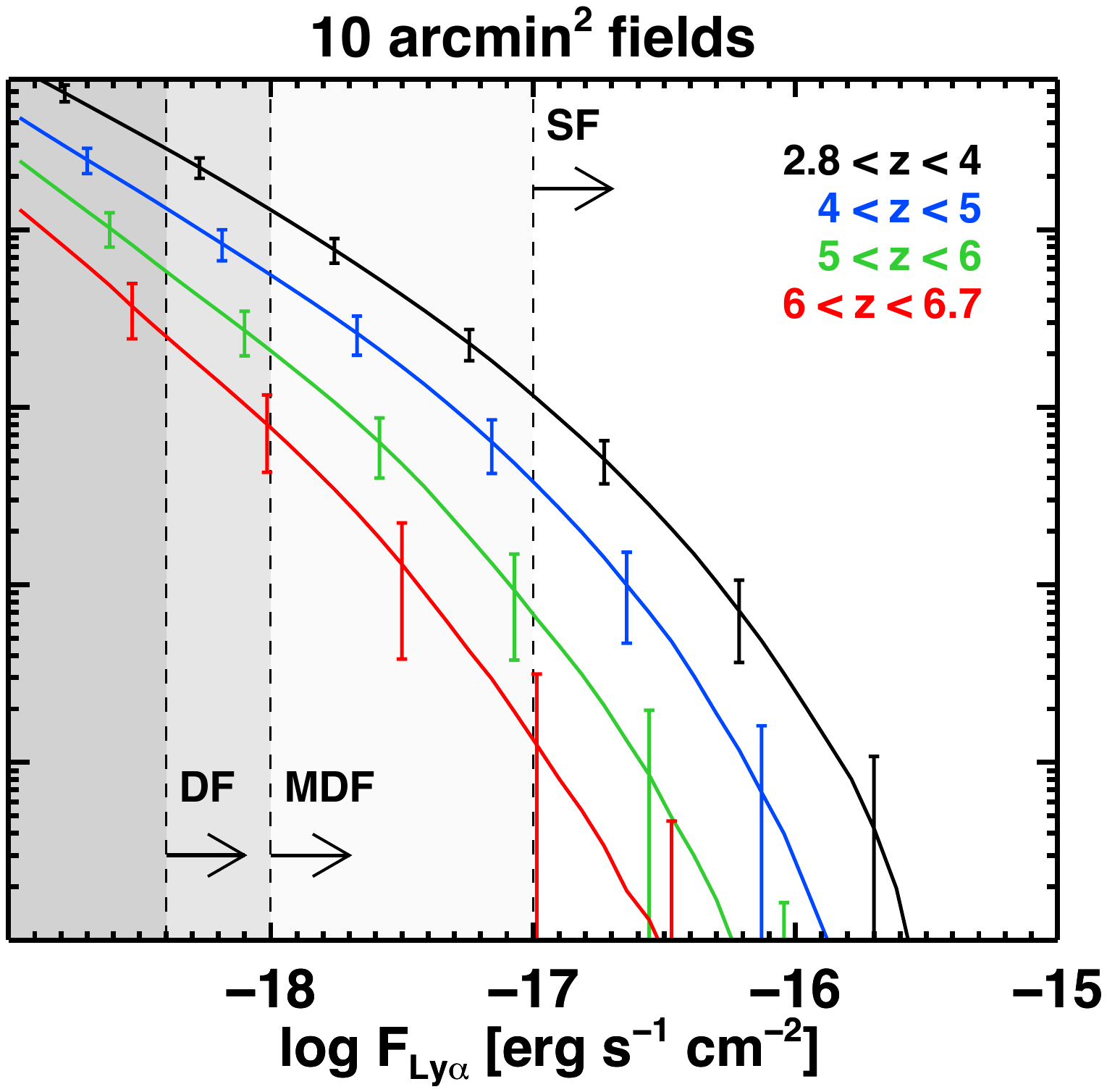}
\end{minipage}
\hskip-1ex
\begin{minipage}[]{0.3\textwidth}
\includegraphics[width=7.5cm,height=10.5cm]{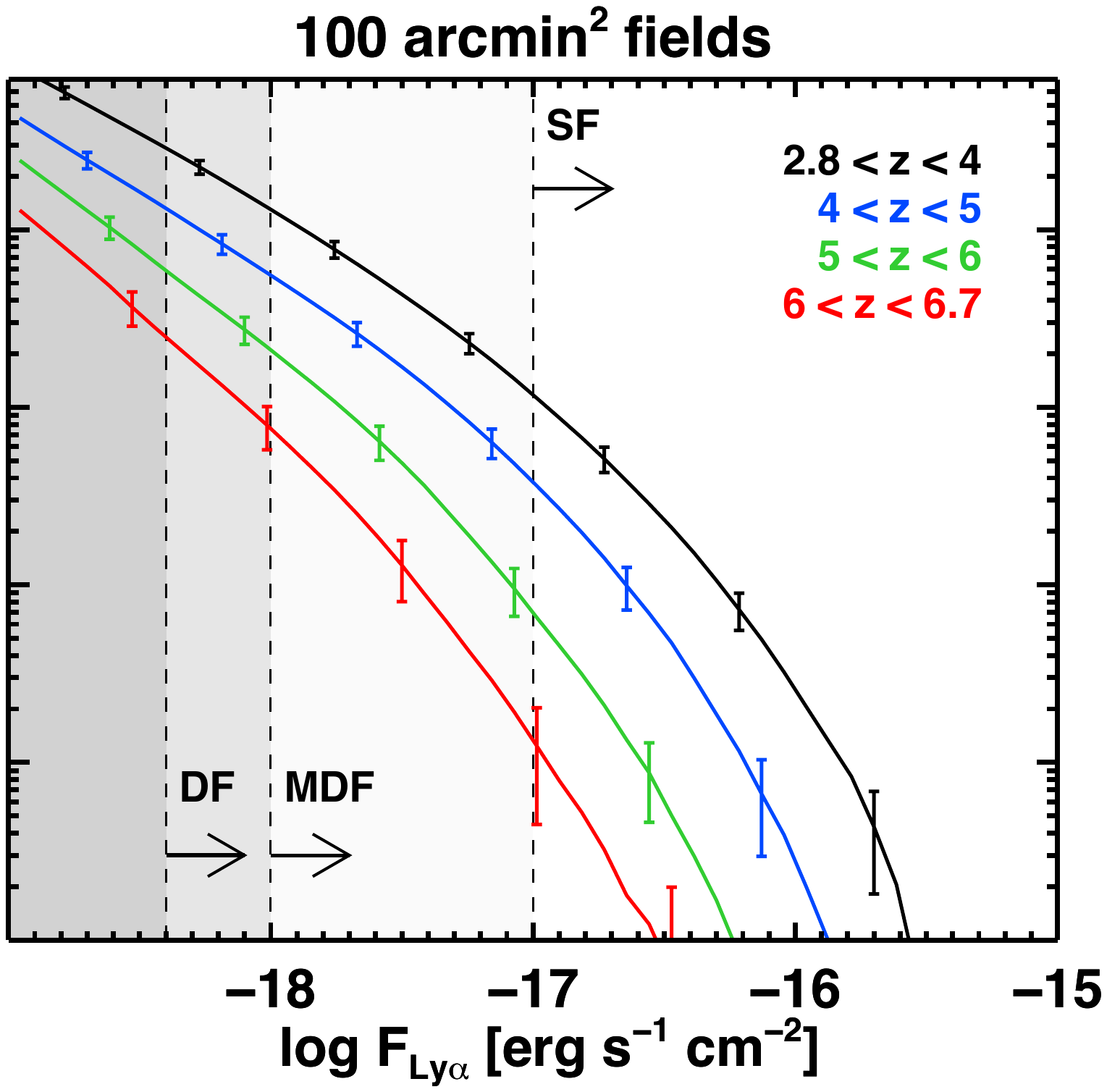}
\end{minipage}
\vskip-18ex 
\caption{Mean number counts in mock fields of 1, 10, and 100 square arcminutes (left, centre, and right panels respectively). The curves show the predicted numbers of LAEs per unit redshift (see legend panel) and square arcminute in four redshift bins: $2.8 < z < 4$, $4 < z < 5$, $5 < z < 6$, and $6 < z < 6.7$ (from top to bottom, as labelled). The error bars represent the standard deviation computed over a large number of lightcones. We add the limiting fluxes for typical Deep, Medium-Deep, and Shallow field surveys to be carried out with VLT/MUSE, labelled DF, MDF and SF respectively.}
\label{fig:lya_counts_all} 
\end{figure*}

In this section, we present the predicted number counts of \lyat-emitting galaxies for each typical MUSE survey, and the contribution of these sources to the cosmic \lya luminosity density and cosmic SFR density as a function redshift.

\subsection{Predicted number counts for typical MUSE surveys}
\label{subsec:counts}

In Figure \ref{fig:z_dist}, we show the redshift distributions that we predict for the three typical surveys we consider in the paper. The redshift range is set by the wavelength range for which MUSE will be able to probe \lya line emitters, i.e. from $z$ $\approx$ 2.8 to $z$ $\approx$ 6.7. The histograms in Figure \ref{fig:z_dist} give the mean expected number of objects as a function of redshift, and the shaded grey areas correspond to the standard deviation (that includes cosmic variance) computed over larger number of realisations of each field.

In Table \ref{tab:surveys_counts}, we present the predicted mean number counts with the associated standard deviations and the median counts, including the 10/90th percentiles. We predict that MUSE would detect as many as $\approx$ 500 sources with \lya fluxes $\geq 4 \times 10^{-19}$ \ergscm{} in 1 \sqarcm between $z$ $\approx$ 2.8 to $z$ $\approx$ 6.7. In the redshift bin $2.8 \leq z \leq 4$, approximatively 315 galaxies could be found in a DF survey, and only 15 are predicted to lie between $z$ $\approx$ 6 and $z$ $\approx$ 6.7. A DF survey would obtain the faintest LAE sample ever observed, pushing down the \lya luminosity function measurement towards the extreme faint end. 

According to our mock catalogs, a MUSE Medium-Deep survey would lead to more than 2,000 LAE detections within 10 \sqarcm{}. With about ten hours exposure per pointing, the \lya detection limit will reach $\approx 10^{-18}$ \ergscm{} for the MDF, which is of the same order as previous surveys by \citet{rauch08}, \citet{cassata2011a}, and \citet{dressler2015a} whose samples contain 27, 217, and 210 LAEs respectively. We predict that $\approx$ 1,500 sources would be found in a MDF survey at $2.8 < z < 4$, and 500 at $4 < z < 5$, which would outnumber all existing spectroscopic surveys of faint LAEs. At $5 < z < 6$, we expect a bit less than two hundreds detections down to $10^{-18}$ \ergscm{} in the MDF. In addition, we expect 45 LAEs at $z$ $\gtrsim$ 6. To sum up, we can expect a MDF survey to yield statistical samples in all the redshift ranges discussed here, allowing MUSE to put reliable constraints on the slope of the faint end of the \lya LF, and its evolution, from $z$ $\approx$ 2.8 to $z$ $\approx$ 6.7. 

\begin{table*}
\vskip2ex 
\hskip10ex
\centering
\vskip1ex
\begin{tabular}{ccccccccc}
\hline
\hline
\multirow{2}{*}{} &
      \multicolumn{2}{c}{Deep field} & &
      \multicolumn{2}{c}{Medium-Deep field} & &
      \multicolumn{2}{c}{Shallow field} \\
      \hline
      & \: {${\sigma}_{{\raisebox{-1pt}{\tiny{\rm V,rel}}}}$ (\%)} \: & \: {$\sigma_{{\raisebox{-1pt}{\tiny{\rm P,rel}}}}$ (\%)} \: \: & & \: {$\sigma_{{\raisebox{-1pt}{\tiny{\rm V,rel}}}}$ (\%)} \: & \: {$\sigma_{{\raisebox{-1pt}{\tiny{\rm P,rel}}}}$ (\%)} \: \: & & \: {$\sigma_{{\raisebox{-1pt}{\tiny{\rm V,rel}}}}$ (\%)} \: & \: {$\sigma_{{\raisebox{-1pt}{\tiny{\rm P,rel}}}}$ (\%)} \\
\cline{2-3}\cline{5-6}\cline{8-9}
 $2.8 < z < 4$ \: & $16.7$ & $5.6$ & & $14.4$ & $2.7$ & & $14.5$ & $2.8$ \\
 $4 < z < 5$ & \: $24.4$ & $9.1$ & & $21.3$ & $4.4$ & & $20.4$ & $5.5$ \\
 $5 < z < 6$ & \: $33.1$ & $13.7$  & & $28.8$ & $7.4$ & & $30.0$ & $13.1$ \\
 $6 < z < 6.7$ \: & $54.1$ & $25.8$ & & $46.7$ & $14.6$ & & $60.6$ & $37.8$ \\
\hline
\end{tabular}
\vskip3ex 
\caption{Predicted number count uncertainties for the DF, MDF, and SF surveys. $\sigma_{{\raisebox{-1pt}{\tiny{\rm P,rel}}}}$ and $\sigma_{{\raisebox{-1pt}{\tiny{\rm V,rel}}}}$ correspond to the relative Poisson error and the relative cosmic variance \citep{moster11} respectively (see text for more details).}
\label{tab:surveys_cosvar} 
\end{table*}

Finally, more than 1,500 LAEs would be detected between $z$ $\approx$ 2.8 to $z$ $\approx$ 6.7 at fluxes larger than $F_{\rm Ly\alpha} \approx 10^{-17}$ \ergscm{} as part of a typical Shallow-Field survey according to our model. This flux limit is typical of current \lya NB surveys, and therefore, most constraints on the statistical properties of \lyat-emitting galaxies have been derived for such bright LAEs. For instance, the added samples of \citet{ouch08} at $z$ $\approx$ 3.1$\pm0.03$ and $z$ $\approx$ 3.7$\pm0.03$ contain nearly 460 LAE candidates, while approximatively 1200 sources are expected at $2.8 < z < 4$ in the SF survey from our mock catalogues (Table \ref{tab:surveys_counts}). At $5 < z < 6$, we predict 60 LAEs with $F_{\rm Ly\alpha} \gtrsim 10^{-17}$ \ergscm{} within 100 \sqarcm. This is considerably less than in the NB survey of \citet{ouch08} at $z$ $\approx$ 5.7$\pm0.05$ ($\approx$ 400 objects), but comparable to the number of targeted sources in the follow-up observations of \citet{kashikawa2011a} and \citet{hu2010a}, who built some of the largest spectroscopic samples to date at this redshift. Moreover, different NB surveys often use different filters set and selection criteria, while MUSE will build homogeneous samples of LAEs over a very large range of redshift. A SF survey would yield a unique, spectroscopic, large sample of bright LAEs allowing to study the evolution of their statistical (e.g. abundances) and spectral properties from $z$ $\approx$ 2.8 to $z$ $\approx$ 6.7. 

\subsection{Number count uncertainties}
\label{subsubsec:muse_cosvar}

In addition to Table \ref{tab:surveys_counts}, we show our predicted number counts for typical DF, MDF, and SF MUSE surveys in four redshift bins in Figure \ref{fig:lya_counts_all}. For each field, we used mock lightcones of 1, 10, and 100 \sqarcm respectively to compute the mean cumulative projected density of LAEs per unit redshift (curves) and the 1$\sigma$ standard deviation (error bars). We see that the standard deviation, computed from thousands of lightcones, appears to be non-negligible, especially for the DF survey (left panel), and at the bright-end of the MDF and SF surveys (middle and right panels). 

Here, the standard deviation is given by $\sigma = \sqrt{\langle N^2 \rangle - \langle N \rangle^2}$, where N is the number of sources in the mock lightcones in a given redshift range and above a given \lya flux limit. Following \citet{moster11}, we define the \textit{relative cosmic variance}\footnote{Here, we assume that the total variance, $\sigma^2$ is the (quadratic) sum of cosmic variance and Poisson noise: $\sigma_{{\raisebox{-1pt}{\tiny{\rm V}}}}^2 + \sigma_{{\raisebox{-1pt}{\tiny{\rm P}}}}^2$. Relative cosmic variance is then written as $\sigma_{{\raisebox{-1pt}{\tiny{\rm V,rel}}}} = \sigma_{{\raisebox{-1pt}{\tiny{\rm V}}}} / \langle N \rangle = \sqrt{\sigma^2 - \sigma_{{\raisebox{-1pt}{\tiny{\rm P}}}}^2} / \langle N \rangle = \sqrt{(\langle N^2 \rangle - \langle N \rangle^2 - \langle N \rangle)} / \langle N \rangle$ \citep[see][for more details]{moster11}.} as the uncertainty in excess to Poisson shot noise divided by the mean number of counts $\langle N \rangle$, $\sigma_{{\raisebox{-1pt}{\tiny{\rm V,rel}}}} = \sqrt{(\langle N^2 \rangle - \langle N \rangle^2 - \langle N \rangle)} / \langle N \rangle$. Poisson noise normalised to $\langle N \rangle$ can be expressed as $\sigma_{{\raisebox{-1pt}{\tiny{\rm P,rel}}}} = \sqrt{\langle N \rangle}/\langle N \rangle$. Using this simple formalism, we then attempt to quantify the respective contributions of Poisson noise and cosmic variance to number count uncertainties in the MUSE fields.

$\sigma_{{\raisebox{-1pt}{\tiny{\rm P,rel}}}}$ scales like $1 / \sqrt{\langle N \rangle}$, hence it is large for small galaxy samples, and conversely, it tends to 0 when the number of detections is large. The relative cosmic variance $\sigma_{{\raisebox{-1pt}{\tiny{\rm V,rel}}}}$ reflects the uncertainty on the number counts due to field-to-field variation  when probing a finite volume of the sky. 

\begin{figure}
\hskip-6ex
\begin{minipage}[]{0.95\textwidth}
\includegraphics[width=9.cm,height=7.cm]{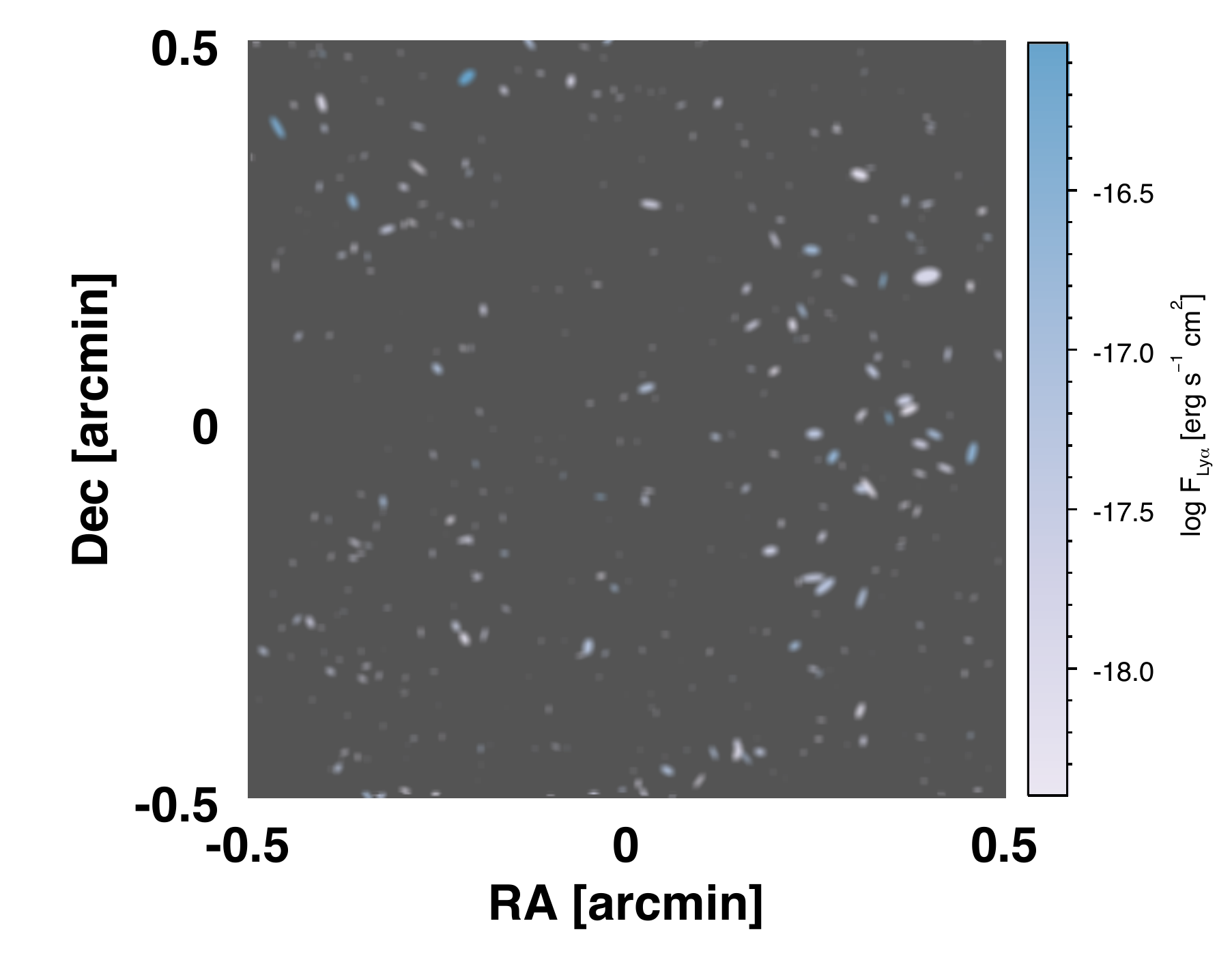}
\end{minipage}
\vskip-2ex 
\hskip-6ex
\begin{minipage}[]{0.95\textwidth}
\includegraphics[width=9.35cm,height=7.3cm]{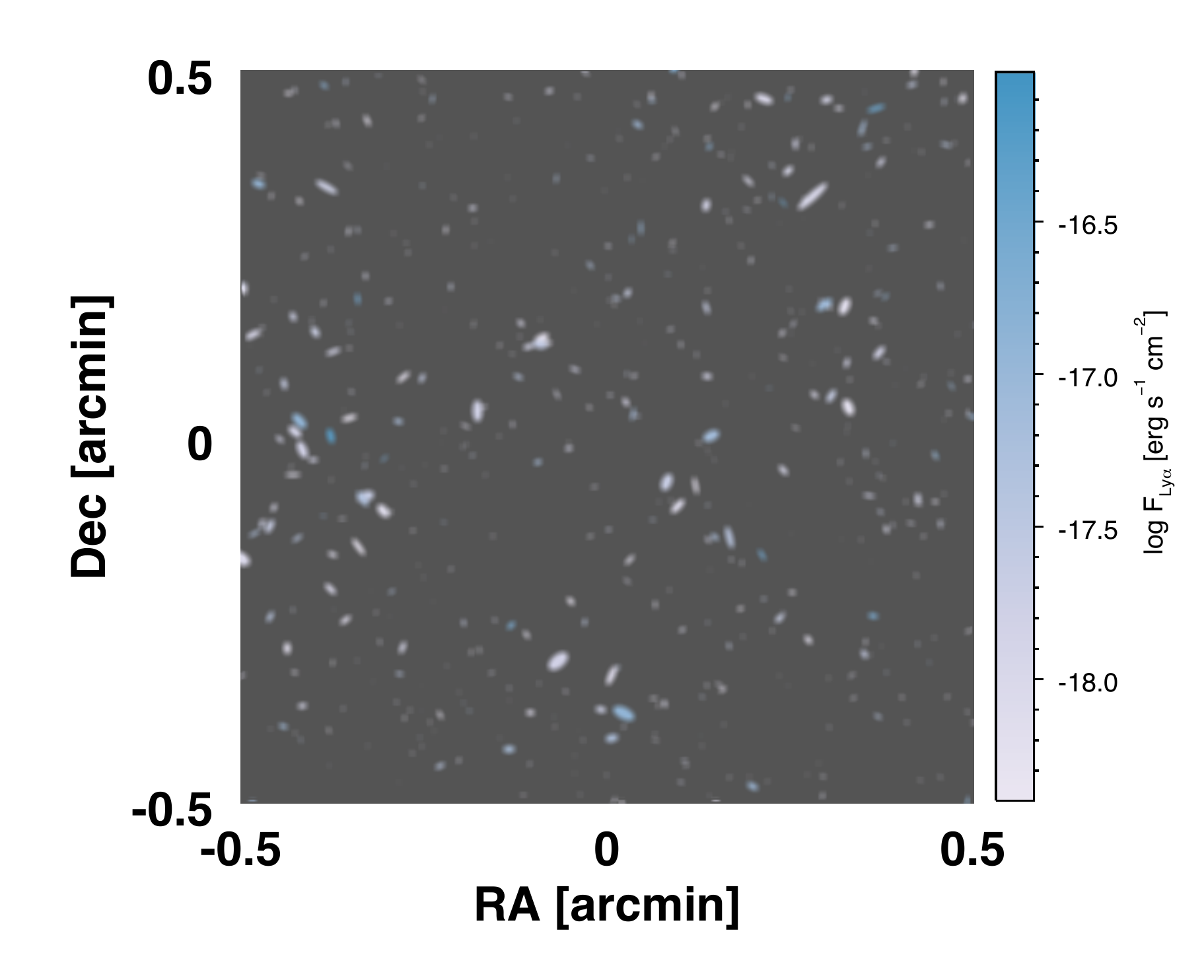}
\end{minipage}
\vskip-3ex 
\hskip-6ex
\begin{minipage}[]{0.95\textwidth}
\includegraphics[width=9.45cm,height=7.3cm]{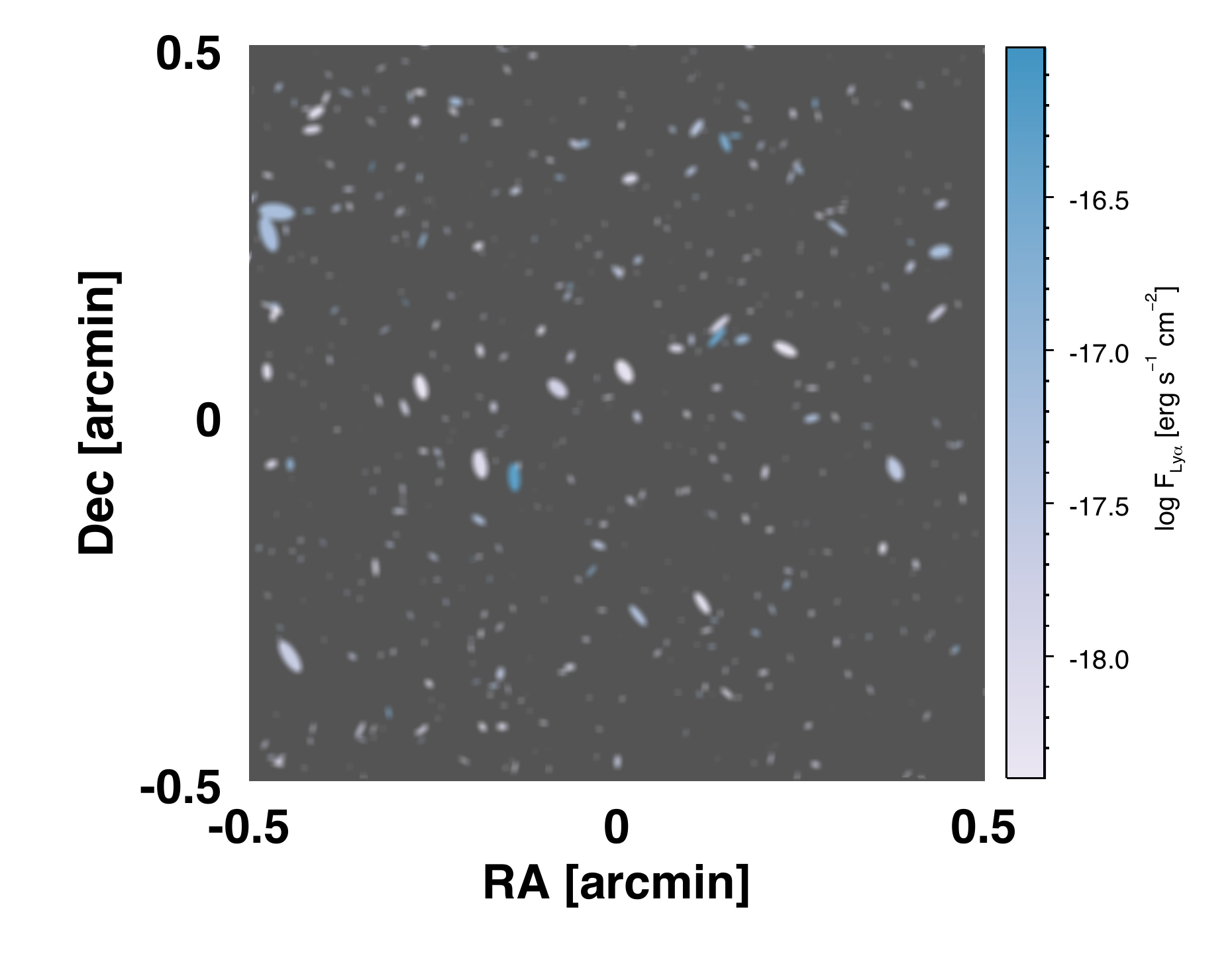}
\end{minipage}
\caption{Mock maps for a typical MUSE Deep Field of size 1 x 1 arcmin$^2$. The three panels illustrate the variance in terms of number counts for different pointings. The middle panel shows a map where the number of galaxies is equal to the median value from 5000 mock fields of 1 \sqarcm. The upper and lower panels correspond to the 10th percentile and 90th percentile respectively. Galaxies have been selected above a threshold of $F_{\rm Ly\alpha} \geq 4 \times10^{-19}$ erg s$^{-1}$ cm$^{-2}$, and lie in the redshift range $2.8 < z < 6.7$.}
\label{fig:maps}
\vskip2ex 
\end{figure}

In Table \ref{tab:surveys_cosvar}, we show the predicted $\sigma_{{\raisebox{-1pt}{\tiny{\rm P,rel}}}}$ and $\sigma_{{\raisebox{-1pt}{\tiny{\rm V,rel}}}}$ in typical DF, MDF and SF surveys at different redshifts. We find that cosmic variance dominates the number count uncertainty in all cases. Its contribution is $3-5$ times larger than the relative Poisson error at $2.8 < z < 4$ in all fields. Both $\sigma_{{\raisebox{-1pt}{\tiny{\rm P,rel}}}}$ and $\sigma_{{\raisebox{-1pt}{\tiny{\rm V,rel}}}}$ values increase with increasing redshift, and at $6 < z < 6.7$, the difference is only a factor of 2-3 as \lya sources are rarer at higher redshift in flux-limited surveys. On the one hand, although the DF survey is very deep, cosmic variance remains large due to the small volume that is probed. As an example, we show in Figure \ref{fig:maps} three mock realisations of a MUSE Deep Field containing respectively 418 (upper panel: 10th percentile), 501 (middle panel: median number) and 590 (lower panel: 90th percentile) LAEs brighter than $4 \times10^{-19}$ erg s$^{-1}$ cm$^{-2}$. On the other hand, a typical SF survey would cover a wider area (100 \sqarcm), but its shallower depth only enables to observe rarer sources, enhancing (i) statistical uncertainties and (ii) cosmic variance as brighter LAEs are located in more massive, rarer, halos than fainter sources on average in our model \citep{garel2015a}. Accordingly, clustering analysis suggest that bright LAEs tend to be more clustered \citep{ouch03,jose2013}. We predict the relative uncertainties to be minimised for a typical MDF survey as it is a trade-off between volume size and flux depth. At $2.8 < z < 4$, $\sigma_{{\raisebox{-1pt}{\tiny{\rm P,rel}}}}$ and $\sigma_{{\raisebox{-1pt}{\tiny{\rm V,rel}}}}$ are about 3\% and 15\% respectively, reaching $\approx$ 15\% and 45\% in the $z$ $=$ 6-6.7 redshift bin. Finally, we note that these values have to be seen as lower limits because of the finite volume of our simulation box.

These simple quantitative estimations suggest that uncertainties on the number counts will be non negligible, and their accurate determination will be needed to derive robust constraints on the \lya LFs.

\subsection{\lya luminosity and Star Formation Rate densities}
\label{subsec:budgets}
MUSE surveys will compile statistical, homogeneous samples of \lyat-emitting galaxies at several limiting fluxes over a large redshift range which will allow to assess the contribution of faint sources to the global LAE population. In the next paragraphs, we therefore present our predictions for cosmic \lya luminosity density and SFR as a function of redshift, that will be probed by typical MUSE surveys.

\begin{figure}
\vskip-20ex
\hskip-8ex
\includegraphics[width=9.3cm,height=10.5cm]{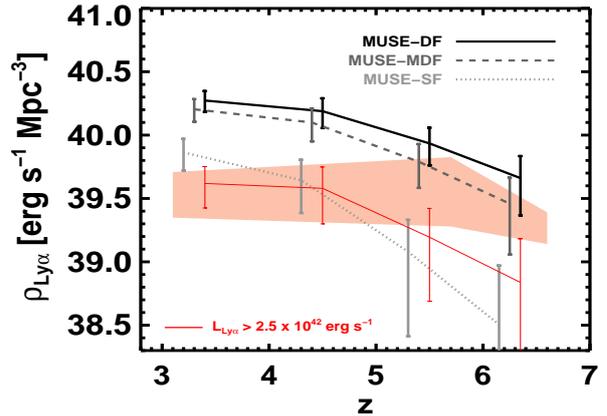}
\vskip-20ex 
\caption{Evolution of the \lya luminosity density. The black, dark grey, and light grey curves represent the \lya luminosity density, $\rho_{\rm Ly\alpha}$, that we expect to probe with typical MUSE Deep Field (DF - $F_{\rm Ly\alpha} \geq 4 \times 10^{-19}$ \ergscm), Medium-Deep Field (MDF - $F_{\rm Ly\alpha} \geq 10^{-18}$ \ergscm), and Shallow Field (SF - $F_{\rm Ly\alpha} \geq 10^{-17}$ \ergscm) surveys respectively. Each curve shows the mean $\rho_{\rm Ly\alpha}$ measured from 5000 lightcones of 10 \sqarcm{} in four redshift bins: $2.8 < z < 4$, $4 < z < 5$, $5 < z < 6$, and $6 < z < 6.7$. The error bars correspond to the 1$\sigma$ standard deviations. The dark grey, and light grey curves have been shifted horizontally by 0.1dex for the sake of clarity. The red curve shows the evolution of the \lya luminosity density using a fixed \lya luminosity threshold of $L_{\rm Ly\alpha} \geq 2.5 \times 10^{42}$ \ergs{} which is typical of current narrow-band wide-field surveys of LAEs, e.g. \citet{ouch08,ouchi2010a} (red shaded area).}
\label{fig:lya_density}
\end{figure}

Figure \ref{fig:lya_density} shows the cosmic \lya luminosity density $\rho_{\rm Ly\alpha}$ in four redshift bins, $2.8 < z < 4$, $4 < z < 5$, $5 < z < 6$, and $6 < z < 6.7$. First, we compare our predictions (red curve) to estimates from narrow-band observations \citep[shaded red area;][]{ouch08,ouchi2010a} for which the observed (uncorrected for dust) \lya luminosity function is integrated down to $L_{\rm Ly\alpha}$ $=$ $2.5 \times 10^{42}$ \ergs. The model agrees well with the data at $z$ $=$ 3-5 but seems a factor of two lower at higher redshift. As shown on the Figure 2 of \citet{garel2015a}, our model reproduces reasonably well the observed \lya luminosity functions from $z$ $=$ 3 to 7, but it slightly underpredicts the abundances of LAEs reported by \citet{ouch08,ouchi2010a} at $z$ $\approx$ 6 (possibly due to high contamination in narrow-band LAE samples at this redshift), hence the difference between the model and the data in Figure \ref{fig:lya_density}. 

Next, we present the redshift evolution of $\rho_{\rm Ly\alpha}$ as predicted by our model for the MUSE Deep Field (DF; solid black line), Medium-Deep Field (MDF), and Shallow Field (SF) surveys. Here, we computed $\rho_{\rm Ly\alpha}$ by summing up the contribution of galaxies in our mock catalogues above the limiting \lya flux of each MUSE survey using lightcones of 10 \sqarcm. First, we see that the SF survey (dotted light grey line) should be recovering a \lya luminosity density roughly similar to what we predict for current narrow-band surveys (red curve). This is not surprising because the SF \lya sensitivity flux limit ($F_{\rm Ly\alpha} \geq 10^{-17}$ \ergscm) corresponds to luminosities of $\approx 8 \times 10^{41}$ \ergs{} at $z$ $\approx$ 3 and $\approx 4 \times 10^{42}$ \ergs{} at $z$ $\approx$ 6, which is of the same order as in typical NB surveys \citep{shima06,ouch08,shioya,hu2010a}. 

Second, we compare the predictions for the various MUSE surveys with one another. A typical MDF survey (dashed dark grey line) would be able to detect \lya line fluxes as low as $10^{-18}$ \ergscm, that is ten times fainter than in a SF survey. We clearly notice that the cosmic \lya luminosity density probed by a MDF is expected to be much larger than for a SF survey and than what is currently available in narrow-band surveys. For instance, between the SF and MDF surveys, we expect a gain in terms of $\rho_{\rm Ly\alpha}$ of a factor of $\approx$ 2 at $z$ $=$ 3 and $\approx$ 6 at $z$ $=$ 6. With even longer exposure, a typical DF survey will reach \lya fluxes down to $4 \times 10^{-19}$ \ergscm{} and our model predicts an additional gain of 25 \% to 70 \% at $z$ $=$ 3 and 6 compared to the MDF. 

\begin{figure}
\vskip-1ex
\hskip-4ex
\includegraphics[width=8.65cm,height=5.95cm]{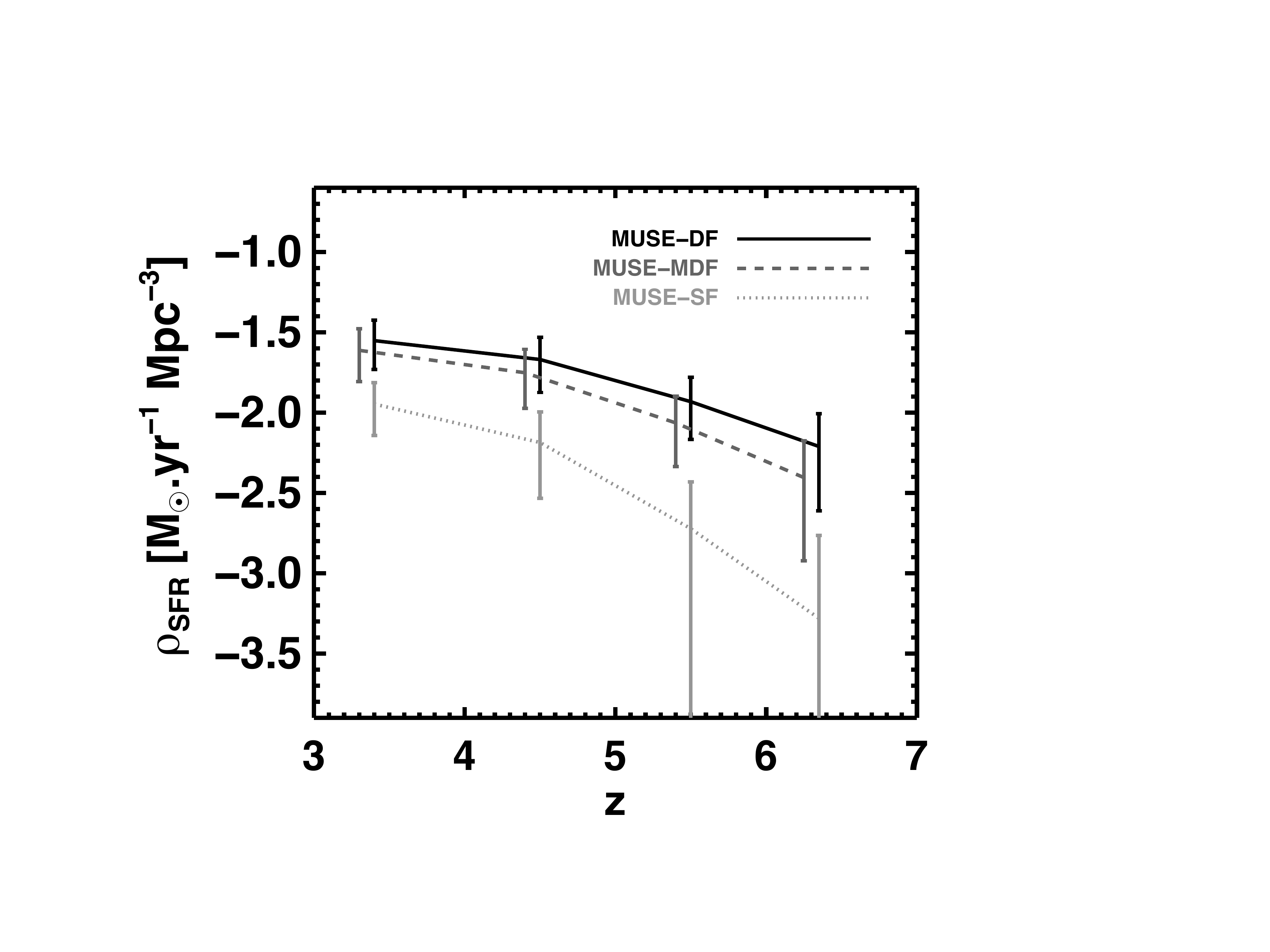}
\vskip-1ex 
\caption{Evolution of the SFR density. The black, dark grey, and light grey curves represent the SFR density, $\rho_{\rm SFR}$, that we expect to probe with the \lyat-emitting galaxies in typical MUSE Deep Field (DF - $F_{\rm Ly\alpha} \geq 4 \times 10^{-19}$ \ergscm), Medium-Deep Field (MDF - $F_{\rm Ly\alpha} \geq 10^{-18}$ \ergscm), and Shallow Field (SF - $F_{\rm Ly\alpha} \geq 10^{-17}$ \ergscm) surveys respectively. Each curve shows the mean $\rho_{\rm Ly\alpha}$ measured from 5000 lightcones of 10 \sqarcm{} in four redshift bins: $2.8 < z < 4$, $4 < z < 5$, $5 < z < 6$, and $6 < z < 6.7$. The error bars correspond to the 1$\sigma$ standard deviations. The dark grey curve has been shifted horizontally by 0.1dex for the sake of clarity.}
\label{fig:sfr_density}
\vskip-2ex 
\end{figure}

Similar trends are seen in Figure \ref{fig:sfr_density} where we plot the predicted cosmic SFR density, $\rho_{\rm SFR}$, to be probed by typical MUSE surveys. Again, we see that deeper \lya surveys are expected to unveil sources that make a significant contribution to the cosmic SFR density compared to existing samples of brighter LAEs. Compared to the SF survey, we predict that \lyat-emitting galaxies to be found in the MDF (DF) survey are likely to increase the global SFR budget by a factor of 2 at $z$ $\approx$ 3 and a factor of 7 at $z$ $\approx$ 6 (x2.5 at $z$ $\approx$ 3 and x10 at $z$ $\approx$ 6 for the DF survey).

Overall, we predict that the faint LAEs to be found in MUSE Deep and Medium-Deep surveys make a larger contribution to the global cosmic \lya luminosity density and SFR density compared to brighter galaxies seen in the SF survey or current wide-field NB surveys. The values quoted above remain somehow dependent on the exact faint-end slope of the \lya luminosity function. The LF being still non-constrained at such extremely low fluxes, our predictions for the DF survey will need to be tested, in particular by MUSE surveys themselves. At $z$ $\approx$ 3 and 6, the number counts predicted by our model reasonably agree with the data of \citet{rauch08} and \citet{dressler2015a} (see Figure \ref{fig:lya_counts_all}), which reached \lya fluxes of approximatively $10^{-18}$ \ergscm, so we expect our predictions for the MDF survey to be reliable enough.\\

In conclusion, it appears that a MUSE survey over 10 \sqarcm down to $F_{\rm Ly\alpha} \geq 10^{-18}$ \ergscm, i.e. a MDF survey, represents an optimal strategy to probe a large fraction of SFR density and to minimise cosmic variance as it seems to provide the best trade-off between scientific gain and telescope time.

\section{The role of LAEs in the hierarchical context}
\label{sec:desc-prog}

In CDM cosmology, galaxy formation is described within the hierarchical clustering scenario in which dark matter halos grow through the accretion of smaller structures. Hybrid models of galaxy formation, e.g. GALICS, are based on this scheme, and they use cosmological $N$-body simulations to follow the evolution of the DM density field. The identification of virialised halos at each simulation output timestep, and the reconstruction of the history of these halos are stored in order to compute the baryonic physics as a post-processing step, and then describe the evolution of galaxies. In this context, the hybrid method is thus an extremely powerful tool to study the formation and merging history of a population of galaxies. 

In this section, we perform a merger tree analysis to investigate the connection between the host halos of high-redshift LAEs and nowadays halos. We identify in our simulation the $z$ $=$ 0 descendants of high-redshift \lya sources to be detected by the various MUSE surveys, and conversely, the progenitors of local objects, and in particular the building blocks of Milky Way (MW)-like halos. In the following, we will focus on the progenitor/descendant link between $z$ $=$ 0 objects and the host halos of LAEs at two epochs which somehow bracket the wavelength range where \lya will be detectable by MUSE, $z$ $=$ 3 and $z$ $=$ 6. 

\subsection{The host halos of LAEs at high-redshift}
\label{subsec:mh_hiz}

\begin{figure}
\vskip-20ex
\hskip-4ex 
\includegraphics[width=9cm,height=12.5cm]{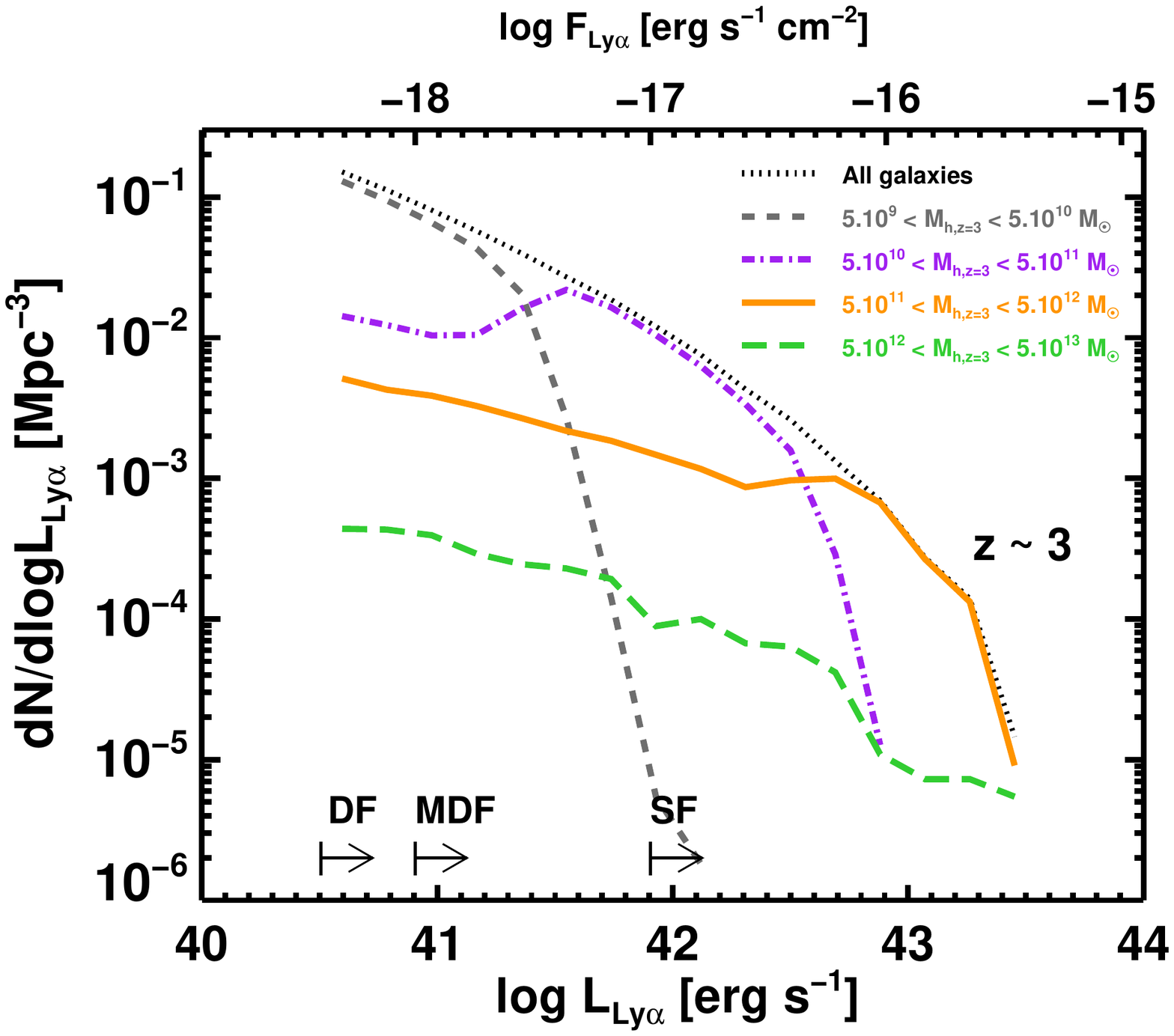}
\vskip-38ex 
\hskip-4ex 
\includegraphics[width=9cm,height=12.5cm]{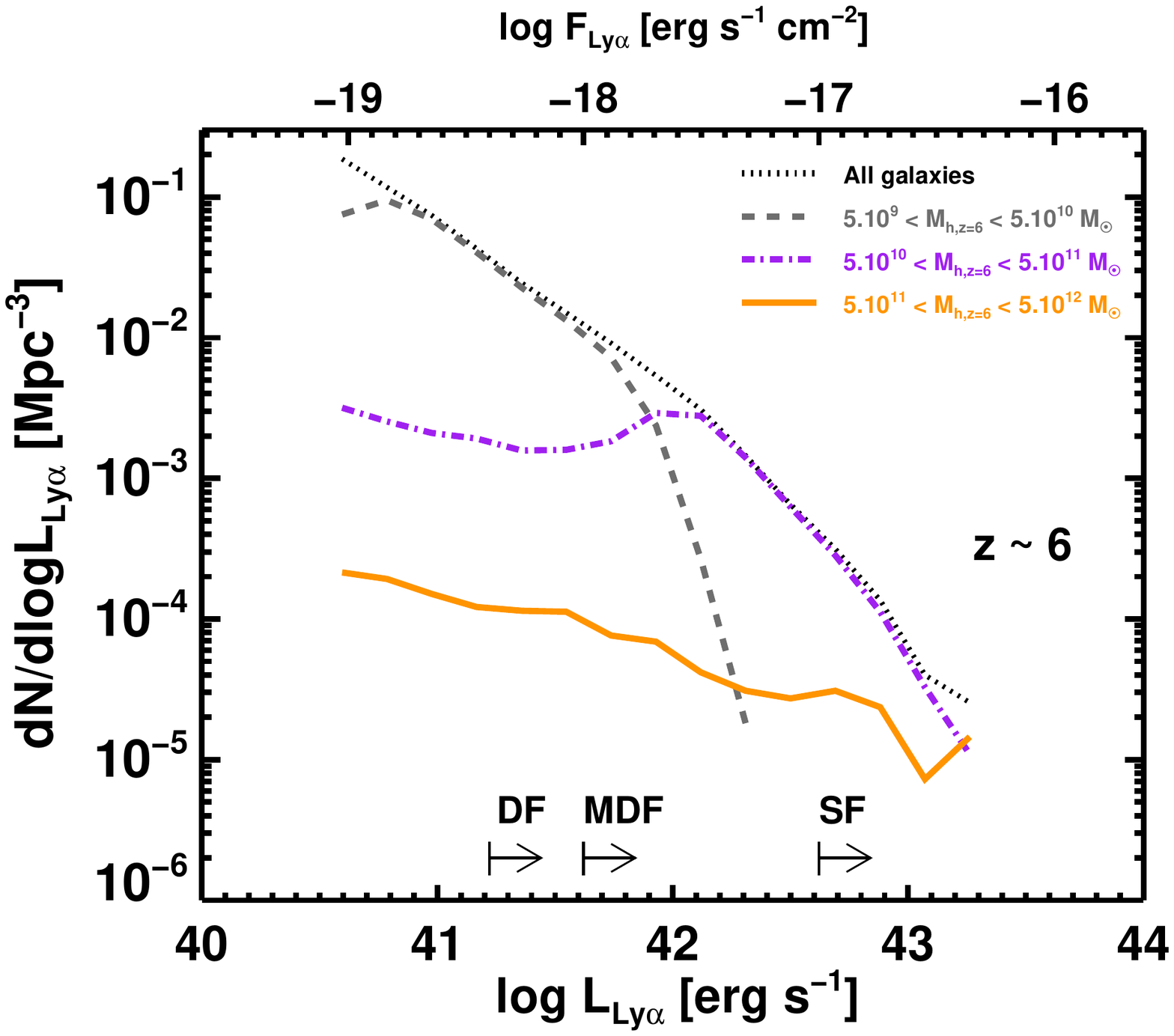}
\vskip-24ex 
\caption{\lya luminosity function at $z$ $=$ 3 (top panel) and $z$ $=$ 6 (bottom panel). The dotted curve shows the total \lya LF, while the other curves distinguish between the mass of the host halos of LAEs, as labelled.}
\label{fig:lyalf_mh_hiz}
\end{figure}

\begin{figure*}
\vskip-18ex 
\hskip-7ex
\includegraphics[width=1.0\textwidth,height=12.8cm]{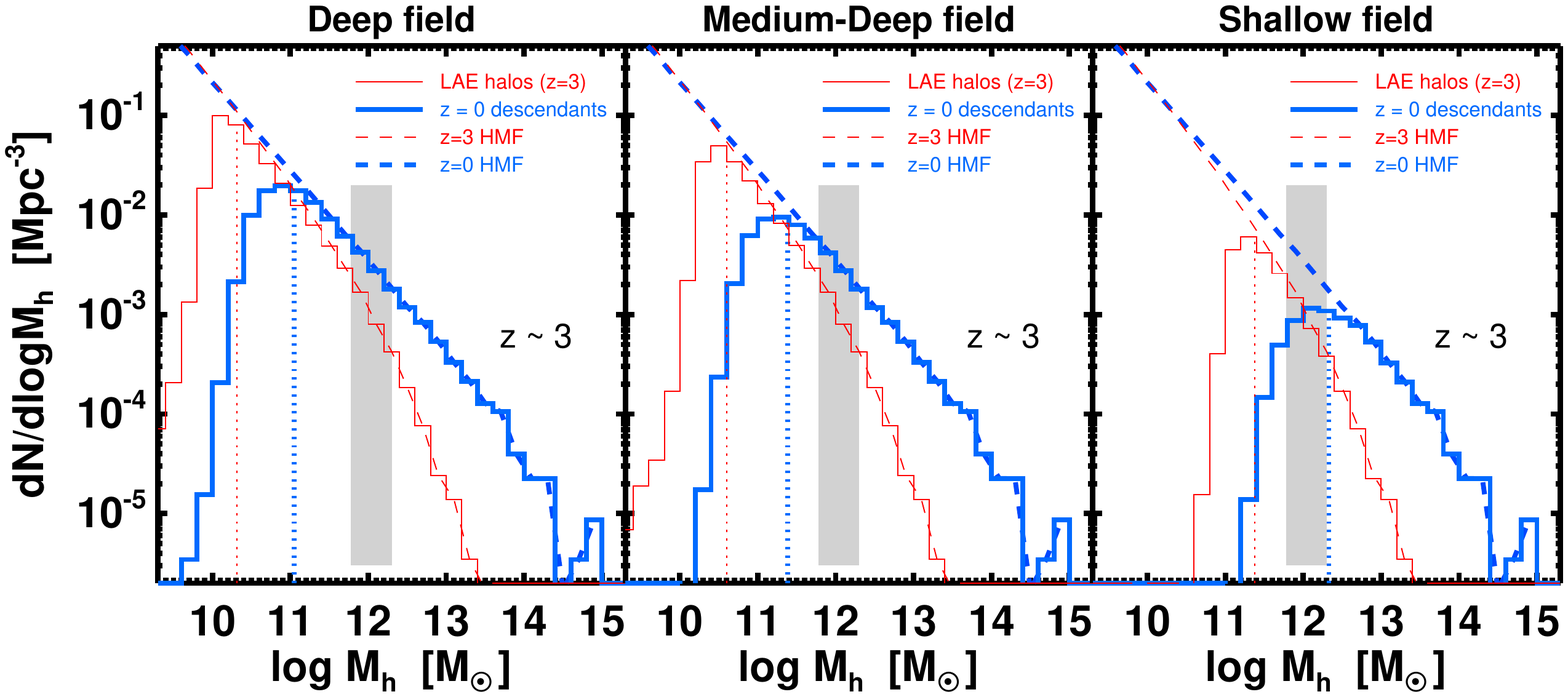}
\vskip-46ex 
\hskip-7ex
\includegraphics[width=1.0\textwidth,height=12.8cm]{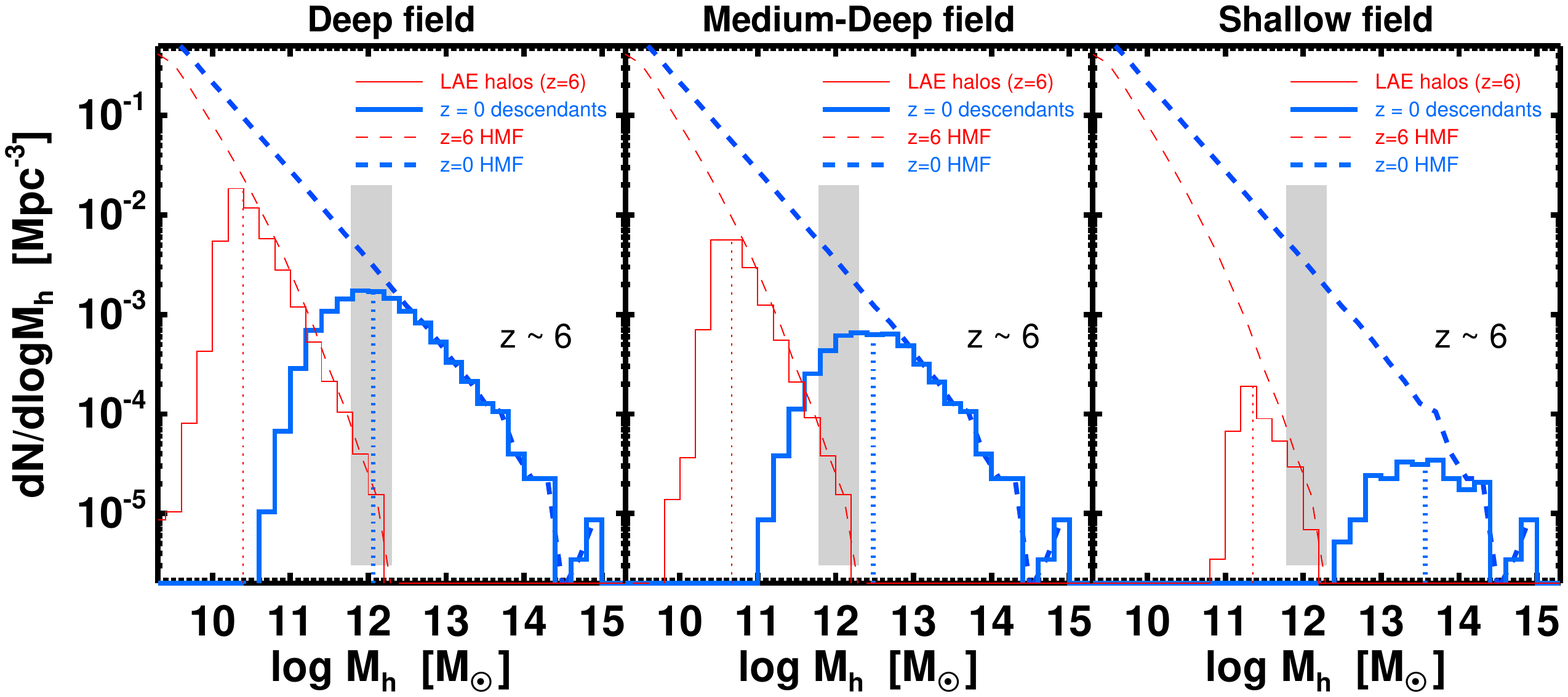}
\vskip-26ex 
\caption{Halo mass distributions of the $z$ $=$ 0 descendants of LAEs at $z$ $\approx$ 3 (top panels) and  $\approx$ 6 (bottom panels). The left, middle, and right panels correspond to LAEs selected in typical Deep Field ($F_{\rm Ly\alpha} \geq 4 \times 10^{-19}$ \ergscm), Medium-Deep Field (middle: $F_{\rm Ly\alpha} \geq 10^{-18}$ \ergscm), and Shallow Field (bottom: $F_{\rm Ly\alpha} \geq 10^{-17}$ \ergscm) surveys. The thin red histograms represent the distribution of the LAE host halos, while the thick blue histograms show their descendants at $z$ $=$ 0. The median masses of each distribution are represented by a vertical dotted line. The thin red and thick blue dashed lines illustrate the total halo mass function at $z$ $=$ 3/6 and $z$ $=$ 0 respectively. The mass estimate of the Milky-Way halo is shown by the grey shaded area \citep[$6 \times 10^{11} < {\rm M}_{\rm h,z=0} < 2 \times 10^{12}$ \msun;][]{battaglia2005}.}
\label{fig:hiz_z0_mh_dist}
\end{figure*}

Here, we explore the dynamical range spanned by LAEs at high redshift as a function of \lya luminosity as predicted by our model. In Figure \ref{fig:lyalf_mh_hiz}, we plot the \lya luminosity functions at $z$ $=$ 3 and $z$ $=$ 6 and we highlight the contribution of subsamples of LAEs split by host halo mass. The first thing to note is that LAEs located in low-mass halos make the faint-end of the \lya LF (short-dashed grey curve), while more massive halos host brighter LAEs (dot-dashed purple, solid orange and long-dashed green curves). Hence, at a given redshift, deeper surveys probe lower mass halos. This is simply because more massive halos accrete more gas, so the galaxies they host have higher SFR, hence higher intrinsic \lya luminosity. In each range of halo mass, the highest \lya luminosity allowed is set by the maximal gas accretion rate taking place in most massive halos. According to our model, LAEs currently seen by NB surveys ($L_{\rm Ly\alpha} \gtrsim 10^{42}$ \ergs) are predominantly hosted by halos with masses of $5 \times 10^{10-12}$ \msun. We expect the majority of faint sources in typical MUSE DF and MDF surveys to inhabit much less massive halos, i.e. $5 \times 10^{9-10}$ \msun.

Second, for a given halo mass range, we see that the \lya LF extends to lower luminosities. In our model, we do not identify and neither follow substructures, so each halo may contain more than one galaxy. Massive halos usually host one central galaxy and many satellites. As the gas supply from diffuse accretion only feeds the central galaxy of a given halo, satellites are more likely to display a fainter intrinsic emission than the central source. The extending tail towards low \lya luminosities is then mainly populated by the large number of satellites. Intrinsically \lyat-bright, central, galaxies make an additional, though minor statistically speaking, contribution to this. As extensively discussed in \citet{garel2015a}, the attenuation of the \lya line due to resonant scattering is small in low-mass LAEs because of their low dust content. However, the \lya escape fraction can be very low in more massive, intrinsically \lyat-bright, objects with large \hi column density and dust opacity, redistributing these galaxies at the faint end of the LF.

\subsection{The descendants of the LAE host halos}

Using the information stored in the merger trees, we can now investigate the link between the host halos of the high redshift sources to be detected by typical MUSE surveys and their descendants in the local Universe. Figure \ref{fig:hiz_z0_mh_dist} shows the halo mass distributions of LAEs (thin red histograms) at $z$ $\approx$ 3 (top panel) and $z$ $\approx$ 6 (bottom panel) in the three surveys (Deep, Medium-Deep, and Shallow fields). Unsurprisingly, the brightest \lya galaxies at high redshift, as those in the SF survey, are hosted by the most massive halos (see Section \ref{subsec:mh_hiz}). When fainter sources are considered (i.e. with the DF and MDF surveys), the host halos sample the lower mass end of the halo mass function (HMF - thin red dashed lines). It is interesting to point out that the median mass of LAE halos, illustrated by the vertical dotted lines, evolves weakly from $z$ $\approx$ 6 to $z$ $\approx$ 3 for all three samples considered here. At both redshifts, the (log) median halo mass is approximatively 10.3, 10.7 and 11.3 \msun{} in the DF, MDF, and SF surveys respectively\footnote{For the halo mass distributions shown in Figure \ref{fig:hiz_z0_mh_dist}, we only count halos (resp. halo descendants) which contain at least one galaxy (resp. one progenitor galaxy) brighter than $F_{\rm Ly\alpha}^{\rm limit}$. Given that massive halos are more likely to host more than one galaxy, the median halo masses that we quote would be higher if we were associating one halo to each LAE instead.}.

In each panel of Figure \ref{fig:hiz_z0_mh_dist}, we also plot the descendant halo distributions at $z$ $=$ 0 (thick blue histograms) for each corresponding LAE sample. Again, we see that the descendants of the halos of the brightest high-redshift LAEs make the high-mass end of the $z$ $=$ 0 HMF (thick blue dashed line), whereas the hosts of fainter \lya sources evolve into less massive halos on average. The descendants of LAEs in the DF and MDF surveys at $z$ $\approx$ 3 span a mass range from $\approx 10^{10}$ to $10^{15}$ \msun, with a median value around $10^{11}$ \msun. The brighter sources of the SF survey are predicted to end up in halos more massive than $10^{11}$ \msun{} at $z$ $=$ 0, with a median mass of $\approx 2 \times 10^{12}$ \msun{} which corresponds to the upper limit estimate of the Milky-Way (MW) halo mass (grey stripe) reported by \citet{battaglia2005}.

At $z$ $\approx$ 6, we predict that a typical SF survey would probe halos that have very massive descendants at $z$ $=$ 0 ($M_{\rm h}^{\rm med} \approx 5 \times 10^{13}$ \msun), e.g. group/cluster galaxy halos. The DF and MDF surveys are expected to probe LAEs which evolve into lower mass halos at $z$ $=$ 0 ($\gtrsim 10^{11}$ \msun), with median masses of the same order as the MW dark matter halo \citep[i.e. $M_{\rm h} \approx 10^{12}$ \msun;][]{battaglia2005,mcmillan2011,phelps2013,kafle2014}.

\subsection{The high-redshift progenitors of $z$ $=$ 0 halos}

\begin{figure}
\vskip-20ex 
\hskip-8ex 
\includegraphics[width=9cm,height=12cm]{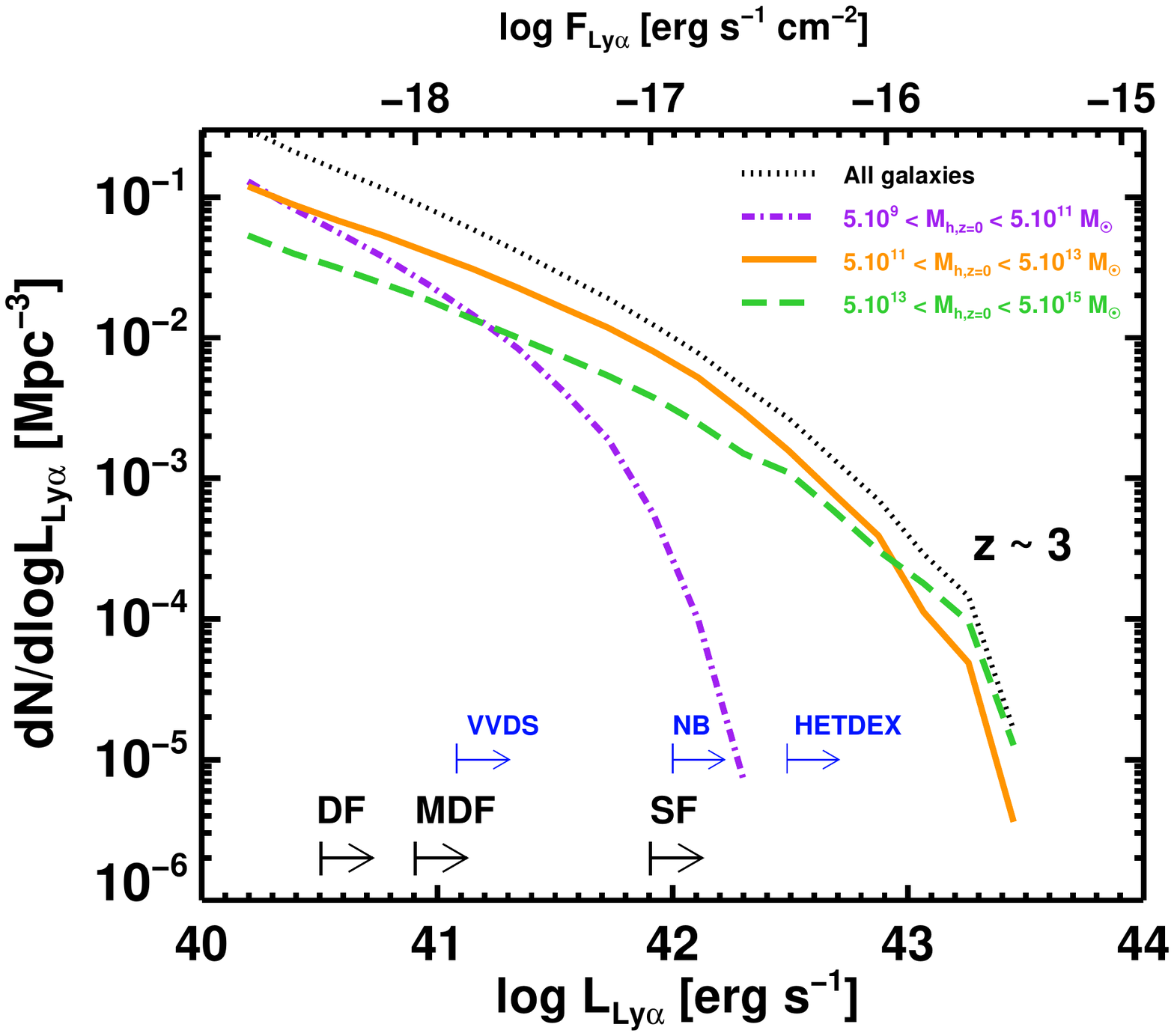}
\vskip-36ex 
\hskip-8ex 
\includegraphics[width=9cm,height=12cm]{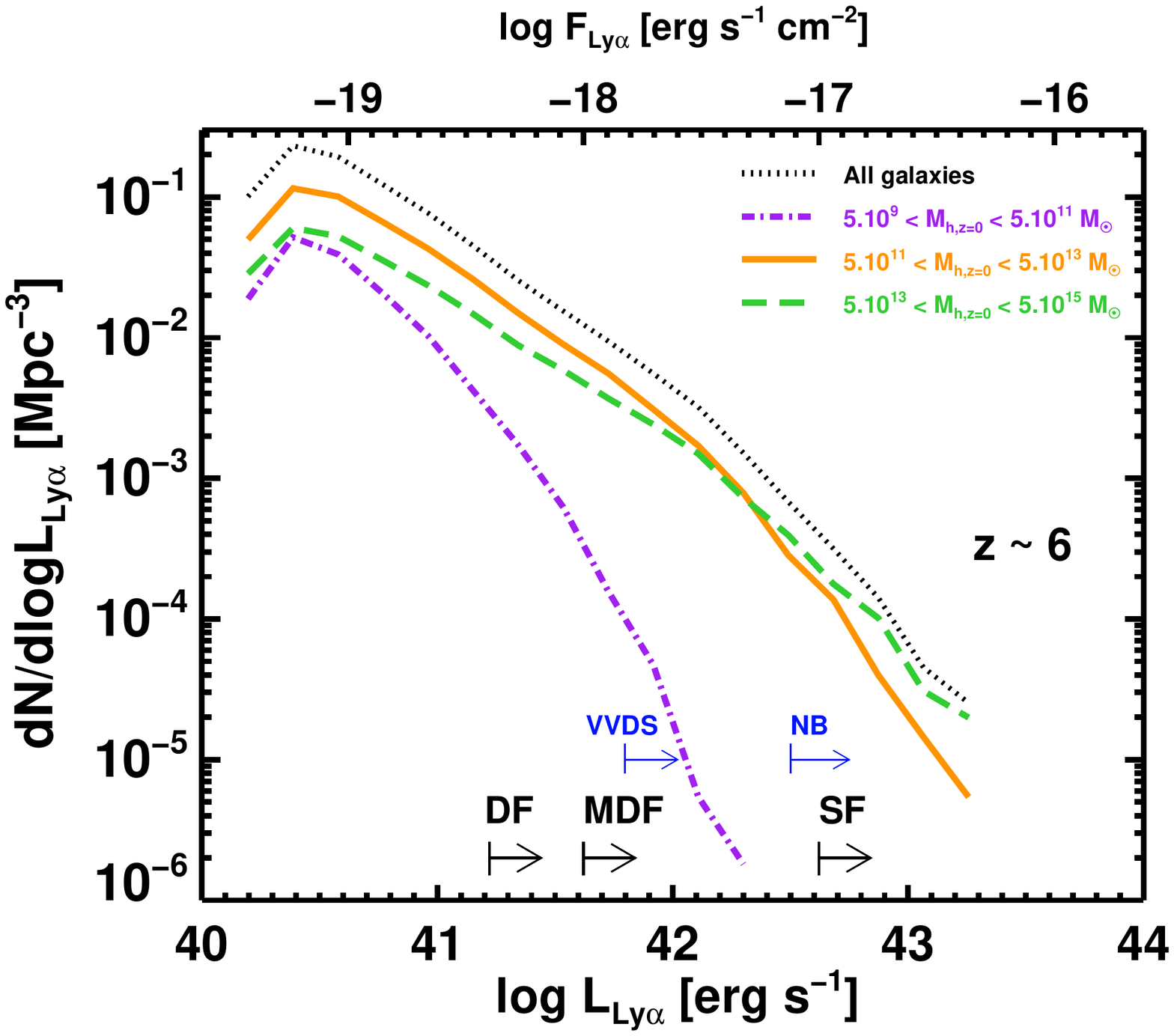}
\vskip-22ex 
\caption{\lya luminosity function at $z$ $=$ 3 (top panel) and $z$ $=$ 6 (bottom panel). The dotted curve shows the total \lya LF. The other curves correspond to the \lya luminosity distributions of LAEs residing in the progenitors of $z$ $=$ 0 halos divided in three mass ranges (see legend). Black arrows with labels illustrate the \lya detection limits of typical MUSE Deep, Medium-Deep, and Shallow field surveys. For comparison, we include the minimum \lya fluxes (blue arrows) reached by the VVDS \citep[][]{cassata2011a}, by the HETDEX pilot survey \citep[][]{blanc2011a}, and by current narrow-band surveys \citep[using the thresholds of][at $z$ $=$ 3.1 and $z$ $=$ 5.7]{ouch08}.}
\label{fig:lyalf_mhz0}
\end{figure}

Having discussed the local descendants of LAE host halos at different \lya luminosities in the previous section, we now attempt to assess how LAEs trace the progenitors of $z$ $=$ 0 halos. This is illustrated in Figure \ref{fig:lyalf_mhz0} where we show the \lya LFs at $z$ $=$ 3 (top panel) and $z$ $=$ 6 (bottom panel) for three halo mass ranges at $z$ $=$ 0. In both panels, the dotted black curves give the total \lya LF, whereas the dot-dashed purple, solid orange, and long-dashed green curves correspond to the distribution of the progenitors of halos with masses of $5 \times 10^{9} < {\rm M}_{\rm h,z=0} < 5 \times 10^{11}$,   $5 \times 10^{11} < {\rm M}_{\rm h,z=0} < 5 \times 10^{13}$, and  $5 \times 10^{13} < {\rm M}_{\rm h,z=0} < 5 \times 10^{15}$ \msun.

According to our model, the progenitors of halos in the lowest mass bin are mainly hosting faint LAEs at high redshift (i.e. $L_{\rm Ly\alpha} \lesssim 10^{42}$ \ergs). These objects are nearly never detected in typical NB surveys at $z$ $=$ 3-6 or in the HETDEX spectroscopic pilot survey \citep[z $\lesssim$ 3.8][]{blanc2011a}, and are unlikely to be probed in a MUSE Shallow field survey. Typical Deep and Medium-Deep surveys would probe these faint LAEs, adding up to the existing samples of \citet{rauch08}, \citet{cassata2011a}, and \citet{dressler2015a}. 

The most massive halos at $z$ $=$ 0 (green curves), corresponding mainly to the hosts of massive early-type galaxies \citep{van-den-bosch2003,mandelbaum2006,yang2009}, are predicted to be made up of the brightest LAEs at high redshift. The bulk of their progenitors is however composed of fainter \lya sources, that are either (i) satellite galaxies in massive halos at high redshift, or (ii) central galaxies in low-mass halos which were accreted to form very massive halos towards $z$ $=$ 0 through hierarchical merging. The \lya distribution of the progenitors of the medium-mass halos (orange curves) spans a similar range, from the highest luminosities towards the faint end, but is steeper than for very massive halos. These intermediate-mass halos are thought to be predominantly the locus of L$^{\star}$, late-type galaxies like our Galaxy, as the range $5 \times 10^{11} < M_{\rm h,z=0} < 5 \times 10^{13}$ \msun{} broadly encompasses the halo mass of a MW-like galaxy, estimated to be $0.8^{+1.2}_{-0.2} \times 10^{12}$ \msun{} \citep{battaglia2005}. Nevertheless, we see from Figure \ref{fig:lyalf_mhz0} that the distributions of LAEs residing in the progenitors of $z$ $=$ 0 objects vary quickly as a function of halo mass, so the orange curve might not represent accurately the predicted progenitors distribution of MW-like objects. We will then concentrate on the progenitors of MW-like halos in the next section, and we will compare our results with other theoretical studies \citep[e.g.][]{gawiser2007a,salvadori2010} in Section \ref{sec:discussion}.

\subsection{The high-redshift progenitors of MW-like halos}
\label{subsec:mw_progs}

\begin{figure}
\vskip-4ex 
\hskip-6ex 
\includegraphics[width=9.55cm,height=7.67cm]{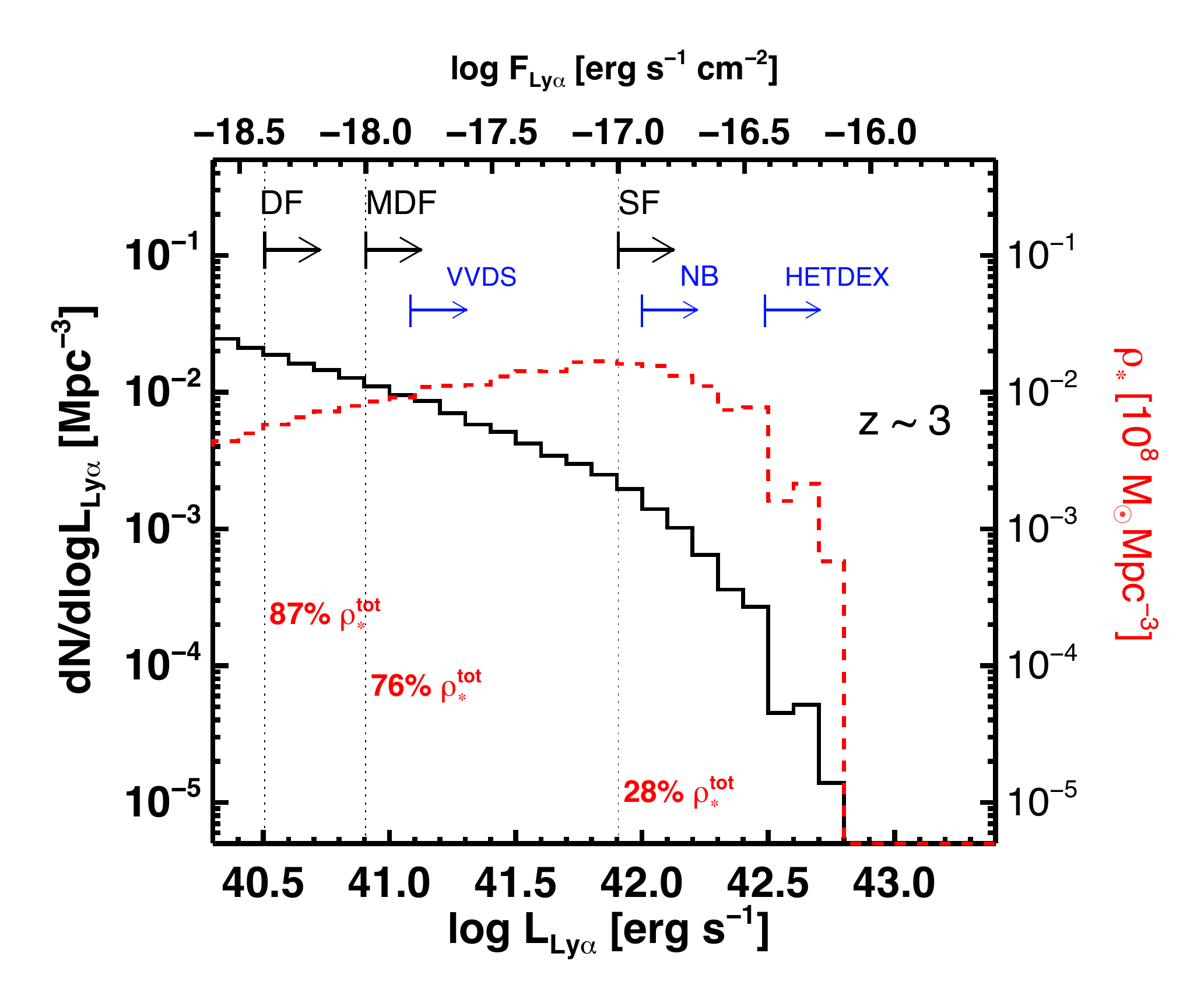}
\vskip1ex 
\hskip-6ex
\includegraphics[width=9.72cm,height=7.47cm]{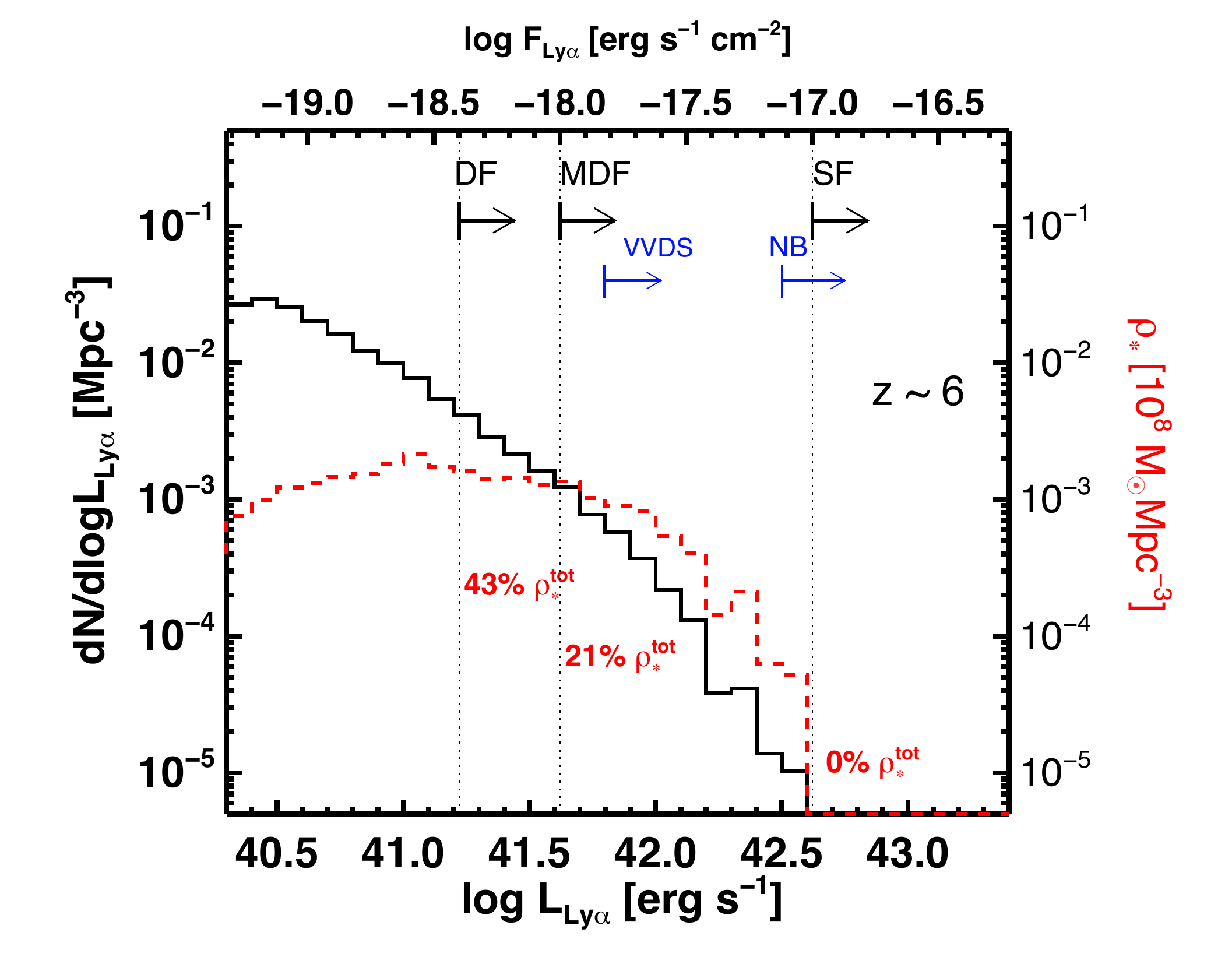}
\vskip-2ex
\caption{Distribution of the LAEs residing in the progenitor halos of MW-like halos at $z$ $\approx$ 3 (top panel) and $z$ $\approx$ 6 (bottom panel). In each panel, the black histogram shows the \lya luminosity distribution of the galaxies hosted by the progenitors of MW-like halos, while the red histogram gives the stellar mass density per $L_{\rm Ly\alpha}$ bin. The arrows show the \lya flux limits of the three typical MUSE fields (black) and others existing surveys (see caption of Figure \ref{fig:lyalf_mhz0}).  The red labels represent the fraction of the total stellar mass density sitting in the progenitors of MW-like halos, $\rho_{\rm *}^{\rm tot}$, that can be probed by LAEs in the different typical MUSE surveys.}
\label{fig:mwprog_z0}
\vskip-2ex
\end{figure}

In Figure \ref{fig:mwprog_z0}, we plot the \lya luminosity distribution (black histogram) at $z$ $\approx$ 3 and $z$ $\approx$ 6 of the LAEs residing in the progenitors of $z$ $=$ 0 halos with $6 \times 10^{11} < M_{\rm h,z=0} < 2 \times 10^{12}$ \msun{}, that we define as MW-like halos in what follows. We first note that current NB surveys are only able to probe the progenitors of MW-like halos which host LAEs with \lya luminosities $\gtrsim$ $10^{42}$ \ergs{} at $z$ $\approx$ 3 and $z$ $\approx$ 6. The vast majority of the progenitors of MW-like halos contain LAEs with fainter luminosities, which number density keeps increasing towards lower values, even below the MUSE DF limit. In the bottom panel of Figure \ref{fig:mwprog_z0} (z $\approx$ 6), the apparent flattening of the distribution at $L_{\rm Ly\alpha} \approx 3 \times 10^{40}$ \ergs{} is due to the limit of resolution of our simulation, and the curve would start decreasing at lower luminosities (see Section \ref{subsec:resol}). At $z$ $\approx$ 3 (top panel), a similar effect would be seen at $L_{\rm Ly\alpha} \lesssim 2 \times 10^{40}$ \ergs. In practice, this means that we \textit{miss} galaxies located in halos less massive than our resolution limit, and the number distribution of LAEs should keep increasing down to lower luminosities if we were using a higher resolution simulation. Even though these very faint LAEs are obviously more numerous than the sources to be detected by MUSE surveys, they consist of low mass objects, forming stars a very low rate, and they represent a small fraction of the overall SFR and stellar mass budget.

To illustrate this point, we also show on Figure \ref{fig:mwprog_z0} the stellar mass density in the high redshift progenitors of MW-like halos per bin of log$L_{\rm Ly\alpha}$, $\rho_{\rm *}$ (red histogram). Given that stellar mass is well correlated to SFR (see Figure \ref{fig:sfr_ms}), and that the intrinsic \lya intensity is directly proportional to SFR to first order \citep[see e.g. Eq. 8 of][]{barnes2014}, it is not surprising that the brightest LAEs make a significant contribution to the stellar mass density. As shown on Figure \ref{fig:mwprog_z0}, $\rho_{\rm *}$ increases faster than the number density from high to low \lya luminosities. This is especially true at $z$ $\approx$ 3, where $\rho_{\rm *}$ reaches a maximum at $L_{\rm Ly\alpha} \approx 10^{42}$ \ergs, and starts declining towards fainter \lya luminosities. This roughly corresponds to the SF survey limit ($F_{\rm Ly\alpha} \geq 10^{-17}$ \ergscm), and the model predicts that 28\% of the total stellar mass density (SMD) sitting in the progenitors of MW-like halos can be probed in this case. A significantly higher fraction is expected to be recovered from faint LAEs in typical MDF and DF surveys (i.e. 0.76$\rho_{\rm *}^{\rm tot}$ and 0.87$\rho_{\rm *}^{\rm tot}$ respectively). This implies that these deep surveys could probe the bulk of the $z$ $=$ 3 progenitors of local galaxies like ours according to our model. At $z$ $\approx$ 6, the progenitors of the MW-like halos will not be traced by LAEs in a SF survey. However, we expect the LAE sample of a MDF survey to contain about 21\% of the total SMD in the progenitors of MW-like halos. Moreover, almost half of the stars present in the $z$ $=$ 6 progenitors of MW-like halos should be sitting in LAEs detectable in a typical Deep-Field survey.

As mentioned earlier, the resolution limit of our simulation implies that our sample of LAEs is not complete below a given \lya flux, as we miss galaxies which should form in halos less massive than $M_{\rm halo}^{\rm min}$. The real value of $\rho_{\rm *}^{\rm tot}$ is then unknown, so the absolute contributions to the SMD quoted in the previous paragraph must be viewed as upper limits. Determining $\rho_{\rm *}^{\rm tot}$ accurately is quite uncertain since we would need to make assumptions about the number density of extremely faint galaxies and the halo mass at which galaxy formation is prevented \citep[e.g. due to photoheating from the ionising background;][]{okamoto08}. 

Nevertheless, if we look at the relative contribution to the SMD between the different typical MUSE surveys, we are no longer affected by mass resolution effects. We then compare the SMD probed by typical MUSE surveys relatively to the NB surveys of \citet{ouch08} at $z$ $\approx$ 3 ($F_{\rm Ly\alpha} \geq 1.2 \times 10^{-17}$ \ergscm) and $z$ $\approx$ 6 ($F_{\rm Ly\alpha} \geq 8 \times 10^{-18}$ \ergscm). At $z$ $\approx$ 3, the SF, MDF, and DF surveys are predicted to recover a stellar mass content in LAEs hosted by the progenitors of MW-like halos $\approx$ 1.25, 3, and 4 times larger than the NB survey of \citet{ouch08}. At $z$ $\approx$ 6, we find that \citet{ouch08} only probe the very-bright end of the \lya LF so these numbers go up to 100 and 200 for the MDF and DF surveys respectively. Comparing instead with the VVDS survey which obtained the faintest existing sample at this redshift, we expect the fraction of the total mass density in LAEs located in the progenitors of MW-like halos to be $\approx$ 1.5 and 3.5 times larger for typical MDF and DF surveys respectively.

\section{Discussion}
\label{sec:discussion}

\begin{table*}
\vskip2ex
Fraction of MW-like halos with at least one progenitor at $z$ $\approx$ 3 and $z$ $\approx$ 6.
\begin{center}
\begin{tabular}{ccccccc}
    \hline
    \hline
    & & \:  $F_{\rm Ly\alpha} \geq 0^{(1)}$ \: & \:  DF$^{(2)}$ \: & \:  MDF$^{(3)}$ \: & \:  SF$^{(4)}$ \: &  \: $L_{\rm Ly\alpha} \geq 10^{42} \:{}^{(5)}$\\
    \hline
    \multicolumn{1}{c}{\multirow{2}{*}{\lya flux/lum. cut only}} \:  \: & $z$ $\approx$ 3  \: & 0.98 & 0.97 & 0.96 & 0.30 &0.20 \\
   & $z$ $\approx$ 6  \: & 0.97 & 0.50 & 0.16 & 0.00 & 0.025\\
    \hline
     \multicolumn{1}{c}{\multirow{2}{*}{\lya flux/lum. cut \& $M_{\rm 1500}$ $<$ -18}} \:  \: & $z$ $\approx$ 3  \: & 0.44 & 0.43 & 0.43 & 0.30 & 0.19 \\
   & $z$ $\approx$ 6  \: & 0.09 & 0.09 & 0.08 & 0.00 &0.025\\
   \hline
\end{tabular}
\vskip2ex
\caption{Column (1) gives the fraction of MW-like $z$ $=$ 0 halos with one or more progenitor halos at $z$ $=$ 3 and $z$ $=$ 6, irrespectively from the \lya flux of the galaxy they host (i.e. $F_{\rm Ly\alpha}$  $\geq$ 0 \ergscm). Columns (2), (3) and (4) correspond to LAEs detectable in typical Deep Field (DF), Medium-Deep Field (MDF), and Shallow Field (SF) surveys. Column (5) corresponds to LAEs detectable with an observed \lya luminosity greater than $10^{42}$ \ergs.
The first two rows show the fractions of MW-like halos which progenitors have a LAE selected above the quoted \lya flux/luminosity limits only. The two last rows show the fractions for LAEs which are also detectable as LBG in typical dropout surveys \citep[i.e. with an absolute rest-frame UV magnitude at 1500 \AA{} brighter than $-18$;][]{bouwens,van-der-burg2010,duncan2014}.}
\label{table:frac_mw_progs}
\end{center}
\vskip-2ex
\end{table*}

The role of high redshift LAEs in the mass assembly of local galaxies has been discussed in a few previous studies, based either on the redshift evolution of the observed LAE bias, cosmological simulations, or a combination of them. \citet{gawiser2007a} performed a clustering analysis on 162 $z$ $=$ 3.1 LAEs from the sample of \citet{gronwall07}, and they derived a median halo mass $M_{\rm h}^{\rm med} \approx 10^{11}$ \msun. For comparison, \citet{kovac2007a} find $M_{\rm h}^{\rm med} \approx 1.5-3 \times 10^{11}$ \msun{} at $z$ $=$ 4.5, and \citet{ouchi2010a} report that the host dark matter halos of LAEs remain in the range $10^{11\pm1}$ \msun{} from $z$ $\approx$ 3 to 7. For similar LAE selection ($L_{\rm Ly\alpha} \gtrsim$ a few $10^{42}$ \ergs{} and $EW_{\rm Ly\alpha} \gtrsim 20$ \AA), we find that $M_{\rm h}^{\rm med}$ increases from 1 to 3 $\times$ $10^{11}$ \msun{} from $z$ $\approx$ 6 to $z$ $\approx$ 3, in broad agreement with the observations \citep[see Section 6 of][for more details]{garel2015a}. A more robust way to assess the expected $L_{\rm Ly\alpha}-M_{\rm h}$ relation is to quantitatively examine the spatial distribution of LAEs and compare with observational data. To this aim, we will investigate the two-point correlation functions of LAEs in a future study (Garel et al., in preparation).

Using merger trees from the MilliMillenium simulation, \citet{gawiser2007a} identify the $z$ $=$ 0 descendants of LAEs at $z$ $\approx $ 3 to have a median halo mass of $\approx 1.2 \times 10^{12}$ \msun. This reasonably matches our predictions for a MUSE SF survey ($M_{\rm h}^{\rm med} \approx 2 \times 10^{12}$ \msun; top right panel of Figure \ref{fig:hiz_z0_mh_dist}), which is expected to detect similar LAEs as the ones investigated by \citet{gawiser2007a}. 
Part of the difference might be due to the different cosmology assumed in the studies (based on $WMAP$-1 and $WMAP$-5 releases respectively). In addition, \citet{gawiser2007a} used the minimum LAE host halo mass derived from their clustering analysis, $3 \times 10^{11}$ \msun, to perform the merger tree study, whereas we use the full information provided by our model, i.e. the \lya luminosities of galaxies and their dark matter host halos. Using the same data as \citet{gawiser2007a}, \citet{walker-soler2012} developed abundance-matching models of LAEs to track their evolution in the Millennium-II simulation. They also report that descendants of $z$ $\approx$ 3 LAEs selected above $L_{\rm Ly\alpha} \gtrsim 10^{42}$ \ergs{} have halo masses typical of L$^{*}$-galaxies, i.e. $\approx 10^{12}$ \msun.

A complementary question is to wonder if high redshift LAEs are located in the main progenitors of present day MW-like halos. In Section \ref{subsec:mw_progs}, we tracked the progenitors of MW-like halos at $z$ $=$ 3 and 6 using our merger trees, and we found that the brightest sources in these halos have $L_{\rm Ly\alpha} \approx 5 \times 10^{42}$ \ergs, while most progenitors of MW-like halos host faint LAEs. Similar results are reported by \citet{yajima2012} who combined a cosmological hydrodynamical simulation with 3D radiative transfer calculations of the 60 most massive progenitors at $z$ $\lesssim$ 10 (see their Figure 7).  The study of \citet{yajima2012} focussed on one single MW-like galaxy (and their progenitors) in a zoomed-in region and their initial conditions were set especially to model a MW-size galaxy at $z$ $=$ 0. Contrary to them, we have identified all MW-like host halos according to their mass in our simulation and looked at their high-z building blocks, which allows us to investigate their properties in a statistical way. We discussed in Section \ref{subsec:mw_progs} their predicted \lya luminosity distribution and stellar mass density. From our model, we can also try to estimate, for a given LAE survey, what fraction of MW-like halos will have high redshift progenitors that are detectable through the LAEs they contain.  From Table \ref{table:frac_mw_progs}, the fraction of MW-like halos \textit{with at least one LAE host halo as progenitor} at $z$ $\approx$ 3 in the DF, MDF surveys is very high, i.e. 0.97 and 0.96 respectively. This fraction is $\approx$ 4 times larger than for the Shallow Field or typical NB surveys. At $z$ $\approx$ 6, we predict that about half of present day MW-like halos will have a progenitor hosting a LAE in the DF survey. 

For a \lya detection threshold of $L_{\rm Ly\alpha} \gtrsim 10^{42}$ \ergs, \citet{salvadori2010} found that up to 68\% of MW-like halos have at least one LAE host halo as a progenitor at $z$ $\approx$ 6, while we find the percentage to be less than 3\%. The origin of the discrepancy is not obvious but it might come from the different definition of the $z$ $=$ 0 MW-like object used in this paper and in the study of \citet{salvadori2010}. Here, we used a hybrid model of galaxy formation that can match the \lya luminosity functions from $z$ $\approx$ 3 to 7 and we searched for galaxies located in halos at high redshift that are the progenitors of local halos, only selected from their mass ($6 \times 10^{11} < M_{\rm h,z=0} < 2 \times 10^{12}$ \msun). The model of \citet{salvadori2010}, based on the extended Press-Schechter theory, was instead adjusted to reproduce the $z$ $=$ 0 properties our Galaxy (e.g. stellar mass and metallicity) and its local environment, which corresponds to a high-density region. As they are investigating a highly biased region of the Universe, their predicted LAE abundance at $z$ $\approx$ 6 is much larger than the \textit{mean} number density as observed in current NB \lya surveys. 

In spite of the differences between the results of \citet{salvadori2010} and ours, which suggest that the contours of the population of MW-like progenitors might highly depend on how we define a MW-like galaxy and its environment, it is interesting to note that both models predict that almost all progenitors of MW-like halos traced by LAEs with $L_{\rm Ly\alpha} \gtrsim 10^{42}$ \ergs{} should also be probed in typical LBG surveys with $M_{\rm 1500}$ $\lesssim$ -18. We find that this is also true for all LAEs in the Shallow Field (see Table \ref{table:frac_mw_progs}). This seems very consistent with the work of \citet{gonzalez2012}, based on the Durham model, who finds that a MW-like galaxy has a 95\% (70\%) probability of having at least one LBG with $M_{\rm 1500}$ $\lesssim$ -18.8 as a progenitor at $z$ $\approx$ 3.5 (z $\approx$ 6.5). According to our model, only a smaller fraction of the \lya sources expected in deeper surveys, such as the DF and MDF, should have $M_{\rm 1500}$ $\lesssim$ -18, although they should be detectable in very deep UV-selected surveys \citep[e.g.][]{bouwens2015a}.\\

As discussed in Section \ref{subsec:lya_model}, a noticeable outcome of our \lya RT modelling in expanding shells is that the IGM becomes transparent to \lya photons emerging from galaxies. Assuming alternative scenarios in which most of the \lya flux emerges from galaxies close to the line centre (e.g. a Gaussian profile centered on $\lambda_{\rm Ly\alpha}$, or even a blue-shifted line in the presence of gas infall for instance), it would no longer be the case, especially at $z$ $\gtrsim$ 6 when reionisation is not necessarily complete yet. The impact on the visibility of LAEs would then depend on many factors, such as the exact form of the intrinsic \lya line, feedback, star formation rate, source clustering, or the structure, the kinematics, and the ionisation state of the local IGM \citep[e.g.][]{dijk07a,mcquinn2007b,iliev2008a,dayal2012a,hutter2015a}. As for the present study, should the \lya transmission be much less than unity, LAEs may appear fainter and less progenitors of MW-like halos would be detectable with MUSE compared to the values quoted in Section \ref{subsec:mw_progs}. Similarly, MUSE surveys would thus probe a lower fraction of the global stellar mass budget located in the progenitors of MW-like halos.

Disentangling internal \lya radiative transfer effects and IGM transmission remains a complicated issue, which cannot be easily constrained directly by observations. Nevertheless, theoretical studies have shown that outflows can dramatically alter the shape and the position of the peak of the \lya line \citep[e.g.][]{santos04,verh06,dijkstra2011a}. Observationally, asymmetric profiles as well as velocity offsets between \lya and the systemic redshift are commonly measured both at high and low redshift \citep[e.g.][]{kunth98,shapley03,mclinden,wofford2013a,rivera-thorsen2015a}, which suggests that the IGM is not necessarily the cause of the flux reduction (or suppression) of the blue side of the \lya line and the velocity shift of the peak.

\section{Summary and conclusions}
\label{sec:summary}

In this paper, we presented model predictions for high-redshift \lya galaxies to be observed through a typical wedding cake observing strategy with MUSE from $z$ $\approx$ 2.8 to $z$ $\approx$ 6.7. We used the GALICS hybrid model to describe the formation and evolution of galaxies in the cosmological context and a grid of numerical models to compute the radiative transfer of \lya photons through dusty gas outflows. This model can reasonably reproduce the abundances of \lya emitters and Lyman-Break galaxies \citep{garel2015a}, as well as the stellar mass functions (Section \ref{subsec:data_comparison}), in the redshift range where MUSE will be able to probe the \lya emission line. We built mock lightcones of LAEs corresponding to typical Deep Field (DF), Medium-Deep Field (MDF), and Shallow Field (SF) surveys over 1, 10, 100 \sqarcm, and down to \lya fluxes of $4 \times 10^{-19}$, $10^{-18}$, and $10^{-17}$ \ergscm{} respectively.

A DF survey would yield the faintest statistical sample of LAEs ever observed, allowing to investigate the extreme faint slope of the \lya LF at high redshift. From our mock catalogues, we predict that $\approx$ 500 sources can be found between $z$ $\approx$ 2.8 and $z$ $\approx$ 6.7. At $F_{\rm Ly\alpha} \gtrsim 10^{-18}$ \ergscm, our model agrees well with the abundances of faint LAEs reported by \citet{rauch08} and \citet{dressler2015a} which suggest a steep faint-end slope of the \lya LF. MUSE is expected to compile a large sample of such faint sources,  as we predict $\approx$ 2,000 LAEs to be detected in a typical MDF survey. Furthermore, 1500 LAEs should be discovered in 100 \sqarcm with a shallower survey at fluxes greater than $\approx 10^{-17}$ \ergscm. Overall, we find that the main source of uncertainty will be cosmic variance, as it is often the case in small-volume, pencil-beam, surveys. In addition, our results suggest that the very faint galaxies to be seen in MUSE surveys, and usually missed by current optical surveys, will contribute significantly to the cosmic star formation rate budget at $z$ $\approx$ 3-7.

Based on our N-body dark matter simulation, we performed a merger tree analysis to assess the role of LAEs, and especially faint ones, in the hierarchical scenario of structure formation. We thus explored the link between the host halos of MUSE LAEs at high redshift and halos in the local Universe. On the one hand, we predict that bright LAEs ($F_{\rm Ly\alpha} \gtrsim 10^{-17}$ \ergscm) evolve, on average, into massive halos at $z$ $=$ 0, typical of host halos of massive ellipticals or galaxy groups. On the other hand, we find that faint LAEs at $z$ $\approx$ 3 (z $\approx$ 6) from typical DF and MDF surveys have a median halo mass of $\approx 10^{11}$ \msun{} ($\approx 10^{12}$ \msun), comparable to the halos of sub-L$^{*}$ (L$^{*}$) galaxies at $z$ $=$ 0. Finally, our study predicts that a large fraction of the high-redshift progenitors of MW-like halos can be probed by these surveys. For instance, a survey at $F_{\rm Ly\alpha} \gtrsim 4 \times 10^{-19}$ \ergscm{} is expected to probe the bulk of the global stellar mass budget enclosed in the $z$ $\approx$ 3 progenitors of MW-like host halos. 

In this paper, we have shown that deep surveys, e.g. with MUSE, can efficiently probe the population of faint \lyat-emitting galaxies at high redshift. The understanding of the formation and evolution of these sources appears to be essential to get insight into the mass assembly of local objects, such as the MW. In a future study, we will keep investigating the physical and spectral properties of galaxies in the early Universe fed by forthcoming MUSE data, as well as optical HST surveys \citep[e.g.][]{bouwens2015a}  and spectroscopic redshift surveys \citep[e.g.][]{le-fevre2015a}. \\

Mock catalogues and LAE number count predictions from Figure \ref{fig:counts1deg} and \ref{fig:lya_counts_all} are available at: \href{http://cral.univ-lyon1.fr/labo/perso/thibault.garel/}{\tt http://cral.univ-lyon1.fr/labo/perso/thibault.garel/}.

Additional information is available upon request at: \href{mailto:thibault.garel@univ-lyon1.fr}{thibault.garel@univ-lyon1.fr}.
 
\section*{Acknowledgements}
We thank the anonymous referee for his/her comments. TG is grateful to the LABEX Lyon Institute of Origins (ANR-10-LABX-0066) of the Universit\'e de Lyon for its financial support within the programme "Investissements d'Avenir" (ANR-11-IDEX-0007) of the French government operated by the National Research Agency (ANR). TG acknowledges support from an Australian Research Council SuperScience Fellowship. The authors thank L\'eo Michel-Dansac for the $N$-body simulation used in this study. This work was granted access to the HPC resources of CINES under the allocation 2012-c2012046642 made by GENCI (Grand Equipement National de Calcul Intensif).

\bibliographystyle{mn2e}
\bibliography{biblio_muse}

\begin{thebibliography}{}
 \providecommand{\href}[2]{#2}
  \providecommand{\doi}[1]{\href{http://dx.doi.org/#1}{doi:#1}}
  \providecommand{\eprint}[1]{\href{http://arxiv.org/abs/#1}{arXiv:#1}}

\bibitem[\protect\citeauthoryear{{Bacon} et~al.,}{{Bacon}
  et~al.}{2010}]{bacon2010a}
{Bacon} R.  et~al., 2010, in Society of Photo-Optical Instrumentation Engineers
  (SPIE) Conference Series. p.~8, \doi{10.1117/12.856027}

\bibitem[\protect\citeauthoryear{{Bacon} et~al.,}{{Bacon}
  et~al.}{2006}]{bacon06}
{Bacon} R.  et~al., 2006, The Messenger, 124, 5

\bibitem[\protect\citeauthoryear{{Barnes}, {Garel} \& {Kacprzak}}{{Barnes}
  et~al.}{2014}]{barnes2014}
{Barnes} L.~A.,  {Garel} T.,    {Kacprzak} G.~G.,  2014, \pasp, 126, 969

\bibitem[\protect\citeauthoryear{{Battaglia} et~al.,}{{Battaglia}
  et~al.}{2005}]{battaglia2005}
{Battaglia} G.  et~al., 2005, \mnras, 364, 433

\bibitem[\protect\citeauthoryear{{Birnboim} \& {Dekel}}{{Birnboim} \&
  {Dekel}}{2003}]{birnboim2003}
{Birnboim} Y.,  {Dekel} A.,  2003, \mnras, 345, 349

\bibitem[\protect\citeauthoryear{{Blaizot}, {Guiderdoni}, {Devriendt},
  {Bouchet}, {Hatton} \& {Stoehr}}{{Blaizot} et~al.}{2004}]{blaizot04}
{Blaizot} J.,  {Guiderdoni} B.,  {Devriendt} J.~E.~G.,  {Bouchet} F.~R.,
  {Hatton} S.~J.,    {Stoehr} F.,  2004, \mnras, 352, 571

\bibitem[\protect\citeauthoryear{{Blaizot}, {Wadadekar}, {Guiderdoni},
  {Colombi}, {Bertin}, {Bouchet}, {Devriendt} \& {Hatton}}{{Blaizot}
  et~al.}{2005}]{blaizot05}
{Blaizot} J.,  {Wadadekar} Y.,  {Guiderdoni} B.,  {Colombi} S.~T.,  {Bertin}
  E.,  {Bouchet} F.~R.,  {Devriendt} J.~E.~G.,    {Hatton} S.,  2005, \mnras,
  360, 159

\bibitem[\protect\citeauthoryear{{Blanc} et~al.,}{{Blanc}
  et~al.}{2011}]{blanc2011a}
{Blanc} G.~A.  et~al., 2011, \apj, 736, 31

\bibitem[\protect\citeauthoryear{{Bouwens} et~al.,}{{Bouwens}
  et~al.}{2009}]{bouwens09}
{Bouwens} R.~J.  et~al., 2009, \apj, 705, 936

\bibitem[\protect\citeauthoryear{{Bouwens}, {Illingworth}, {Franx} \&
  {Ford}}{{Bouwens} et~al.}{2007}]{bouwens}
{Bouwens} R.~J.,  {Illingworth} G.~D.,  {Franx} M.,    {Ford} H.,  2007, \apj,
  670, 928

\bibitem[\protect\citeauthoryear{{Bouwens} et~al.,}{{Bouwens}
  et~al.}{2015}]{bouwens2015a}
{Bouwens} R.~J.  et~al., 2015, \apj, 803, 34

\bibitem[\protect\citeauthoryear{{Caputi}, {Cirasuolo}, {Dunlop}, {McLure},
  {Farrah} \& {Almaini}}{{Caputi} et~al.}{2011}]{caputi2011a}
{Caputi} K.~I.,  {Cirasuolo} M.,  {Dunlop} J.~S.,  {McLure} R.~J.,  {Farrah}
  D.,    {Almaini} O.,  2011, \mnras, 413, 162

\bibitem[\protect\citeauthoryear{{Cassata} et~al.,}{{Cassata}
  et~al.}{2011}]{cassata2011a}
{Cassata} P.  et~al., 2011, \aap, 525, A143

\bibitem[\protect\citeauthoryear{{Cattaneo}, {Dekel}, {Devriendt}, {Guiderdoni}
  \& {Blaizot}}{{Cattaneo} et~al.}{2006}]{cattaneo2006a}
{Cattaneo} A.,  {Dekel} A.,  {Devriendt} J.,  {Guiderdoni} B.,    {Blaizot} J.,
   2006, \mnras, 370, 1651

\bibitem[\protect\citeauthoryear{{Davis}, {Efstathiou}, {Frenk} \&
  {White}}{{Davis} et~al.}{1985}]{davis85}
{Davis} M.,  {Efstathiou} G.,  {Frenk} C.~S.,    {White} S.~D.~M.,  1985, \apj,
  292, 371

\bibitem[\protect\citeauthoryear{{Dayal}, {Ferrara} \& {Gallerani}}{{Dayal}
  et~al.}{2008}]{dayal08}
{Dayal} P.,  {Ferrara} A.,    {Gallerani} S.,  2008, \mnras, 389, 1683

\bibitem[\protect\citeauthoryear{{Dayal}, {Ferrara} \& {Saro}}{{Dayal}
  et~al.}{2010}]{dayal10b}
{Dayal} P.,  {Ferrara} A.,    {Saro} A.,  2010, \mnras, 402, 1449

\bibitem[\protect\citeauthoryear{{Dayal} \& {Libeskind}}{{Dayal} \&
  {Libeskind}}{2012}]{dayal2012a}
{Dayal} P.,  {Libeskind} N.~I.,  2012, \mnras, 419, L9

\bibitem[\protect\citeauthoryear{{Dayal}, {Maselli} \& {Ferrara}}{{Dayal}
  et~al.}{2011}]{dayal2011a}
{Dayal} P.,  {Maselli} A.,    {Ferrara} A.,  2011, \mnras, 410, 830

\bibitem[\protect\citeauthoryear{{Dekel} et~al.,}{{Dekel}
  et~al.}{2009}]{dekel_nature}
{Dekel} A.  et~al., 2009, \nat, 457, 451

\bibitem[\protect\citeauthoryear{{Devriendt}, {Guiderdoni} \&
  {Sadat}}{{Devriendt} et~al.}{1999}]{devriendt}
{Devriendt} J.~E.~G.,  {Guiderdoni} B.,    {Sadat} R.,  1999, \aap, 350, 381

\bibitem[\protect\citeauthoryear{{Dijkstra}, {Lidz} \& {Wyithe}}{{Dijkstra}
  et~al.}{2007}]{dijk07a}
{Dijkstra} M.,  {Lidz} A.,    {Wyithe} J.~S.~B.,  2007, \mnras, 377, 1175

\bibitem[\protect\citeauthoryear{{Dijkstra}, {Mesinger} \& {Wyithe}}{{Dijkstra}
  et~al.}{2011}]{dijkstra2011a}
{Dijkstra} M.,  {Mesinger} A.,    {Wyithe} J.~S.~B.,  2011, \mnras, 414, 2139

\bibitem[\protect\citeauthoryear{{Dom{\'{\i}}nguez S{\'a}nchez}
  et~al.,}{{Dom{\'{\i}}nguez S{\'a}nchez}
  et~al.}{2011}]{dominguez-sanchez2011a}
{Dom{\'{\i}}nguez S{\'a}nchez} H.  et~al., 2011, \mnras, 417, 900

\bibitem[\protect\citeauthoryear{{Dressler}, {Henry}, {Martin}, {Sawicki},
  {McCarthy} \& {Villaneuva}}{{Dressler} et~al.}{2015}]{dressler2015a}
{Dressler} A.,  {Henry} A.,  {Martin} C.~L.,  {Sawicki} M.,  {McCarthy} P.,
  {Villaneuva} E.,  2015, \apj, 806, 19

\bibitem[\protect\citeauthoryear{{Dressler}, {Martin}, {Henry}, {Sawicki} \&
  {McCarthy}}{{Dressler} et~al.}{2011}]{dressler2011a}
{Dressler} A.,  {Martin} C.~L.,  {Henry} A.,  {Sawicki} M.,    {McCarthy} P.,
  2011, \apj, 740, 71

\bibitem[\protect\citeauthoryear{{Duncan} et~al.,}{{Duncan}
  et~al.}{2014}]{duncan2014}
{Duncan} K.  et~al., 2014, \mnras, 444, 2960

\bibitem[\protect\citeauthoryear{{Efstathiou}}{{Efstathiou}}{1992}]{ef}
{Efstathiou} G.,  1992, \mnras, 256, 43P

\bibitem[\protect\citeauthoryear{{Elsner}, {Feulner} \& {Hopp}}{{Elsner}
  et~al.}{2008}]{elsner2008a}
{Elsner} F.,  {Feulner} G.,    {Hopp} U.,  2008, \aap, 477, 503

\bibitem[\protect\citeauthoryear{{Fontana} et~al.,}{{Fontana}
  et~al.}{2006}]{fontana2006a}
{Fontana} A.  et~al., 2006, \aap, 459, 745

\bibitem[\protect\citeauthoryear{{Forero-Romero}, {Yepes}, {Gottl{\"o}ber},
  {Knollmann}, {Cuesta} \& {Prada}}{{Forero-Romero}
  et~al.}{2011}]{forero-romero2011}
{Forero-Romero} J.~E.,  {Yepes} G.,  {Gottl{\"o}ber} S.,  {Knollmann} S.~R.,
  {Cuesta} A.~J.,    {Prada} F.,  2011, \mnras, 415, 3666

\bibitem[\protect\citeauthoryear{{Furlanetto}, {Zaldarriaga} \&
  {Hernquist}}{{Furlanetto} et~al.}{2006}]{furlanetto2006a}
{Furlanetto} S.~R.,  {Zaldarriaga} M.,    {Hernquist} L.,  2006, \mnras, 365,
  1012

\bibitem[\protect\citeauthoryear{{Garel}, {Blaizot}, {Guiderdoni},
  {Michel-Dansac}, {Hayes} \& {Verhamme}}{{Garel} et~al.}{2015}]{garel2015a}
{Garel} T.,  {Blaizot} J.,  {Guiderdoni} B.,  {Michel-Dansac} L.,  {Hayes} M.,
    {Verhamme} A.,  2015, \mnras, 450, 1279

\bibitem[\protect\citeauthoryear{{Garel}, {Blaizot}, {Guiderdoni}, {Schaerer},
  {Verhamme} \& {Hayes}}{{Garel} et~al.}{2012}]{garel2012a}
{Garel} T.,  {Blaizot} J.,  {Guiderdoni} B.,  {Schaerer} D.,  {Verhamme} A.,
  {Hayes} M.,  2012, \mnras, 422, 310

\bibitem[\protect\citeauthoryear{{Gawiser} et~al.,}{{Gawiser}
  et~al.}{2007}]{gawiser2007a}
{Gawiser} E.  et~al., 2007, \apj, 671, 278

\bibitem[\protect\citeauthoryear{{Gonz{\'a}lez}, {Lacey}, {Baugh}, {Frenk} \&
  {Benson}}{{Gonz{\'a}lez} et~al.}{2012}]{gonzalez2012}
{Gonz{\'a}lez} J.~E.,  {Lacey} C.~G.,  {Baugh} C.~M.,  {Frenk} C.~S.,
  {Benson} A.~J.,  2012, \mnras, 423, 3709

\bibitem[\protect\citeauthoryear{{Gonz{\'a}lez}, {Labb{\'e}}, {Bouwens},
  {Illingworth}, {Franx} \& {Kriek}}{{Gonz{\'a}lez}
  et~al.}{2011}]{gonzalez2011a}
{Gonz{\'a}lez} V.,  {Labb{\'e}} I.,  {Bouwens} R.~J.,  {Illingworth} G.,
  {Franx} M.,    {Kriek} M.,  2011, \apjl, 735, L34

\bibitem[\protect\citeauthoryear{{Gronwall} et~al.,}{{Gronwall}
  et~al.}{2007}]{gronwall07}
{Gronwall} C.  et~al., 2007, \apj, 667, 79

\bibitem[\protect\citeauthoryear{{Haiman} \& {Spaans}}{{Haiman} \&
  {Spaans}}{1999}]{haiman99}
{Haiman} Z.,  {Spaans} M.,  1999, \apj, 518, 138

\bibitem[\protect\citeauthoryear{{Hatton}, {Devriendt}, {Ninin}, {Bouchet},
  {Guiderdoni} \& {Vibert}}{{Hatton} et~al.}{2003}]{hatton}
{Hatton} S.,  {Devriendt} J.~E.~G.,  {Ninin} S.,  {Bouchet} F.~R.,
  {Guiderdoni} B.,    {Vibert} D.,  2003, \mnras, 343, 75

\bibitem[\protect\citeauthoryear{{Hayashino} et~al.,}{{Hayashino}
  et~al.}{2004}]{hayashino2004a}
{Hayashino} T.  et~al., 2004, \aj, 128, 2073

\bibitem[\protect\citeauthoryear{{Hill} et~al.,}{{Hill}
  et~al.}{2008}]{hill2008a}
{Hill} G.~J.  et~al., 2008, in {Kodama} T.,  {Yamada} T.,   {Aoki} K.,  eds,
  Astronomical Society of the Pacific Conference Series Vol. 399, Panoramic
  Views of Galaxy Formation and Evolution. p.~115, \eprint{0806.0183}

\bibitem[\protect\citeauthoryear{{Hu}, {Cowie}, {Barger}, {Capak}, {Kakazu} \&
  {Trouille}}{{Hu} et~al.}{2010}]{hu2010a}
{Hu} E.~M.,  {Cowie} L.~L.,  {Barger} A.~J.,  {Capak} P.,  {Kakazu} Y.,
  {Trouille} L.,  2010, \apj, 725, 394

\bibitem[\protect\citeauthoryear{{Hu}, {Cowie} \& {McMahon}}{{Hu}
  et~al.}{1998}]{hu98}
{Hu} E.~M.,  {Cowie} L.~L.,    {McMahon} R.~G.,  1998, \apjl, 502, L99+

\bibitem[\protect\citeauthoryear{{Hutter}, {Dayal} \& {M{\"u}ller}}{{Hutter}
  et~al.}{2015}]{hutter2015a}
{Hutter} A.,  {Dayal} P.,    {M{\"u}ller} V.,  2015, \mnras, 450, 4025

\bibitem[\protect\citeauthoryear{{Iliev}, {Shapiro}, {McDonald}, {Mellema} \&
  {Pen}}{{Iliev} et~al.}{2008}]{iliev2008a}
{Iliev} I.~T.,  {Shapiro} P.~R.,  {McDonald} P.,  {Mellema} G.,    {Pen} U.-L.,
   2008, \mnras, 391, 63

\bibitem[\protect\citeauthoryear{{Inoue}, {Shimizu}, {Iwata} \&
  {Tanaka}}{{Inoue} et~al.}{2014}]{inoue2014a}
{Inoue} A.~K.,  {Shimizu} I.,  {Iwata} I.,    {Tanaka} M.,  2014, \mnras, 442,
  1805

\bibitem[\protect\citeauthoryear{{Jensen}, {Laursen}, {Mellema}, {Iliev},
  {Sommer-Larsen} \& {Shapiro}}{{Jensen} et~al.}{2013}]{jensen2013a}
{Jensen} H.,  {Laursen} P.,  {Mellema} G.,  {Iliev} I.~T.,  {Sommer-Larsen} J.,
     {Shapiro} P.~R.,  2013, \mnras, 428, 1366

\bibitem[\protect\citeauthoryear{{Jose}, {Srianand} \& {Subramanian}}{{Jose}
  et~al.}{2013}]{jose2013}
{Jose} C.,  {Srianand} R.,    {Subramanian} K.,  2013, \mnras, 435, 368

\bibitem[\protect\citeauthoryear{{Kafle}, {Sharma}, {Lewis} \&
  {Bland-Hawthorn}}{{Kafle} et~al.}{2014}]{kafle2014}
{Kafle} P.~R.,  {Sharma} S.,  {Lewis} G.~F.,    {Bland-Hawthorn} J.,  2014,
  \apj, 794, 59

\bibitem[\protect\citeauthoryear{{Kajisawa} et~al.,}{{Kajisawa}
  et~al.}{2009}]{kajisawa2009a}
{Kajisawa} M.  et~al., 2009, \apj, 702, 1393

\bibitem[\protect\citeauthoryear{{Kajisawa}, {Ichikawa}, {Yamada}, {Uchimoto},
  {Yoshikawa}, {Akiyama} \& {Onodera}}{{Kajisawa} et~al.}{2010}]{kajisawa2010a}
{Kajisawa} M.,  {Ichikawa} T.,  {Yamada} T.,  {Uchimoto} Y.~K.,  {Yoshikawa}
  T.,  {Akiyama} M.,    {Onodera} M.,  2010, \apj, 723, 129

\bibitem[\protect\citeauthoryear{{Kashikawa} et~al.,}{{Kashikawa}
  et~al.}{2011}]{kashikawa2011a}
{Kashikawa} N.  et~al., 2011, \apj, 734, 119

\bibitem[\protect\citeauthoryear{{Kennicutt} Jr.}{{Kennicutt}}{1983}]{kenn83}
{Kennicutt} Jr. R.~C.,  1983, \apj, 272, 54

\bibitem[\protect\citeauthoryear{{Kennicutt}
  Jr.}{{Kennicutt}}{1998}]{kennicutt98}
{Kennicutt} Jr. R.~C.,  1998, \apj, 498, 541

\bibitem[\protect\citeauthoryear{{Kobayashi}, {Totani} \&
  {Nagashima}}{{Kobayashi} et~al.}{2007}]{koba07}
{Kobayashi} M.~A.~R.,  {Totani} T.,    {Nagashima} M.,  2007, \apj, 670, 919

\bibitem[\protect\citeauthoryear{{Kobayashi}, {Totani} \&
  {Nagashima}}{{Kobayashi} et~al.}{2010}]{koba10}
{Kobayashi} M.~A.~R.,  {Totani} T.,    {Nagashima} M.,  2010, \apj, 708, 1119

\bibitem[\protect\citeauthoryear{{Komatsu} et~al.,}{{Komatsu}
  et~al.}{2009}]{komatsu09}
{Komatsu} E.  et~al., 2009, \apjs, 180, 330

\bibitem[\protect\citeauthoryear{{Kova{\v c}}, {Somerville}, {Rhoads},
  {Malhotra} \& {Wang}}{{Kova{\v c}} et~al.}{2007}]{kovac2007a}
{Kova{\v c}} K.,  {Somerville} R.~S.,  {Rhoads} J.~E.,  {Malhotra} S.,
  {Wang} J.,  2007, \apj, 668, 15

\bibitem[\protect\citeauthoryear{{Kudritzki} et~al.,}{{Kudritzki}
  et~al.}{2000}]{kud00}
{Kudritzki} R.  et~al., 2000, \apj, 536, 19

\bibitem[\protect\citeauthoryear{{Kunth}, {Mas-Hesse}, {Terlevich},
  {Terlevich}, {Lequeux} \& {Fall}}{{Kunth} et~al.}{1998}]{kunth98}
{Kunth} D.,  {Mas-Hesse} J.~M.,  {Terlevich} E.,  {Terlevich} R.,  {Lequeux}
  J.,    {Fall} S.~M.,  1998, \aap, 334, 11

\bibitem[\protect\citeauthoryear{{Lanzoni}, {Guiderdoni}, {Mamon}, {Devriendt}
  \& {Hatton}}{{Lanzoni} et~al.}{2005}]{lanzoni2005}
{Lanzoni} B.,  {Guiderdoni} B.,  {Mamon} G.~A.,  {Devriendt} J.,    {Hatton}
  S.,  2005, \mnras, 361, 369

\bibitem[\protect\citeauthoryear{{Laursen}, {Sommer-Larsen} \&
  {Andersen}}{{Laursen} et~al.}{2009}]{laursen09}
{Laursen} P.,  {Sommer-Larsen} J.,    {Andersen} A.~C.,  2009, \apj, 704, 1640

\bibitem[\protect\citeauthoryear{{Laursen}, {Sommer-Larsen} \&
  {Razoumov}}{{Laursen} et~al.}{2011}]{laursen2011a}
{Laursen} P.,  {Sommer-Larsen} J.,    {Razoumov} A.~O.,  2011, \apj, 728, 52

\bibitem[\protect\citeauthoryear{{Le Delliou}, {Lacey}, {Baugh}, {Guiderdoni},
  {Bacon}, {Courtois}, {Sousbie} \& {Morris}}{{Le Delliou}
  et~al.}{2005}]{le-delliou2005a}
{Le Delliou} M.,  {Lacey} C.,  {Baugh} C.~M.,  {Guiderdoni} B.,  {Bacon} R.,
  {Courtois} H.,  {Sousbie} T.,    {Morris} S.~L.,  2005, \mnras, 357, L11

\bibitem[\protect\citeauthoryear{{Le F{\`e}vre} et~al.,}{{Le F{\`e}vre}
  et~al.}{2015}]{le-fevre2015a}
{Le F{\`e}vre} O.  et~al., 2015, \aap, 576, A79

\bibitem[\protect\citeauthoryear{{Madau}}{{Madau}}{1995}]{madau}
{Madau} P.,  1995, \apj, 441, 18

\bibitem[\protect\citeauthoryear{{Malhotra} \& {Rhoads}}{{Malhotra} \&
  {Rhoads}}{2004}]{malhotra04}
{Malhotra} S.,  {Rhoads} J.~E.,  2004, \apjl, 617, L5

\bibitem[\protect\citeauthoryear{{Mandelbaum}, {Seljak}, {Kauffmann}, {Hirata}
  \& {Brinkmann}}{{Mandelbaum} et~al.}{2006}]{mandelbaum2006}
{Mandelbaum} R.,  {Seljak} U.,  {Kauffmann} G.,  {Hirata} C.~M.,    {Brinkmann}
  J.,  2006, \mnras, 368, 715

\bibitem[\protect\citeauthoryear{{Marchesini}, {van Dokkum}, {F{\"o}rster
  Schreiber}, {Franx}, {Labb{\'e}} \& {Wuyts}}{{Marchesini}
  et~al.}{2009}]{marchesini2009a}
{Marchesini} D.,  {van Dokkum} P.~G.,  {F{\"o}rster Schreiber} N.~M.,  {Franx}
  M.,  {Labb{\'e}} I.,    {Wuyts} S.,  2009, \apj, 701, 1765

\bibitem[\protect\citeauthoryear{{Mas-Hesse}, {Kunth}, {Tenorio-Tagle},
  {Leitherer}, {Terlevich} \& {Terlevich}}{{Mas-Hesse}
  et~al.}{2003}]{mas-hesse2003a}
{Mas-Hesse} J.~M.,  {Kunth} D.,  {Tenorio-Tagle} G.,  {Leitherer} C.,
  {Terlevich} R.~J.,    {Terlevich} E.,  2003, \apj, 598, 858

\bibitem[\protect\citeauthoryear{{McLinden}, {Finkelstein}, {Rhoads},
  {Malhotra}, {Hibon} \& {Richardson}}{{McLinden} et~al.}{2011}]{mclinden}
{McLinden} E.,  {Finkelstein} S.~L.,  {Rhoads} J.~E.,  {Malhotra} S.,  {Hibon}
  P.,    {Richardson} M.,  2011, in Bulletin of the American Astronomical
  Society. pp 33543--+

\bibitem[\protect\citeauthoryear{{McMillan}}{{McMillan}}{2011}]{mcmillan2011}
{McMillan} P.~J.,  2011, \mnras, 414, 2446

\bibitem[\protect\citeauthoryear{{McQuinn}, {Hernquist}, {Zaldarriaga} \&
  {Dutta}}{{McQuinn} et~al.}{2007}]{mcquinn2007a}
{McQuinn} M.,  {Hernquist} L.,  {Zaldarriaga} M.,    {Dutta} S.,  2007, \mnras,
  381, 75

\bibitem[\protect\citeauthoryear{{McQuinn}, {Lidz}, {Zahn}, {Dutta},
  {Hernquist} \& {Zaldarriaga}}{{McQuinn} et~al.}{2007}]{mcquinn2007b}
{McQuinn} M.,  {Lidz} A.,  {Zahn} O.,  {Dutta} S.,  {Hernquist} L.,
  {Zaldarriaga} M.,  2007, \mnras, 377, 1043

\bibitem[\protect\citeauthoryear{{Moster}, {Somerville}, {Newman} \&
  {Rix}}{{Moster} et~al.}{2011}]{moster11}
{Moster} B.~P.,  {Somerville} R.~S.,  {Newman} J.~A.,    {Rix} H.-W.,  2011,
  \apj, 731, 113

\bibitem[\protect\citeauthoryear{{Murayama} et~al.,}{{Murayama}
  et~al.}{2007}]{murayama07}
{Murayama} T.  et~al., 2007, \apjs, 172, 523

\bibitem[\protect\citeauthoryear{{Nagamine}, {Ouchi}, {Springel} \&
  {Hernquist}}{{Nagamine} et~al.}{2010}]{nag}
{Nagamine} K.,  {Ouchi} M.,  {Springel} V.,    {Hernquist} L.,  2010, \pasj,
  62, 1455

\bibitem[\protect\citeauthoryear{{Nakamura}, {Inoue}, {Hayashino}, {Horie},
  {Kousai}, {Fujii} \& {Matsuda}}{{Nakamura} et~al.}{2011}]{nakamura2011a}
{Nakamura} E.,  {Inoue} A.~K.,  {Hayashino} T.,  {Horie} M.,  {Kousai} K.,
  {Fujii} T.,    {Matsuda} Y.,  2011, \mnras, 412, 2579

\bibitem[\protect\citeauthoryear{{Neufeld}}{{Neufeld}}{1990}]{neufeld1990a}
{Neufeld} D.~A.,  1990, \apj, 350, 216

\bibitem[\protect\citeauthoryear{{Ocvirk}, {Pichon} \& {Teyssier}}{{Ocvirk}
  et~al.}{2008}]{ocvirk08}
{Ocvirk} P.,  {Pichon} C.,    {Teyssier} R.,  2008, \mnras, 390, 1326

\bibitem[\protect\citeauthoryear{{Okamoto}, {Gao} \& {Theuns}}{{Okamoto}
  et~al.}{2008}]{okamoto08}
{Okamoto} T.,  {Gao} L.,    {Theuns} T.,  2008, \mnras, 390, 920

\bibitem[\protect\citeauthoryear{{Orsi}, {Lacey} \& {Baugh}}{{Orsi}
  et~al.}{2012}]{orsi2012a}
{Orsi} A.,  {Lacey} C.~G.,    {Baugh} C.~M.,  2012, \mnras, 425, 87

\bibitem[\protect\citeauthoryear{{Osterbrock} \& {Ferland}}{{Osterbrock} \&
  {Ferland}}{2006}]{osterbrock2006}
{Osterbrock} D.~E.,  {Ferland} G.~J.,  2006, {Astrophysics of gaseous nebulae
  and active galactic nuclei}

\bibitem[\protect\citeauthoryear{{Ouchi} et~al.,}{{Ouchi}
  et~al.}{2008}]{ouch08}
{Ouchi} M.  et~al., 2008, \apjs, 176, 301

\bibitem[\protect\citeauthoryear{{Ouchi} et~al.,}{{Ouchi}
  et~al.}{2003}]{ouch03}
{Ouchi} M.  et~al., 2003, \apj, 582, 60

\bibitem[\protect\citeauthoryear{{Ouchi} et~al.,}{{Ouchi}
  et~al.}{2010}]{ouchi2010a}
{Ouchi} M.  et~al., 2010, \apj, 723, 869

\bibitem[\protect\citeauthoryear{{Pentericci}, {Grazian}, {Fontana},
  {Salimbeni}, {Santini}, {de Santis}, {Gallozzi} \& {Giallongo}}{{Pentericci}
  et~al.}{2007}]{pentericci2007a}
{Pentericci} L.,  {Grazian} A.,  {Fontana} A.,  {Salimbeni} S.,  {Santini} P.,
  {de Santis} C.,  {Gallozzi} S.,    {Giallongo} E.,  2007, \aap, 471, 433

\bibitem[\protect\citeauthoryear{{P{\'e}rez-Gonz{\'a}lez}
  et~al.,}{{P{\'e}rez-Gonz{\'a}lez} et~al.}{2008}]{perez-gonzalez2008a}
{P{\'e}rez-Gonz{\'a}lez} P.~G.  et~al., 2008, \apj, 675, 234

\bibitem[\protect\citeauthoryear{{Phelps}, {Nusser} \& {Desjacques}}{{Phelps}
  et~al.}{2013}]{phelps2013}
{Phelps} S.,  {Nusser} A.,    {Desjacques} V.,  2013, \apj, 775, 102

\bibitem[\protect\citeauthoryear{{Rauch} et~al.,}{{Rauch}
  et~al.}{2008}]{rauch08}
{Rauch} M.  et~al., 2008, \apj, 681, 856

\bibitem[\protect\citeauthoryear{{Reddy}, {Steidel}, {Pettini}, {Adelberger},
  {Shapley}, {Erb} \& {Dickinson}}{{Reddy} et~al.}{2008}]{reddy08}
{Reddy} N.~A.,  {Steidel} C.~C.,  {Pettini} M.,  {Adelberger} K.~L.,  {Shapley}
  A.~E.,  {Erb} D.~K.,    {Dickinson} M.,  2008, \apjs, 175, 48

\bibitem[\protect\citeauthoryear{{Rhoads}, {Malhotra}, {Dey}, {Stern},
  {Spinrad} \& {Jannuzi}}{{Rhoads} et~al.}{2000}]{rhoads00}
{Rhoads} J.~E.,  {Malhotra} S.,  {Dey} A.,  {Stern} D.,  {Spinrad} H.,
  {Jannuzi} B.~T.,  2000, \apjl, 545, L85

\bibitem[\protect\citeauthoryear{{Rivera-Thorsen} et~al.,}{{Rivera-Thorsen}
  et~al.}{2015}]{rivera-thorsen2015a}
{Rivera-Thorsen} T.~E.  et~al., 2015, \apj, 805, 14

\bibitem[\protect\citeauthoryear{{Salmon} et~al.,}{{Salmon}
  et~al.}{2015}]{salmon2015a}
{Salmon} B.  et~al., 2015, \apj, 799, 183

\bibitem[\protect\citeauthoryear{{Salvadori}, {Dayal} \& {Ferrara}}{{Salvadori}
  et~al.}{2010}]{salvadori2010}
{Salvadori} S.,  {Dayal} P.,    {Ferrara} A.,  2010, \mnras, 407, L1

\bibitem[\protect\citeauthoryear{{Santos}}{{Santos}}{2004}]{santos04}
{Santos} M.~R.,  2004, \mnras, 349, 1137

\bibitem[\protect\citeauthoryear{{Sawicki} et~al.,}{{Sawicki}
  et~al.}{2008}]{sawicki2008a}
{Sawicki} M.  et~al., 2008, \apj, 687, 884

\bibitem[\protect\citeauthoryear{{Sawicki} \& {Thompson}}{{Sawicki} \&
  {Thompson}}{2006}]{sawicki}
{Sawicki} M.,  {Thompson} D.,  2006, \apj, 648, 299

\bibitem[\protect\citeauthoryear{{Schaerer}, {de Barros} \&
  {Sklias}}{{Schaerer} et~al.}{2013}]{schaerer2013a}
{Schaerer} D.,  {de Barros} S.,    {Sklias} P.,  2013, \aap, 549, A4

\bibitem[\protect\citeauthoryear{{Schaerer}, {Hayes}, {Verhamme} \&
  {Teyssier}}{{Schaerer} et~al.}{2011}]{schaerer2011a}
{Schaerer} D.,  {Hayes} M.,  {Verhamme} A.,    {Teyssier} R.,  2011, \aap, 531,
  A12

\bibitem[\protect\citeauthoryear{{Shapley}, {Steidel}, {Adelberger},
  {Dickinson}, {Giavalisco} \& {Pettini}}{{Shapley}
  et~al.}{2001}]{shapley2001a}
{Shapley} A.~E.,  {Steidel} C.~C.,  {Adelberger} K.~L.,  {Dickinson} M.,
  {Giavalisco} M.,    {Pettini} M.,  2001, \apj, 562, 95

\bibitem[\protect\citeauthoryear{{Shapley}, {Steidel}, {Pettini} \&
  {Adelberger}}{{Shapley} et~al.}{2003}]{shapley03}
{Shapley} A.~E.,  {Steidel} C.~C.,  {Pettini} M.,    {Adelberger} K.~L.,  2003,
  \apj, 588, 65

\bibitem[\protect\citeauthoryear{{Shimasaku} et~al.,}{{Shimasaku}
  et~al.}{2006}]{shima06}
{Shimasaku} K.  et~al., 2006, \pasj, 58, 313

\bibitem[\protect\citeauthoryear{{Shimizu}, {Yoshida} \& {Okamoto}}{{Shimizu}
  et~al.}{2011}]{shimizu2011a}
{Shimizu} I.,  {Yoshida} N.,    {Okamoto} T.,  2011, \mnras, 418, 2273

\bibitem[\protect\citeauthoryear{{Shioya} et~al.,}{{Shioya}
  et~al.}{2009}]{shioya}
{Shioya} Y.  et~al., 2009, \apj, 696, 546

\bibitem[\protect\citeauthoryear{{Silk}}{{Silk}}{2003}]{silk03}
{Silk} J.,  2003, \mnras, 343, 249

\bibitem[\protect\citeauthoryear{{Song} et~al.,}{{Song}
  et~al.}{2015}]{song2015a}
{Song} M.  et~al., 2015, ArXiv e-prints, (arXiv:1507.05636)

\bibitem[\protect\citeauthoryear{{Springel}}{{Springel}}{2005}]{gadget2}
{Springel} V.,  2005, \mnras, 364, 1105

\bibitem[\protect\citeauthoryear{{Stark}, {Schenker}, {Ellis}, {Robertson},
  {McLure} \& {Dunlop}}{{Stark} et~al.}{2013}]{stark2013a}
{Stark} D.~P.,  {Schenker} M.~A.,  {Ellis} R.,  {Robertson} B.,  {McLure} R.,
   {Dunlop} J.,  2013, \apj, 763, 129

\bibitem[\protect\citeauthoryear{{Steidel}, {Erb}, {Shapley}, {Pettini},
  {Reddy}, {Bogosavljevi{\'c}}, {Rudie} \& {Rakic}}{{Steidel}
  et~al.}{2010}]{steidel2010a}
{Steidel} C.~C.,  {Erb} D.~K.,  {Shapley} A.~E.,  {Pettini} M.,  {Reddy} N.,
  {Bogosavljevi{\'c}} M.,  {Rudie} G.~C.,    {Rakic} O.,  2010, \apj, 717, 289

\bibitem[\protect\citeauthoryear{{Tenorio-Tagle}, {Silich}, {Kunth},
  {Terlevich} \& {Terlevich}}{{Tenorio-Tagle}
  et~al.}{1999}]{tenorio-tagle1999a}
{Tenorio-Tagle} G.,  {Silich} S.~A.,  {Kunth} D.,  {Terlevich} E.,
  {Terlevich} R.,  1999, \mnras, 309, 332

\bibitem[\protect\citeauthoryear{{Tweed}, {Devriendt}, {Blaizot}, {Colombi} \&
  {Slyz}}{{Tweed} et~al.}{2009}]{tweed09}
{Tweed} D.,  {Devriendt} J.,  {Blaizot} J.,  {Colombi} S.,    {Slyz} A.,  2009,
  \aap, 506, 647

\bibitem[\protect\citeauthoryear{{van Breukelen}, {Jarvis} \& {Venemans}}{{van
  Breukelen} et~al.}{2005}]{vanb05}
{van Breukelen} C.,  {Jarvis} M.~J.,    {Venemans} B.~P.,  2005, \mnras, 359,
  895

\bibitem[\protect\citeauthoryear{{van den Bosch}, {Yang} \& {Mo}}{{van den
  Bosch} et~al.}{2003}]{van-den-bosch2003}
{van den Bosch} F.~C.,  {Yang} X.,    {Mo} H.~J.,  2003, \mnras, 340, 771

\bibitem[\protect\citeauthoryear{{van der Burg}, {Hildebrandt} \& {Erben}}{{van
  der Burg} et~al.}{2010}]{van-der-burg2010}
{van der Burg} R.~F.~J.,  {Hildebrandt} H.,    {Erben} T.,  2010, \aap, 523,
  A74

\bibitem[\protect\citeauthoryear{{Verhamme}, {Dubois}, {Blaizot}, {Garel},
  {Bacon}, {Devriendt}, {Guiderdoni} \& {Slyz}}{{Verhamme}
  et~al.}{2012}]{verhamme2012}
{Verhamme} A.,  {Dubois} Y.,  {Blaizot} J.,  {Garel} T.,  {Bacon} R.,
  {Devriendt} J.,  {Guiderdoni} B.,    {Slyz} A.,  2012, \aap, 546, A111

\bibitem[\protect\citeauthoryear{{Verhamme}, {Schaerer}, {Atek} \&
  {Tapken}}{{Verhamme} et~al.}{2008}]{verh08}
{Verhamme} A.,  {Schaerer} D.,  {Atek} H.,    {Tapken} C.,  2008, \aap, 491, 89

\bibitem[\protect\citeauthoryear{{Verhamme}, {Schaerer} \&
  {Maselli}}{{Verhamme} et~al.}{2006}]{verh06}
{Verhamme} A.,  {Schaerer} D.,    {Maselli} A.,  2006, \aap, 460, 397

\bibitem[\protect\citeauthoryear{{Walker-Soler}, {Gawiser}, {Bond}, {Padilla}
  \& {Francke}}{{Walker-Soler} et~al.}{2012}]{walker-soler2012}
{Walker-Soler} J.~P.,  {Gawiser} E.,  {Bond} N.~A.,  {Padilla} N.,    {Francke}
  H.,  2012, \apj, 752, 160

\bibitem[\protect\citeauthoryear{{Wilkins}, {Gonzalez-Perez}, {Lacey} \&
  {Baugh}}{{Wilkins} et~al.}{2012}]{wilkins2012a}
{Wilkins} S.~M.,  {Gonzalez-Perez} V.,  {Lacey} C.~G.,    {Baugh} C.~M.,  2012,
  \mnras, 427, 1490

\bibitem[\protect\citeauthoryear{{Wofford}, {Leitherer} \& {Salzer}}{{Wofford}
  et~al.}{2013}]{wofford2013a}
{Wofford} A.,  {Leitherer} C.,    {Salzer} J.,  2013, \apj, 765, 118

\bibitem[\protect\citeauthoryear{{Yajima}, {Li}, {Zhu} \& {Abel}}{{Yajima}
  et~al.}{2012}]{yajima2012a}
{Yajima} H.,  {Li} Y.,  {Zhu} Q.,    {Abel} T.,  2012, \mnras, 424, 884

\bibitem[\protect\citeauthoryear{{Yajima}, {Li}, {Zhu}, {Abel}, {Gronwall} \&
  {Ciardullo}}{{Yajima} et~al.}{2012}]{yajima2012}
{Yajima} H.,  {Li} Y.,  {Zhu} Q.,  {Abel} T.,  {Gronwall} C.,    {Ciardullo}
  R.,  2012, \apj, 754, 118

\bibitem[\protect\citeauthoryear{{Yamada}, {Nakamura}, {Matsuda}, {Hayashino},
  {Yamauchi}, {Morimoto}, {Kousai} \& {Umemura}}{{Yamada}
  et~al.}{2012}]{yamada2012}
{Yamada} T.,  {Nakamura} Y.,  {Matsuda} Y.,  {Hayashino} T.,  {Yamauchi} R.,
  {Morimoto} N.,  {Kousai} K.,    {Umemura} M.,  2012, \aj, 143, 79

\bibitem[\protect\citeauthoryear{{Yang}, {Mo} \& {van den Bosch}}{{Yang}
  et~al.}{2009}]{yang2009}
{Yang} X.,  {Mo} H.~J.,    {van den Bosch} F.~C.,  2009, \apj, 695, 900

\bibitem[\protect\citeauthoryear{{Zheng}, {Cen}, {Trac} \&
  {Miralda-Escud{\'e}}}{{Zheng} et~al.}{2010}]{zheng2010a}
{Zheng} Z.,  {Cen} R.,  {Trac} H.,    {Miralda-Escud{\'e}} J.,  2010, \apj,
  716, 574

\end{thebibliography}

\label{lastpage}

\end{document}